\documentclass[10pt]{ubthesis}
\usepackage{amsmath}
\usepackage{subfigure}
\usepackage{lscape}
\usepackage{multirow}
\usepackage{multicol}
\usepackage[usenames]{color}
\usepackage{verbatim}
\usepackage{bbm}
\usepackage{graphicx}
\usepackage{amssymb}
\usepackage{amsbsy}
\newdimen\hbigcirc
\newdimen\wbigcirc

\title{Modified Supersymmetric Dark Sectors}

\author{Christopher Samuel Redino}

\conferral{May 14th, 2015}

\dept{Physics}
\degree{Doctor of Philosophy}

\begin{document}

\begin{titlepage}
\maketitle
\end{titlepage}

\begin{ubfrontmatter}
\makecopyright

    \newenvironment{dedication}
        {\vspace{6ex}\begin{center}}
        {\par\end{center}}
    \begin{dedication}
    This thesis would not exist without the support of my advisor, Doreen Wackeroth. She allowed me the freedom to pursue my own research interests from the start, even though the subject matter was a bit outside her usual domain. We both learned a great deal together through the course of this work, and she always showed faith in me, even when I was most uncertain about the eventual outcome of my efforts. The rest of the high energy physics and cosmology group at UB has greatly influenced my development as a physicist, leading towards this thesis. Discussions with Will Kinney, Dejan Stojkovic, and Sal Rappochio have helped me to focus my attitude and attention as a scientist, in how to choose interesting problems to study, how to approach them in an objective way, and also to let go of our fantastic ideas when they just don't fit the data.
I must thank Howard Baer, whose name appears in my references more than any other author and who also contributed valuable input through our conversations. I  also would like to thank Ian Woo Kim for his assistance with evchain and Benjamin Fuks for his assistance with Feynrules.

    \end{dedication}

\tableofcontents
\listoffigures

\begin{abstract}
SUSY models with a modified dark sector require constraints to be
reinterpreted, which may allow for scenarios with low tuning. A modified
dark sector can also change the phenomenology greatly. The addition
of the QCD axion to the Minimal Supersymmetric Standard Model (MSSM) 
solves the strong CP problem and also
modifies the dark sector with new dark matter candidates. While SUSY
axion phenomenology is usually restricted to searches for the axion itself
or searches for the ordinary SUSY particles, this work focuses on
scenarios where the axion's superpartner, the axino may be detectable
at the Large Hadron Collider (LHC) in the decays of neutralinos displaced from the primary
vertex. In particular this work focuses on the KSVZ axino. The decay
length of neutralinos in this scenario easily fits the ATLAS detector
for SUSY spectra expected to be testable at the 14 TeV LHC.
This signature of displaced decays to axinos is compared to other
well motivated scenarios containing a long lived neutralino which
decays inside the detector. These alternative scenarios can in some
cases very closely mimic the expected axino signature, and the degree
to which they are distinguishable is discussed. The cosmological viability of such a scenario is also considered briefly.
\end{abstract}

\end{ubfrontmatter}

\begin{ubmainmatter}
\chapter{Introduction}
\label{ch:intro}

	Aside from gravity, the known forces of nature are described by the Standard Model (see \cite{Herrero:1998eq} for a review), which was developed over time by various theorists and the work of a great many experimentalists.  Before we can hope to push the boundaries of this understanding, we must take stock of what knowledge we have.
 The Standard Model is a quantum field theory (QFT). A quantum field theory is a generalization of quantum mechanics that incorporates the effects of special relativity. The quantization of fields, rather than just the states of particles, allows for the creation and annihilation of particles, which is the basic principal on which collider experiments are based on. The creation of new particles in collider experiments drove the development of the Standard Model.  The reason that the Standard Model is “just” an example of a QFT is that quantum field theory in general describes the propagation and interaction of various fields of integer or half integer spin. The fields of the Standard Model have the further requirement that they be renormalizable, that is to say, the quantum corrections to interactions and masses, when evolved to a very high scale should be convergent, which limits the types of operators allowed for interactions. The requirement of renormalizability also puts constraints on the types of spins a particle is allowed to have. Spins greater or equal to 2 generally lead to interactions that are not renormalizable in 3+1 dimensions, this is one of the difficulties in quantizing gravity, because we predict the graviton should have a spin of 2. Of course these requirements of renormalizability make assumptions about the overall theory, new physics at a higher scale can possible find ways around these constraints.
	Much of the Standard Model's predictive success comes from the way the interactions are organized under a guiding principle and that principle is the gauge invariance of fields.  The Standard Model Lagrangian respects gauge symmetries. The overall symmetry of the Standard Model is SU(3)$_c$ x SU(2)$_I$ x U(1)$_Y$, corresponding to the gauge groups for color ($c$), weak isospin ($I$) and hypercharge ($Y$).  The content of the Standard Model consists of the particles that constitute matter (in the form of fermions), and the particles that mediate the forces between them (in the form of vector bosons). These symmetries reflect the three forces of nature (apart from gravity). Electromagnetism is communicated via photons between particles with a U(1)$_{em}$ electric charge. The so called weak interaction which governs nuclear decays (and nuclear interactions in the sun) comes from the exchange of the W and Z bosons between particles with a SU(2)$_I$ weak isospin. The strong nuclear force, which holds together nuclear matter is derived from the interactions of vector gluons (octets under SU(3)$_c$ )and matter fields that have a SU(3)$_c$  color charge. The matter particles have different charges under these symmetries: there are quarks, which have charges under all three groups but are distinguished as being the only fermions with color charge (triplets under SU(3)$_c$ and there are also the leptons, which are singlets under SU(3)$_c$. The quarks come in two types, up and down, distinguished by their electric charge, with up types having +2/3 and down types having -1/3 in units of $e$, the charge of the electron. The leptons also come in two types. The charged leptons have +1 under U(1)$_{em}$ and their associated neutrinos are singlets under U(1)$_{em}$.  So the overall organization of the Standard Model fermions  in terms of gauge symmetries is  left handed triplets under SU(3)$_c$ and  left handed doublets under SU(2)$_I$. There are right handed quarks and leptons  in the Standard Model as well, associated with their left handed partners, but they are organized in electroweak singlets, as the weak interaction violates parity. All empirical evidence indicates that the W boson only interacts with the left handed particles, so that if there is a right handed neutrino then it is a ``sterile" particle with no weak interactions. The reason for this asymmetry is not predicted by the Standard Model, but is simply stated by it. In addition to this level of organization we find in nature that there are three generations of these left handed doublets and right handed singlets. So there are three up type quarks: up (u), charm (c) and top (t), and  three down types of quarks: down (d), strange (s) and bottom (b).  There are three types of electrons and three corresponding flavors of neutrinos: electron, muon and tau. The three generations for a given type all have the same charges and are only distinguished by mass. When an interaction can just as easily contain a substitute particle from a different generation, there is some observed preference to particles of a particular “flavor” and this preference is described by a mixing angle.  Again, the Standard Model cannot predict this structure of generations, but only describe it as observed. The values of the mixing angles themselves are not predicted by the Standard Model, they must be observed, such as the components of the CKM matrix which describes the relative strength of different flavor changing weak decays.  In addition to all of this structure in the Standard Model, every particle is also joined by its anti-particle with all its charges opposite.
	The last piece of the Standard Model, the Higgs boson came about from a naively incorrect prediction of the Standard Model. In the absence of the Higgs field, we would assume the W and Z bosons to be massless, because putting a mass term in ``by hand" would violate the SU(2)$_I$ symmetry of the Lagrangian. To remedy this, an additional complex scalar doublet is added to the Standard Model, the Higgs field.  The Higgs Lagrangian is symmetric under U(1)$_Y$ x SU(2)$_I$  but its ground state is not. When a ground state is chosen, the Higgs field undergoes spontaneous symmetry breaking. As a complex doublet the field has four degrees of freedom. Three of these degrees of freedom become the longitudinal components to the charged W$^{\pm}$ and Z bosons, that is to say the W and Z acquire mass through their interaction with the Higgs field. The Lagrangian after spontaneous symmetry breaking remains symmetric under U(1)$_{em}$  however, and so the photon remains massless and there is one degree of freedom left over which can be observed as a physical massive scalar, which we call the Higgs boson.  This process by which spontaneous symmetry breaking gives rise to masses for the W and Z bosons is called the Higgs mechanism.  The fermions of the Standard Model also naively have zero mass, but can acquire a mass through Yukawa couplings to the Higgs field. The quarks and charged leptons acquire their masses in this way, but in the Standard Model the neutrinos remain massless (though experiments lead us to believe they do indeed have mass). 
 We know that the Higgs boson itself is also massive, though the Standard Model does not predict a value for this mass. The value of the Higgs mass is of great importance because of its scalar nature. The Higgs boson is the first fundamental (i.e. not composite or a pseudo particle) particle to be discovered that is a scalar, and scalars are more sensitive to quantum corrections than fermions, which are protected by chiral symmetry, or vector bosons, which are protected by gauge symmetry. These corrections can be dangerous as the Higgs mass quickly becomes very large. The leading term of these correction is proportional to the square of a cut off scale where new physics enters and the Standard Model breaks down. If there is no other new physics then this scale is expected to be the Planck scale ($10^{19}$ GeV) when the effects of quantum gravity become relevant. The huge discrepancy in scales between gravity and the forces of the Standard Model is called the hierarchy problem. That the Higgs should somehow remain at a low mass in the presence of such possible corrections introduces a question of fine tuning. When considering the Standard Model alone this is not an issue, but if we are to believe there are any new particles between the weak scale and the Planck scale they should have a large effect on the Higgs mass unless there is some sort of cancellation between the corrections, either coincidently or guided by some symmetry principle in the new physics. Now that the Higgs have been discovered the Standard Model is complete and we must ask ourselves how to move forward. 
	
Particle physics is currently at a critical point. The Standard Model is now complete with the discovery of the Higgs boson \cite{Herrero:1998eq} \cite{Aad:2012tfa}, and we must decide as a community where to focus our efforts for the next steps.  The Higgs boson was the last piece to the puzzle we have been putting together for decades, but we know the final picture of the Standard Model is not a perfect reflection of nature. The shortcomings of the Standard Model are not  just simple  inaccuracies, or measurements requiring more precision, but also include very large unpredicted effects and particles that remain unaccounted for.  Some of these short comings, such as the unexplained existence of neutrino oscillations and by extension, unexplained neutrino mass are rather recent in the lifetime of particle physics as a human endeavor, and yet, other problems, like the existence of dark matter and its elusive particle identity are very nearly as old as the first particle colliders.
	 The progression of science necessarily requires wading into the unknown, but the next steps involve an uncertainty which particle physics has not known for several decades. We knew that to discover the mechanism of electroweak symmetry breaking, the searches would have to probe near the electroweak scale.  The last two particles discovered before the Higgs were likewise expected, the tau neutrino and the top quark to complete the generational structure observed in the Standard Model. Completing the Standard Model has been an effort guided by theory since the mid-1970s, but the successor to the Standard Model may not be as generous in providing hints. The issue of dark matter for example, must inevitably be confronted, as there is too much evidence on large scales for its existence, but its particle identity could accommodate current constraints with couplings and mass scales over several orders of magnitude. To put it bluntly, there are too many possibilities to be explored exhaustively. We know there is physics beyond the Standard Model, with dark matter being just one concrete example of it, but to proceed further, to find a place to begin our searches, we must rely on guiding principles which are perhaps less rigorous and more subjective in nature. These guiding principles, such as symmetry, unification, and naturalness have informed the progression of physics in the past, but are in no way guaranteed to be manifest in physics beyond the Standard Model. These qualities in a physical theory may be considered by some to be purely aesthetic, but a theory that has them has more explanatory power in a more economic way, and at the very least they provide a starting point for our searches beyond the Standard Model. 
	The most studied framework for new physics beyond the Standard Model that embodies these guiding principles is supersymmetry (SUSY). Supersymmetry remains one of the most popular frameworks for physics
beyond the Standard Model (SM) near the weak scale (see, e.g., \cite{Signer:2009dx,Martin:1997ns} for a review), motivated primarily
by naturalness and an apparent dark matter solution. 
The lightest neutralino in the Minimal Supersymmetric SM (MSSM) acting as a WIMP gives
approximately the correct relic abundance. The so-called ``WIMP miracle'' in SUSY is often
overstated however, in that there can still exist a tuning of SUSY
parameters in order to get the correct abundance and SUSY models with
the correct abundance only constitute a very small fraction of SUSY
model space, and models that over or under predict the dark matter (DM) are
common \cite{Hooper:2013qjx,Huang:2014xua,Baer:2012mv,Amsel:2011at}.
Simply by virtue of reducing the available parameter space, the
requisite of a dark matter solution can exclude more natural scenarios
in SUSY \cite{Baer:2012mv}. Models that do achieve the correct relic
abundance often do so at the cost of tuning, such as by having a very
massive Higgsino \cite{Hisano:2006nn} or a specific mixture of wino
and bino (the well tempered
neutralino) \cite{King:2006tf}. Furthermore, regardless of relic
abundance, when only concerned with detectability, in a generic
parameter space direct detection probes progressively more finely
tuned models \cite{Perelstein:2011tg,Amsel:2011at}. Co-annihilation
scenarios can provide long lasting holdouts in direct detection, but
to continue evading detection long enough this also introduces tuning \cite{Hooper:2013qjx}. It is important to note here that a conspiracy
of parameters to give the correct relic abundance (such as a
co-annihilation or a tempering effect) can introduce a tuning that is
separate from any consideration of electroweak naturalness
\cite{King:2006tf}. To summarize, the WIMP miracle is in general an
exaggeration when applied to SUSY.

The additional constraints and tuning introduced by accommodating dark
matter in a SUSY model can be avoided by extending the dark sector
beyond just neutralinos, but it is desirable to do this in a way that
is minimal and well motivated in of itself. Existing studies on
radiatively natural SUSY and on scanning the 19 parameter
phenomenological MSSM (PMSSM) often make the relaxed assumption that
the models only need not exceed the correct DM abundance, and any
deficit can be made up by an additional species, usually assumed to be
an axion \cite{CahillRowley:2012cb,Baer:2012cf}.  Axions and
axion-like particles (ALPS) are popular candidates for an extended
dark sector because they have motivation beyond their properties as dark matter
particles. Extremely light, weakly coupled ALPS can arise in string
theories \cite{Svrcek:2006yi}, but the most famous axion is likely the
QCD axion first proposed by Peccei and Quinn \cite{Peccei:1979}, which
will be the axion referred to in this thesis. The Peccei-Quinn axion's
origins are independent of any considerations of dark matter, and their motivation in a separate issue of tuning in the Standard Model.

While SUSY scenarios with a complementary axion may be well motivated,
their phenomenology can be ambiguous. Particular scenarios may have
predictions about the SUSY spectra and the relevant collider signals,
but the axion itself must be detected separately, so a collider signal
for such a scenario is not specific to an axion as the extra dark
matter. Furthermore, if the axino is the lightest SUSY particle (LSP),
then the lighest ordinary SUSY particle (LOSP) WIMP may not be detectable at all in direct or indirect
dark matter searches \cite{Baer:2010gr,Baer:2010kw}. Usually, the only
proposed scenario where the axino itself could be directly observed is
when the next-to-lightest SUSY particle (NLSP) is a charged sparticle
so that it leaves a charged track in its decay to the axino at a
collider experiment \cite{Brandenburg:2005he}. The work here takes its
motivation from the interest in models such as those studied in \cite{Baer:2009vr,Baer:2011hx,Bae:2013hma,Baer:2011uz,Baer:2010wm},
but with the goal of more predictive collider phenomenology,
specifically to study scenarios where the axino itself has a collider
signature (without a charged NLSP). In particular we focus on the KSVZ axino
which may be detectable
at the 14 TeV Large Hadron Collider (LHC) in the decays of neutralinos displaced from the primary
vertex. The decay
length of neutralinos in this scenario easily fits the ATLAS detector
for SUSY spectra expected to be testable at the 14 TeV LHC.
We compare this signature of displaced decays to axinos to other
well motivated scenarios containing a long lived neutralino which
decays inside the detector, including neutralino decays to light gravitinos or neutralino decays via RPV.
To make the collider phenomenology
possible at all requires certain assumptions about the axion model and
the SUSY spectra, which makes this scenario distinct from those
already studied but nonetheless it is a predictive scenario with the
possibility of low tuning, and compatibility with an attractive
cosmology.

The scenario explored in this thesis provides a dark matter solution by invoking both supersymmetry and the Peccei-Quinn mechanism (with the resulting axion), so in the next few chapters we motivate this scenario with a brief background on  dark matter (\ref{ch:darkmatter}), supersymmetry (\ref{ch:susy}), and axions (\ref{ch:axions}).  In
chapter~\ref{ch:background} we discuss what assumptions are necessary
for an axion model so that collider phenomenology is possible, and
what is the cost of these assumptions.  In chapter~\ref{ch:signal}, we
discuss the proposed signal and examples of SUSY benchmark models with
parameters that put this signal within reach at the 14 TeV LHC. In
chapter~\ref{ch:results} we compare this signal in detail to other
similar possibilities from gravitinos and $R$ parity violation (RPV) and in chapter~\ref{ch:width} we explore the dependence of the neutralino width on the particular choice of model. A
few remarks about how the scenario under study could be accommodated
in a viable cosmology model can be found in
chapter~\ref{ch:cosmo}. Lastly I conclude by considering the
limitations of this work and how it can be expanded in the future.

\chapter{Dark Matter}
\label{ch:darkmatter}

\section{Evidence for Dark Matter}
\label{sec:dmevidence}

	The particle identity of dark matter is a standing problem in modern physics. Though the evidence, models and searches are summarized here, a more complete review can be found in \cite{Garrett:2010hd}. The problem of dark matter has been known since 1933, when Fritz Zwicky discovered that the outer member galaxies of the Coma cluster were traveling too fast to only be under the influence of the gravitational force from the cluster's mass alone \cite{1933AcHPh.6.110Z}. Zwicky's inference was that there was additional unseen matter making up the ``missing mass" and so the idea of dark matter was first proposed. It should not be immediately obvious that this is a problem to particle physics, and certainly not immediate that it is a challenge to the Standard Model, but it is far from the only observation we have of the existence of this dark matter. Following observation of the Coma cluster, other galaxy clusters were shown to possess a similar feature in their velocity dispersions, with the outer galaxies traveling too quickly, as if there was a large amount of additional mass in the galaxy cluster, not in the form of luminous stars. The effect is also seen on different scales: not only in galaxy clusters, but the rotation curves of individual galaxies, with the outer stars orbiting the galactic centers more quickly than one would naively infer if all or most of the matter was luminous. These observations alone are not enough to peg this as a problem of particle physics however, and in the past there were strong competitive theories to the generic idea of ``dark matter", in the form of modified theories of gravity. While a subset of these theories are still possible, they are progressively more constrained with time, and with further observations of different kinds it becomes apparent that even if a modified theory of gravity contributes to the effects seen, it is most likely in addition to the effect of a missing mass.

	Beyond the kinematics of galaxies and clusters of galaxies, dark matter is also observed through the gravitational lensing of distant objects, again showing there is more mass than what is visible. This is perhaps the most striking in the bullet cluster \cite{Markevitch:2003at}, which is actually the merger of two clusters. As the two clusters pass through one another there are three different populations of matter that can be imaged separately and differentiated. The ordinary luminous matter from stars and the galaxies they compose comes from objects that are point like on the distance scale of a galaxy cluster, and so they pass though one another unaffected. In addition to the stars there is an expected amount of ``dark" matter from the intra-cluster medium, ordinarily not luminous, but which emits X-rays as the two huge gas clouds are compressed into one another and interact electromagnetically. A third population of matter can also be detected from the effects of gravitational lensing. The center of the lensing effect should be around the center of mass, which in the case of the bullet cluster would be expected to be in the center of the colliding intra-cluster gas, but instead the two centers of mass of the clusters are shown to  have passed through each other, indicating little or no interaction with the baryonic matter. Observation of colliding clusters of galaxies like the bullet cluster, or like the ``train wreck cluster", Abell 520 \cite{2012ApJ} are particularly difficult to explain with theories of modified gravity (though attempts are still made).

	Even if observations like the bullet cluster are taken as evidence enough that there must indeed be missing mass, regardless of whether or not gravity is modified, there are still many possibilities for what this dark matter can be, not all of which require the introduction of new particles. Black holes, brown dwarfs, and rogue planets are all examples of ordinary, well understood matter which is expected to be ``dark", and which certainly contributes to a small fraction of the dark matter causing the effects above. Such objects are collectively referred to as massive compact halo objects (MACHOs) \cite{Bennett:1995nm}. If these dark objects formed structures  and existed in sufficient numbers they could explain the effects of galactic kinematics and also, being constituted of compact stellar size objects would still be consistent with observations of colliding structures like the bullet cluster. These dark clusters as dark matter candidates are referred to as robust associations of massive baryonic objects or RAMBOs \cite{Moore:1994gk}. While MACHOs and RAMBOs were popular dark matter candidates for a time, further observations made it clear that they simply did not exist in sufficient numbers to explain the observed dark matter. Furthermore, there are other observations which show the dark matter observed in the universe is in a dominant fraction, non-baryonic, ruling out these types of theories (or at least  constraining their contribution to the total dark matter abundance to be very small). 

	Dark matter is observed on various distance scales through the kinematics of galaxies and clusters of galaxies as well as by gravitational lensing, but there is evidence of dark matter on different time scales also, not only in the recent universe, but in the early universe as well, showing that dark matter is present throughout most of the history of the observable universe. If dark matter is present in enough abundance to effect the kinematics of galaxies and clusters, it should also affect the way these structures form. Simulations show not only how much dark matter must be present to produce the observed structure, but also how ``hot" i.e how relativistic the dark matter must be. These simulations have shown that the majority of dark matter should be cold (non-relativistic) at the time of structure formation. Observing structure formation in the universe as a whole, i.e. measuring the galactic power spectrum, puts constraints on what fraction of the universe's energy density is composed of matter, $\Omega_{m}$. This measurement, plus information from big bang nucleosynthesis (BBN), which constraints the fraction of the universe's energy budget which is baryonic matter  $\Omega_{b}$, can be used to show that the majority of matter in the universe is not baryonic in nature. Drawing these conclusions from BBN involves a degree of uncertainty, because BBN predicts the elemental abundances in the primordial universe, but we can only directly observe the elemental abundance today and the relationship between these quantities is confounded by various astrophysical processes. This shortcoming in measuring $\Omega_{b}$ can be circumvented by measuring the cosmic microwave background (CMB) and hence observing the early universe directly. The shape of power spectrum of the CMB is determined by the temperature anisotropies of photons, which are the result of density perturbations in the early universe. The evolution of these density perturbations up to the time of last scattering (when the CMB was emitted) depends on the relative fractions of baryonic matter and dark matter, so fits to the CMB power spectrum provide much more accurate bounds on $\Omega_{b}$ and  $\Omega_{m}$, with the Planck satellite in 2015 reporting \cite{Planck:2015xua} $\Omega_{b}$ = 0.022 and  $\Omega_{m}$ =  0.316. With the dark matter fraction being constituted nearly entirely of a non-baryonic species it seems inevitable that a new particle, outside of the Standard Model must be introduced. The only possible way around this is if Standard Model neutrinos could provide a suitable candidate, but due to their very light mass, and only having the weak interaction and gravity to slow/cool them, they are typically only considered as hot dark matter, which we know cannot be the dominant constituent because of large scale structure as discussed above.

\section{Models of Dark Matter}
\label{sec:dmmodels}
	Despite all the evidence for its existence, very little is known about the specific particle nature of dark matter. The observations themselves provide hints to the nature of dark matter, but objectively we have almost no model independent limits on the number of new species, their masses,  the size and types of couplings they possess, or whether or not they are protected from decay by symmetries. We do not even know how the dark matter abundance we observe came to exist in the first place. The dark matter can be produced thermally, in which case it was in thermal equilibrium with other particles in the early universe and had a simliar number density to other particles before the universe cooled enough that the dark matter could no longer be produced. The amount of dark matter remaining in the universe self annihilates, its number density decreasing as the universe continues to expand and the probablity of two particles meeting up for an annhilation event becomes less and less likely.  Once annihilation events become sufficiently rare the amount of dark matter in the univese is relatively constant, and the relic abundance is said to have undergone ``freeze out".  This is not the only way a dark matter abundance can be produced however, and there are non-thermal ways, from the decay of heavy particles or from phase transitions in the early universe. This leaves a very vast space of ideas for models, which has been explored imaginatively for decades, but is still not exhausted.
	
	 The simplest theories assume only one new non-baryonic cold species for the sake of simplicity, but as described above we know this non-baryonic species shares at least a small portion of the dark matter abundance with hot Standard Model neutrinos and dark massive compact objects. There is no compelling reason to believe there cannot be more than one new species, and when this is the case many of the hints we gather from the observations in the last section become less useful. Structure formation, baryon acoustic oscillations and cluster mergers like the bullet cluster all seem to indicate that dark matter is largely non-interacting. The only certainty is that it interacts gravitationally, as all the observations we do have are due to gravitational effects. If dark matter does truly interact by gravitation alone, as a truly sterile species, then any hope for detection seems to evaporate. The usual assumption is that the only interactions dark matter has (in addition to gravity) are weak interactions. This assumption is not made simply because it provides a possible detection channel, but also because if dark matter is thermally produced, then weak interactions give an annihilation cross section of the right size such that the thermal relic abundance of dark matter approximately matches what is observed. Massive dark matter particles with weak interactions are often refered to as weakly interacting massive particles or WIMPS, and this coincidence of weak scale interactions for dark matter giving the correct abundance is know as the ``WIMP miracle". Getting the correct abundance of course also depends on the mass of the wimp, and the specifics concerning the annihilation channels available. While WIMPs are perhaps the most studied class of dark matter, there are many other other possibilities and many types of interactions which dark matter may have. With sufficiently light dark matter, even strong interactions for dark matter are possible, and dark matter candidates in such models are called strongly interacting massive particles (SIMPS)\cite{Hochberg:2014dra}. Even though observations such as that of the bullet cluster show dark matter to be largely non-interacting, this does not exclude the possibility that the dark matter contains a sub population with self interactions, which could lead to dark atoms and eventually to sub structures in dark matter halos. An example of such a model is partially interacting dark matter (PIDM) \cite{Fan:2013yva}, but other models are possible. 
	
	Another attribute of dark matter that is inferred from observation is its relative stability. Some models make this stability absolute by introducing a new symmetry, such that the dark matter is protected against decays, this is often the case for WIMPs in model with supersymmetry as will be discussed in chapter \ref{ch:susy}.  Absolute stability is not a requirement however, and dark matter can have decays, so long as it is stable at the order of the lifetime of the universe. An interesting class of models known as dynamical dark matter (DDM) \cite{Dienes:2011sa} consists of not one or several species of new particles, but rather an ensemble of new states in a dark sector, where no single species may be stable, and decays between different members of the ensemble are possible. Even if some of the species decay quickly, so long as all decays remain within the dark sector the total ``dark" abundance will remain the same. A less exotic example of a model where the dark matter is not strictly stable is axions. Axions subvert many of the standard assumptions about dark matter, in that they can interact with photons, are not protected by a symmetry and  typically are not thermally produced in models. The one reason axions can have all these bizzare characteristics is because their couplings are suppressed by a very large new scale. More will be said about axions in chapter \ref{ch:axions}, but for now it should at least be added that they belong to a broader class of dark matter candidates which are sometimes called extremely weakly interacting particles (E-WIMPs)\cite{Choi:2005vq}. Gravitinos (supersymmetric partners to gravitons) are also E-WIMPs as their interactions are supressed by the Planck scale.
	
	While simulations of large scale structure show that the dark matter cannot be very relativistic (hot), this does not mean it has to be entirely cold and there are models of ``warm" dark matter \cite{Bode:2000gq}, where the dark matter has thermal velocites, just not as great as dark matter which is considered ``hot". Sterile neutrinos are a common candidate for warm dark matter, but some of the other examples above such as WIMPs and gravitinos can be ``warm" depending on their mass, and how they were produced in the early universe. There is also the possibility of mixed dark matter, dark matter composed of multiple species, with some small hot fraction and a dominant cold fraction. It was already stated above that the Standard Model neutrinos can be such a hot component, but even the role of SM neutrinos depends on the greater cosmology. Modifying the thermal history of the universe between models can lead to very different scenarios of dark matter. By lowering the reheat temperature Standard Model neutrinos were shown to be viable warm dark mater candidates  in \cite{Giudice:2000dp}. Warm dark matter is typically very light with Standard Model neutrinos as an example, and non thermally produced axions can be extremely light ($\mu$ eV or even smaller), but there are also very massive examples of dark matter candidates.  Super massive ``wimpzillas" can be produced gravitationally during a phase transition in the early universe and can be as heavy as the GUT (grand unified theory) scale, as large as $10^{16}$ GeV \cite{Kolb:1998ki}.
	
	The models mentioned above show the great diversity of possibilities for dark matter models. With such a vast landspace of ideas it helps to have a motivation for the model outside of just its ablity to fit our observations and constraints. Chapter 3 will cover supersymmetry in more depth, but it is sufficient for now to state that while supersymmetry does provide dark matter candidates it has various motivations outside of it. Models of axions also have a strong motivation outside of dark matter considerations and this will be addressed in chapter 4. Models of asymmetric dark matter or ADM \cite{Petraki:2013wwa} are motivated in their attempts to also provide a solution to the problem of baryogenesis. It is an observational fact that we observe far more matter than antimatter in the universe, and ADM seeks to explain this fact by transfering asymmetries in the dark sector to asymmetries in the visible sector. Besides focusing on models with theoretical motivation beyond just a dark matter solution, studies in this vast space of ideas can be further narrowed by focusing on those ideas for which we can make testable predictions. Especially intersting are those models where the properties of the dark matter, its mass and couplings, can be measured directly. Methods of dark matter detection will of course depend on the classes of models considered, but the most common strategies are reviewed in the next section.

\section{Searches for Dark Matter}
\label{sec:dmsearches}
	
	Of the various types of dark matter candidates mentioned in the previous section, WIMPs are perhaps the most widely studied, and the usual dark matter candidates from SUSY models are also WIMPs. If these WIMPs are thermally produced in the early universe then there are in principal three methods of detection which can be summarized by viewing the generic Feynman diagram in figure ~\ref{fig:dm_detect} from different sides. The contents of the blob in the center of the diagram will be model dependent, they may be Standard Model particles, or something more exotic from a new sector. There also may be more than one mediator, from multiple processes that contribute to a detection signal. The rotation of the diagram for different methods should not be taken too literally, as the SM part of the diagram is likely to change, the annihilation products of indirect detection are unlikely to be the same particles used as a target for scattering in direct detection which are also unlikely to be the particles constituting the colliding beams in an accelerator. The mediator may also be changed when rotating the diagram for different search methods. Some of these methods wil work for certain non-wimp candidates, and there are also specific models of wimps for which some of these methods will not be feasible.

\begin{figure}
\centerline{\includegraphics[scale=0.3]{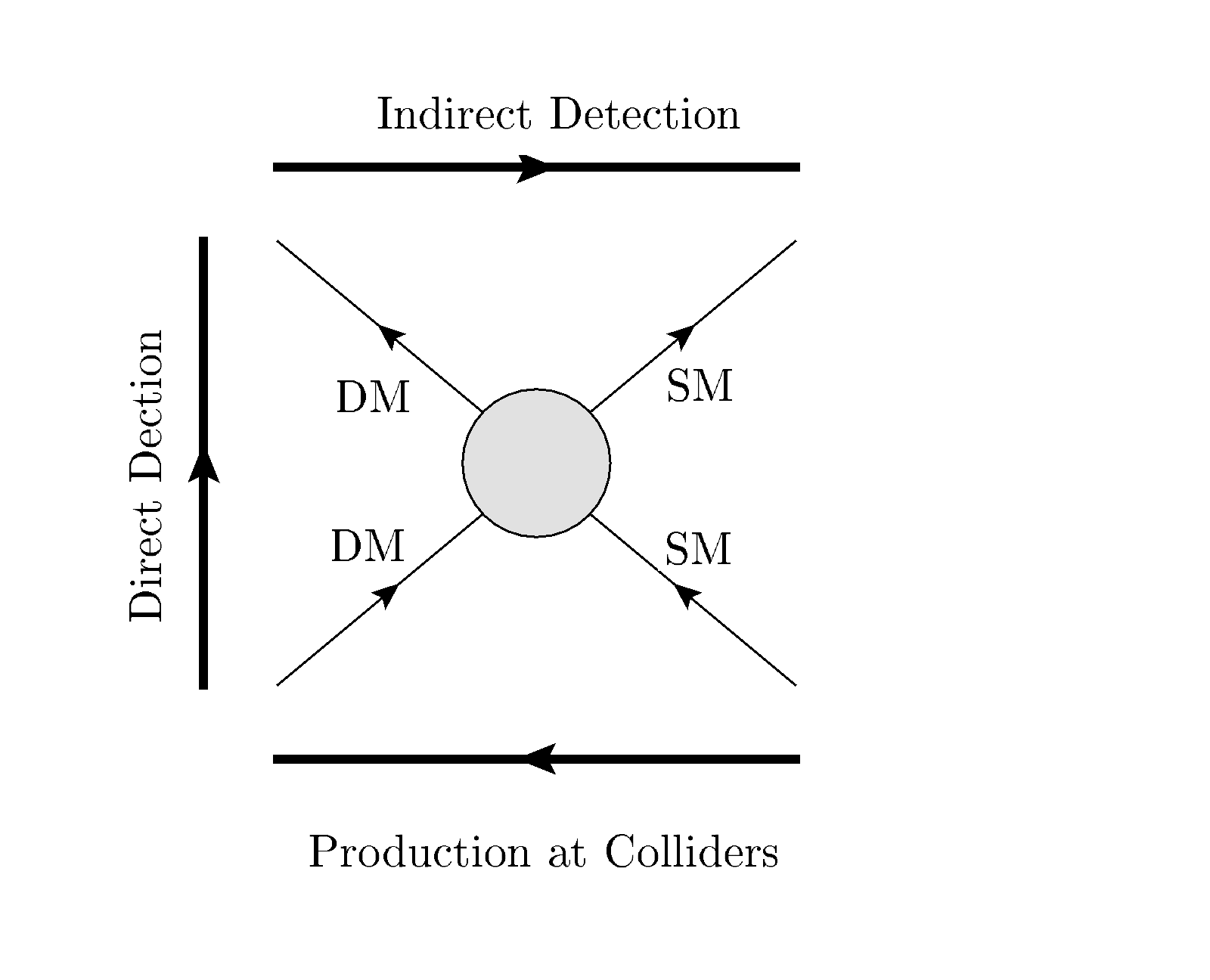}} \caption{The three principle ways to observe dark matter interacting with the Standard Model.}\label{fig:dm_detect}
\end{figure}
 	
	If dark matter self-annihilates at a sufficient rate then the resulting Standard Model products may be observable in what it is called indirect detection.  Dark matter seems to be ubiquitous, appearing in every large structure we observe, and supposedly the earth is adrift in the dark matter halo of our own galaxy, and yet there are regions of greater dark matter density, which will have a higher annhilation rate and therefore a greater signal. Dark matter at the very least interacts gravitationally, so large gravitational wells should make for good sources for indirect detection. One excellent source would be the core of our own galaxy. The coupling and number density will have to be of a sufficient size and the Standard Models annhilation products will have to be distinguishble over the background of other cosmic sources. Typically these types of searches look for positrons or photons. The difficulty with positrons is that their charge means their flight path en route to the detector will be deflected by magnetic fields and  we cannot extrapolate what the source was, so we cannot tell if the signal is strongest in the direction of the galactic core, or if a positron excess is coming from some other poorly understood astrophysical source. One expected characteristic of a positron signal would be that it should show a cut off, that is the excess should drop off above the mass energy of a DM pair. A photon signal is considered much cleaner, and ideally could be observed in the form of a monochromatic line. A single narrow peak photon signal could come from (for example) a DM pair annhilating to two photons, in which case the photon energy can tell us the cold dark matter mass. Such a signal could also put constraints on the size of the dark matters couplings, though for a full understanding the branching fraction to the signal channel must be known, as there could be other annhilation products that don't appear in signal channels. Detection of photons or positrons from space is obviously best done by a satellite experiment, such as Fermi LAT \cite{Ackermann:2015tah}, but earth based indirect detection is also possible, if for example the annihilation products are neutrinos then existing neutrino observatories on earth such as IceCube \cite{Sullivan:2012hf} can put constraints on certain models of dark matter. To quote specific  limits from such experiments would be highly model dependent, needless to say the only currently known signals are tentative at best, and may vanish in the future with more data, a better understanding of systematics or a better understanding of astrophysical sources. 

	As already stated, the earth should itself be inside the dark  matter halo of the milky way, so a second way to detect dark matter is to simply wait for a rare event where a dark matter particle scatters off of some detector material on earth. This method is known as direct detection.  The physics of direct detection is governed by the equation for the rate of particles detected per recoil energy of scattering events,
\begin{equation}
\frac{dN}{dE_{r}}=\frac{\sigma\rho}{2\mu^{2}m_{\chi}}F^{2}\intop_{\mathrm{v_{min}(Er)}}^{\mathrm{v_{esc}}}\frac{f(v)}{v}dv \;  ,
\end{equation}

where $\sigma$ is the scattering cross section of the DM with the target nucleus, $\rho$ is the density of dark matter in the halo, $\mu$ is the reduced mass of the DM and the target nucleus, $F$ is the nuclear form factor and $f(v)$ is the distribution of dark matter velocities in the halo, which is not precisely known, and this is where some degree of assumption enters the analysis. To get the correct rate, one must integrate over the velocity distribution, from the lowest velocity that gives a recoil event with energy Er, up to the velocity at which the DM escapes the galactic halo. Recoil energies of nuclei scattering off of dark matter are expected to be very small, so detectors must be very sensitive,  very cold and well shielded to reduce noise as much as possible. Because of the expected rarity of such events a large detector volume is preferable: more material means more events. The choice of target material when designing an experiment depends on what type of dark matter one hopes to detect. The choice of target nuclei can effect how detector sensitivity scales with the DM mass and also can have an impact on whether the detector is more sensitive to spin-independent or spin-dependent interactions. Which cross section, the spin-independent, or spin-dependent, depends on what mediators are available for scattering events, and will depend on the dark matter model. There are in principle three effects of  a recoil event which are used to generate a signal. Phonons produced in the detector material can be measured as a very small change in temperature. Ionization of target material can be measured by applying a bias and causing charged particles to drift to a collection plate. Photons produced in the collision can be collected by simple photodetectors. In practice, a detector is usually designed to make use of at least two of the three effects, so that the combined information (such as the relative timing between different parts of the signal) can be used to veto false events. 
	
	Simply measuring events that pass vetoes would be exciting, but for direct detection the signal is expected to display a predictable pattern (if enough events are seen). As the earth orbits around the sun our trajectory through the dark matter ``wind" changes its heading, so that the relative velocity of dark matter in the halo with respect to the detector will be modulated with an annual frequency. While the precise distribution of dark matter is unknown to us, and it does affect the rate of events, the annual modulation is rather model independent. Two experiments, CoGeNT \cite{Aalseth:2012if}and DAMA \cite{Bernabei:2013xsa}, have already claimed to see this annual modulation and several others have claimed to see small signals. Systematics and elimination of backgrounds are very important however, and  there is a great deal of skepticism about these signals. At first glance, the tentative signals from the different experiments are in disagreement with each other, but their interpretation is highly model dependent, and for the correct models with the correct dark matter distributions the signals can be made to agree \cite{Kelso:2011gd}. More important than their disagreement with one another however, these tentative signals are not seen by successive generations of detectors which should be more sensitive across the mass/coupling range in question. In particular, the successive experiments of the XENON collaboration \cite{Beltrame:2013bba} have all failed to detect any events in the signal region of the light dark matter detected by CoGeNT, DAMA and others. Future direct detection experiments are expected to probe a wide range of masses, and down to very small couplings, covering very large chunks of parameter space, including a very large chunk of model space for SUSY WIMPs. The eventual limit to these experiments comes from what is called the ``neutrino floor". Once these types of detectors become sensitive enough, the effects of ambient neutrinos scattering off the detectors will actually become a significant background to the desired signal and more creative detector technologies will have to be developed to probe further \cite{Grothaus:2014hja}.

	The last method of dark matter detection discussed in this section is by direct production at a particle collider. This is the method with which the original work of this thesis is concerned with, as will be explored in later chapters. In general dark matter produced in a particle collider is not strictly speaking ``detected" and passes right through all layers of the detector. If this was the end of the story then there would be no triggerable signal, but these types of events, where dark matter could be produced, will inevitably also involve some ordinary Standard Model particles. Even in the simplest case of a two to two pair production of dark matter particles, there is still the possiblity of Standard Model particles in the form of initial state radiation, leading to mono jet, mono lepton or mono photon events with missing energy. Which events are most likely depends on both the model of dark matter and the type of collider, but triggering on mono-anything is a good model independent search strategy for dark matter and can be used to set limits on various couplings in an effective field theory approach.
	
	 For lepton colliders missing energy is a useful variable since the energy of colliding beams is known precisely but in a hadron collider this is confounded by convolution with parton distribution functions, so that at best it is only meaningful to talk about missing transverse energy. The transverse momenta of an event is the momenta in the direction perpendicular to the beam axis, and should total zero by the demands of momentum conservation. For massless particles the transverse momentum and the transverse energy are the same. The missing transverse energy  is simply the difference between the transverse energy expected from the event and what was actually measured by the detector. 
	
	When discussing models of dark matter in previous sections, it was mentioned that new symmetries can be introduced to protect the dark matter from decay. If this new symmetry is manifest across a whole new sector, with the dark matter being the lightest example, then this can provide more varied search channels for dark matter, and with a richer phenomenology. SUSY with $R$ parity is the most famous example of this, but the same princple can be applied to any new sector that is collectively charged under a new parity. If a pair of new heavy  states are produced in a collider and they are odd under a new parity then they cannot decay completely to Standard Model particles which are assumed to be even under this parity. This means that the cascade decay of such particles will eventually terminate in their lightest stable member, which, if neutral can be a good dark matter candidate. The Standard Model byproducts of these decays will determine what the signal is. At a hadron collider this is usually expected to produce jets and missing transverse energy. 

	These are just the basic three methods for detecting dark matter and it is by no means exhaustive. Next to SUSY, axions are also a very popular dark matter candidate, and interestingly enough they are expected to provide no signal by any of these methods. Chapter~\ref{ch:axions} is dedicated to this one type of model of dark matter alone, as some details of the models will be important for the original work done in later chapters.

\chapter{Supersymmetry}
\label{ch:susy}

\section{Theoretical Motivation for Supersymmetry}
\label{sec:susytheory}
Supersymmetry at its simplest is just another symmetry; a feature of possible theoretical models that often has appealing consequences. Usually when people discuss supersymmetry what they really mean is models with supersymmetry near the weak scale, such as the Minimal Supersymmetric Standard Model (MSSM).  Without specifying any specific model though, supersymmetry alone, just as a symmetry, has its own theoretical motivations. Long before the MSSM was considered seriously, self consistent models of supersymmetry were interesting to study soley for these theoretical reasons. The first supersymmetric field theory to work in four dimensional space time was proposed by Wess and Zumino in \cite{Wess:1974tw}. Symmetry has always had a prominent role in physics, even if its role has not always been emphasized by the physicists at that time. When Newton unified terrestrial and celestial mechanics, he was recognizing an invariance in gravity, that is to say, he discovered a symmetry. With Noether’s theorem the role of symmetry in classical mechanics was made concrete, defining what is meant by a conserved quantity classically. Einstein's special relativity is basically the application of Poincar\'e symmetry to mechanics. The success and predictive power of the Standard Model of particle physics is deeply rooted in the success of gauge theory, which requires the Lagrangians for fields to respect local symmetries. Finding larger symmetries, and seeing what physical theories can realize these symmetries is a motivation in of itself because of the past success of theories rooted in symmetry. 

	As  relativistic theories, all quantum field theories respect Poincar\'e symmetry, but they can also respect the symmetries of various gauge groups as well. In a certain mathematical sense though, the addition of gauge groups to a field theory is uninteresting. In going from a theory with translational and rotational symmetry (such as classical mechanics) to a theory with full Poincar\'e symmetry (such as special relativity) the old generators have non-trivial commutation relations with the new generators of the Lorentz boosts. In expanding a theory with gauge symmetries however all of the new generators for the gauge groups are guaranteed to commute with the generators of the Poincar\'e group, i.e. $[T^{a},P^{\mu}]=0$ and  $[T^{a},M^{\mu\nu}]=0$ holds for any gauge group generator $T$,  where $ M^{\mu\nu}$ is the generator of rotations and boosts, and $P^{\mu}$ is the generator for translations. Expanding the symmetry group of quantum field theory beyond the Poincar\'e group in a non trivial way may just be a neat mathematical trick, but the last time such an expansion was realized in a physical theory it was our leap from classical mechanics to relativistic mechanics, and an attempt to follow this pattern of success is clearly an encouraging motivation. The generators of supersymmetry transform a bosonic field into a fermionic field: $Q_{\alpha}|bos>=|ferm>_{\alpha}$ , and vice versa: $Q_{\alpha}|ferm>^{\alpha}=|bos>$ , and these generators are  the only kind that can expand the Poincar\'e symmetry in a non-trival way, by mixing the new generators with the old:$[M^{\mu\nu},Q_{\alpha}]=-i(\sigma^{\mu\nu})_{\alpha}^{\beta}Q_{\beta}$ . A supersymmetric theory can actually contain multiple supersymmetries, that is, multiple sets of supersymmetric generators, but for the remainder of the thesis the focus will be on those models with only one supersymmetry, as these are the most minimal, and the most phenomenologically viable.

\section{Content of the MSSM}
\label{sec:susycontent}

	Beyond the purely mathematical motivation for supersymmetry there are many practical reasons why particular supersymmetric theories are appealing. Most of these appealing features are consequences of the many new fields added.  For a field theory to respect supersymmetry the objects of the theory must be invariant under the action of the generators, and clearly the Standard Model fields are not. All the fields of the Standard Model are either bosonic or fermionic, so they will all be transformed by the SUSY generators, so for a theory to be supersymmetric the fundamental objects are no longer these fields but rather superfields are introduced. A superfield contains an equal number of bosonic and fermionic field degrees of freedom, and with the same quantum numbers, so that under operation from the SUSY generators the same superfield is reproduced.  The new states introduced in the superfields, those not found in the Standard Model, are referred to as superpartners, or more generally as sparticles.  For a supersymmetric theory to contain all of the field content of the Standard Model there are two types of superfields required, chiral superfields (both left and right) and vector superfields. Chiral superfields contain fermions and scalars, this is where the matter fermions of the Standard Model can be found, along with new scalar partners. The quarks of the Standard Model have their new superpartners, the scalar squarks. There are scalar partners for both left and right handed quarks, and these are usually referred to as left and right handed squarks respectively, even though in the literal sense a scalar particle has no chirality. Likewise the charged leptons of the Standard Model are joined by sleptons and the neutrinos have their sneutrinos. Vector superfields contain vector fields and fermion fields, and this is where the vector bosons of the Standard Model can be found. The fermionic partners to the gauge bosons are collectively called gauginos. The partner to the gluon is the gluino, and the electroweak gauginos are the bino and winos (corresponding to the SM fields prior to electroweak symmetry breaking). The Higgs sector in supersymmetric theories is slightly more complicated than in the Standard Model, as the structure of supersymmetry requires two Higgs doublets, one to couple to up type quarks, and one to couple to down types. Trying to do this with a single Higgs doublet as in the Standard Model would violate supersymmetry and also introduce a gauge anomaly. As scalars, the Higgs can be found in chiral superfields, and their fermionic superpartners are known as Higgsinos. The Minimal Supersymmetric Standard Model  (MSSM) is a field theory with only the minimum superfield content required to contain all of the Standard Model fields.
	
	With the field content of the MSSM established it is possible to take inventory of the particle content as well. After electroweak symmetry breaking, as in the Standard Model, some of the degrees of freedom of the Higgs field will become the longitudinal degrees of freedom for the weak bosons, resulting in masses for the W's and the Z. In the Standard Model a single complex Higgs doublet has three degrees of freedom eaten, leaving one physical Higgs scalar. In the MSSM, there are originally eight degrees of freedom in the two complex doublets, and after electroweak symmetry breaking this leaves five physical Higgses. In addition to a Standard Model like Higgs there is also a heavier neutral scalar, a neutral pseudo scalar, and also two charged scalars. All of these Higges have a corresponding Higgsino. There is also mixing of the new particle states in the MSSM. The neutral Higgsinos and the neutral gauginos mix, resulting in four neutral fermionic states called neutralinos. The charged Higgses and the charged gauginos also mix, and the two resulting states are called charginos. The particle content of the MSSM, arranged by its superfields is summarized in figure \ref{fig:mssm} \cite{Signer:2009dx}

	Defining superfields is convenient because it allows the particle content of the theory to be expressed compactly. Likewise, defining functions of these superfields can allow us to have compact representations of the interactions and dynamics of the theory. The superpotential $W$ is an analytic function of the superfields, from which the interaction terms (aside from gauge interactions) in the Lagrangian can be derived. The gauge interactions can also be derived from functions of the superfields. The K\"{a}hler potential, $K$ is a real function of the superfields and from it the kinetic terms can be derived.  The specific form of the  K\"{a}hler potential depends on the method of supersymmetry breaking, but the generic superpotential for the MSSM is given by

\begin{equation}
W=\epsilon_{ij}(y_{ab}^{U}Q_{a}^{j}U_{b}^{c}H_{u}^{i}+y_{ab}^{D}Q_{a}^{j}D_{b}^{c}H_{d}^{i}+y_{ab}^{L}L_{a}^{j}E_{b}^{c}H_{d}^{i}+\mu H_{u}^{i}H_{d}^{j})+W_{RPV}\;  ,
\end{equation}
where $i$,$j$ are SU(2) indices and $a$,$b$, are indices for the generation. The term $W_{RPV}$ will be discussed in more detail later on, but usually it is made to vanish in the MSSM. From the superpotential and  K\"{a}hler potential, the main parts of the MSSM Lagrangian can be derived. The overall structure of the MSSM Lagrangian is given by

 \begin{equation}
\mathcal{L}_{MSSM}=\mathcal{L}_{kinetic}+\mathcal{L}_{gauge}+\mathcal{L}_{scalar}+\mathcal{L}_{break}\;  ,
\end{equation}
where the kinetic and gauge parts are analogous to the standard model, but now with interactions for the new fields as well. The scalar part of the Lagrangian involves the interactions derived from the superpotential, including new Yukawa interactions. The last remaining part, the $\mathcal{L}_{break}$ term, breaks supersymmetry, and would not exist if the symmetry was perfect, but it must be included in realistic models for reasons that will be discussed shortly.

\begin{figure}
\centerline{\includegraphics[width=4.5in]{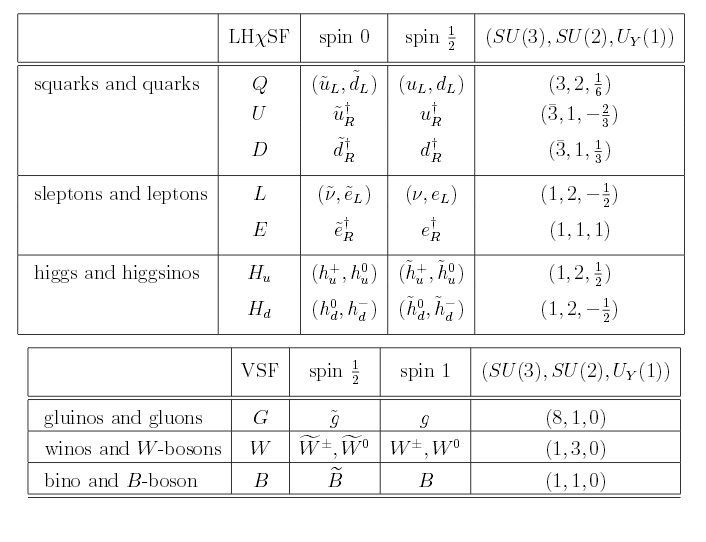}}
\caption{Summary of the particle content of the MSSM, arranged into left handed chiral superfields (LF$\chi$SF and vector superfields (VSF) as they appear in \protect\cite{Signer:2009dx}. Note that the new neutral fermionic states  (aside from the gluino which has color charge) mix to form the neutralinos ($\tilde{\chi}_{n}^{0}$ ), and the new charged fermionic states mix to form the charginos ($\tilde{\chi}_{n}^{+}$ )} \label{fig:mssm}
\end{figure}

\section{Practical Motivations for the MSSM}
\label{sec:susypractical}

\begin{figure}
\centerline{\includegraphics[width=4.5in]{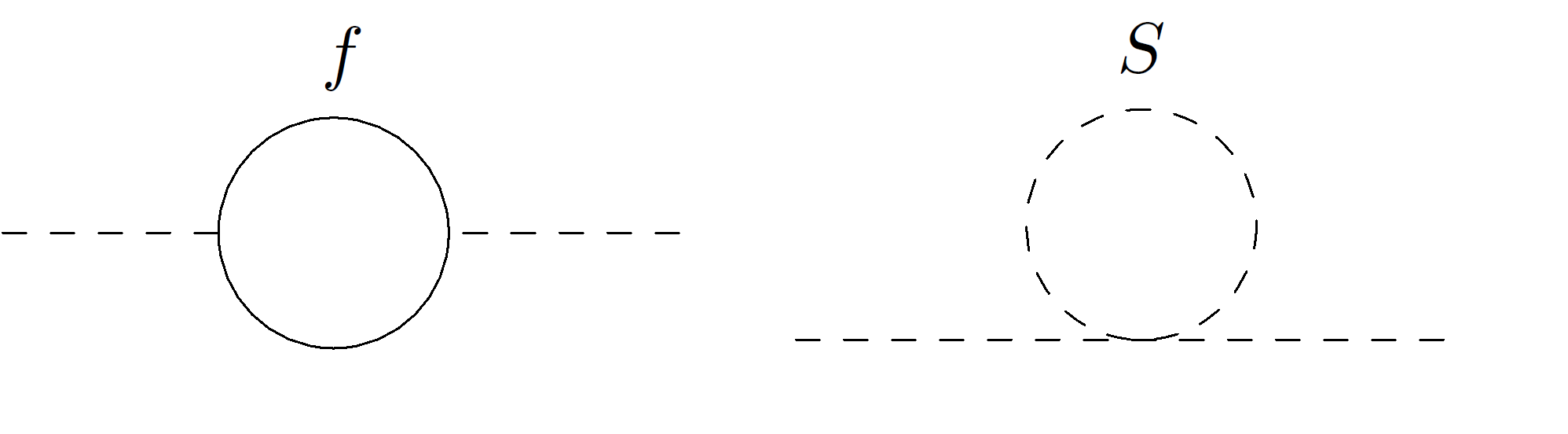}}
\caption{Feynman diagrams for the one loop corrections to the Higgs mass from fermions (denoted by $f$) and other scalars (denoted by $S$). Supersymmetry guarantees that the leading term (quadratically divergent in the cutoff scale) from any fermion loop will be canceled exactly by the corresponding scalar loop.} \label{fig:higgs}
\end{figure}

	If the MSSM were truly supersymmetric, that is, if every term of the full Lagrangian was invariant under operation by the SUSY generators, then all of the superpartners, the new scalars, the gauginos and the higgsinos, would be the same mass as there coresponding Standard Model partners. This is obviously not the case, as none of these states have been detected, and the superpartners should couple with the same strength as there Standard Model partners. This means that if supersymmetry is realized at all, then it must be broken below some scale. The way this is usually done in practice is to introduce soft mass terms in the SUSY Lagrangian. These terms break supersymmetry and give the new supersymmetric states masses greater than their SM partners. Above the scale of SUSY breaking the masses are equal again.  In theory, the scale of SUSY breaking can be anywhere, and if it is sufficiently high then it becomes impossible to produce SUSY partners at a collider, but there is strong theoretical motivation for SUSY near the weak scale. 

The Higgs boson, as a scalar, is much more strongly sensitive to loop corrections, such as those in figure  than other Standard Model particles: fermions in the SM are protected by chiral symmetry and vector bosons by gauge symmetry. The loop corrections to the Higgs mass are quadratically divergent, with the leading contributions from a single fermion going as, 

\begin{equation}
\delta M_{H}^{2}=\frac{g^{2}}{16\pi^{2}}\Lambda^{2}+\frac{g^{2}}{16\pi^{2}}m^{2}\mathrm{ln}\frac{\Lambda^{2}}{m^{2}}+. . . \;  ,
\end{equation}
where $\Lambda$ is the cut off scale where Standard Model physics breaks down. Such corrections would lead one to believe the Higgs mass should be quite large, but perturbative unitarity requires the Higgs to be light, on the order of hundred GeV, and this is what is observed for the measured Higgs mass. To have such a light Higgs despite such potentially large corrections naively requires a large cancellation of parameters. This unnatural fine tuning of parameters to keep the Higgs mass small is likely the most famous example of a hierarchy problem, so much so that it is often referred to as \textit{the} hierarchy problem.  The tuning may just be an unfortunate coincidence of nature, but a ``natural" solution, that is one that does not rely on coincidence for its explanatory power, would be to introduce a symmetry between bosons and fermions (supersymmetry), such that their contributions to the Higgs mass corrections will cancel.  Note that the leading divergent term does not have any dependence on the mass, so that supersymmetry at any scale will remove these terms which are quadratically dependent on the cutoff scale. The sub leading term, the one that goes as the log of the cutoff scale, does contain the mass however, so as the mass of superpartners becomes larger and larger the tuning is reintroduced. To keep these logarithmic corrections small, superpartners must not be too much heavier than their Standard Model counterparts and this is the primary motivation for supersymmetry at the weak scale. How much tuning is considered ``unnatural" is a subjective issue, so thereis a blurry line where supersymmetry at colliders begins to lose its motivation.
	In addition to the logarithmic corrections which depend on mass, there is an additional source of tuning in the supersymmetric Higgs sector. The parameter $\mu$ appears in the MSSM Lagrangian as a Higgs mass term. Though this term is not sensitive to higher order corrections, it should be near the weak scale for the Higgs to remain light, though there is no a priori reason for the $\mu$  parameter to be of this size; this is known as the ``mu problem". Stated another way, for the Higgs mass to remain light in supersymmetry requires a coincidence of scales:  both the soft mass terms that lead to logarithmic corrections and the $\mu$ parameter which determines the ``bare" Higgs mass. There is no guarantee that the scale of these parameters, $\mu$ and the soft masses are connected, but there are models that can address this issue, two of which will be mentioned later on.

	While the hierarchy problem is the primary reason SUSY at the weak scale is desirable, the dark matter solution from SUSY often requires weak scale soft masses. There are a variety of ways in which SUSY can provide a dark matter solution, and the rest of this thesis will explore one of the more non-standard ways, but first, the most common solution can be described. The typical case is to require the lightest superpartner (LSP) to be neutral, so either a sneutrino or a neutralino. The interactions of the neutralinos and the sneutrinos just mirror those of their Standard Model partners so they are WIMPs, and with sparticle masses around the weak scale, ~10GeV to ~1TeV the relic abundance can be approximately correct, depending on the relative weight of the various annihilation channels in a particular model. Sneutrinos are usually disfavored however, as they are already constrained by direct detection experiments, described in the previous chapter. For the LSP WIMP neutralino to be a viable dark matter candidate there is an additional requirement however, as the simplest MSSM Lagrangian does not guarantee stability of any of its particles, even the lightest superpartners. To protect the LSP neutralino from decay an additional symmetry is added to the MSSM, known as $R$ parity, defined as,

\begin{equation}
P_{R}=(-1)^{3(B-L)+2s} \;  ,
\end{equation}
where $B$ is baryon number, $L$ is lepton number, and $s$ is the spin. Requring $R$ parity is what causes the $W_{RPV}$ term in the superpotential to vanish. In this way all Standard Model particles have even $R$ parity and all sparticles have odd $R$ parity. If this symmetry is respected by the MSSM Lagrangian then supersymmetric vertices with a single sparticle are forbidden, making the LSP absolutely stable. $ R$ parity may at first seem like an ad-hoc addition to the MSSM, added for convenience, but it is possible (but not necessary) to embed the MSSM in larger theories where $R$ parity arises naturally, such as \cite{Babu:2002tx}. In addition to providing a dark matter solution, the introduction of $R$ parity has other motivations. Without $R$ parity the MSSM Lagrangian allows for baryon and lepton number violating interactions which would lead to proton decay rates in sharp contrast to the stability of protons that is observed. $R$ parity also allows for predictive collider phenomenology, as mentioned in the previous chapter. Because there are no vertices with a single sparticle allowed, SUSY particles should always be produced in pairs, and no SUSY sparticle will decay entirely to SM states, such that SUSY events should consist of cascade decays of various lengths terminating in a dark matter candidate that leaves the detector. The usual expected signal for this is jets and missing transverse energy, but the details depend on the sparticle mass spectrum and which types of cascades are dominant. There are many different search strategies for SUSY, and instead of attempting to summarize them here, we instead refer to the public search pages from CMS \cite{CMS:2013:Online} and ATLAS \cite{Atlas:2013:Online}. While the MSSM should not make predictions that go against our observations of the stability of protons, this does not mean absolute $R$ parity is required, and $R$ parity violating  (RPV) terms are allowed, so long at they are suppressed enough to be consistent with experiment. Allowing $R$ parity violating interactions will further complicate the collider phenomenology of the MSSM, allowing for an even larger variety of signals.

	While a solution to the Hierarchy problem and a promising dark matter candidate are the primary motivations for SUSY at the weak scale, supersymmetric models in general can have a variety of other motivations which may or may not be realized in each particular case. By introducing many new particles, the running of gauge couplings is altered, and in a typical MSSM spectrum, unlike in the Standard Model, these couplings unify at a large scale, as illustrated in figure \ref{fig:unify}\cite{Martin:1997ns}, and there are many examples of MSSM models which are embedded into GUT theories such as those mentioned in \cite{Babu:2011kj}. The presence of new mixings and possible baryon and lepton violating interactions also leads to the possibility that SUSY models may provide a solution to the problem of baryogenesis. A particularly popular way to do this in SUSY is with thermal leptogenesis  \cite{Davidson:2008bu} using the supersymmetric version of the see-saw mechanism to produce neutrino masses. Theories of leptogenesis first generate the asymmetry in leptons in the early universe and then transfer this to the baryons.

\begin{figure}
\centerline{\includegraphics[width=4.5in]{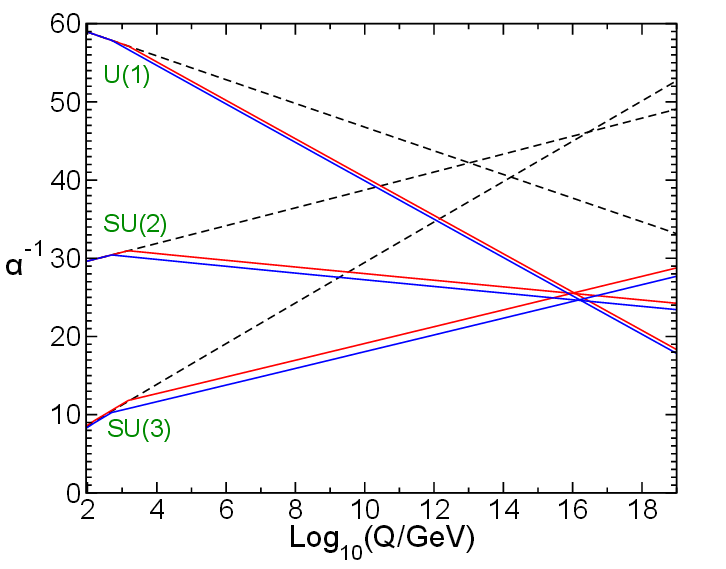}}
\caption{Evolution of the inverse gauge couplings in the Standard Model (dashed lines) and the MSSM (solid lines) as seen in~\protect\cite{Martin:1997ns}. The two colors for the MSSM lines come from varying the chosen input parameters. \label{fig:unify}}
\end{figure}

\section{Models of Supersymmetry}
\label{sec:susymodelsl}

	The mathematical motivations for a generic SUSY theory have been shown, as have some of the more practical motivations from the MSSM, but the MSSM is still not a very specific theory, as has already been hinted at above. The phenomenology of the MSSM changes greatly with different mass hierarchies in the sparticle spectrum, there may or may not be $R$ parity, and the theory may be supplemented by other new physics or embedded in larger theories. The model space of supersymmetry is truly huge, even considering the MSSM alone, as there are hundreds of new parameters from all of the soft masses, mixing angles and new tri-linear couplings between scalars. To begin to approach the phenomenlogy of this model space it helps to consider subcategories of models. One way to sort these models is by the proposing the method by which supersymmetry is broken.  Specifiying the high scale physics theory often leads to relations between the low scale parameters, such that the total number of parameters which are actually free in the theory can be greatly reduced. Minimal supergravity or mSUGRA \cite{Nath:2003zs} breaks SUSY in a hidden sector and communicates this breaking to the visible sector via gravity. Gauge Mediated Supersymmetry Breaking (GMSB) \cite{Giudice:1998bp} models, like mSUGRA, break supersymmetry in a hidden sector, but they communicate this breaking to the visible sector through gauge interactions. In Anomaly Mediated Supersymmetry Breaking (AMSB) \cite{Paige:1999ui} models, the supersymmetry breaking is communicated to the visible sector by a combination of gravity and anomalies. In all of these models the number of free parameters is reduce from  over one hundred to less than ten, making phenomenological predictions more manageable.
	
	 There are also theories which do not explicitly make detailed assumptions about the method of SUSY breaking, but rather they only impose the relationships among the low scale parameters. The Constrainted Minimal Supersymmetric Standard Model (CMSSM) \cite{Ghosh:2012dh} has only five parameters at high scales, a universal mass for new scalars, a universal mass for new fermions, a universal tri-linear coupling, the tangent of $\beta$ (the ratio of Higgs vevs) and the sign of the $\mu$ parameter discussed above. Having so few parameters makes it easier to draw conclusions about the model space as a whole, but can also be overly restrictive, and large portions  of the paramater space of the CMSSM are already disfavored by experiment \cite{Strege:2012bt}. Models with Non-Universal Higgs Mass (NUHM) \cite{Ellis:2002iu} have a few more additional parameters, but can still be considered restrictive. The 19 parameter Phenomenological Minimal Supersymmetric Standard Model or PMSSM \cite{CahillRowley:2012cb} makes no assumptions about the high scale
theory of supersymmetry and only specifies parameters at a low
scale. While it is more flexible than a SUSY model that specifies how
SUSY is broken it can also contain these models as a subset. The only assumptions that enter into the PMSSM are motivated by the minimum requirements of experimental consistency, such as minimal flavor violation. 

	All of the theories discussed in this section are still just models of the MSSM, the minimal case, and non-minimal models are also possible, some of which are highly motivated, such as the Next to Minimal Supersymmetric Standard Model (NMSSM) \cite{Ellwanger:2009dp}. The only addition in the NMSSM is the superfield for an extra Higgs-like singlet, whose superpartner is called the singlino which mixes with the other neutral Higgs and Gauginos so there are now five neutralinos. This additional singlet can dynamically generate the $\mu$ parameter, alleviating some of the remaining tuning in SUSY models, but it also leads to different predictions for how the Higgs mass is calculated, so that some heavier sparticles may still be consistent with a light Higgs without tuning.

One last motivation for supersymmetric models should be mentioned before moving on. As shown in the literature for models with different types of supersymmetric breaking, different high scale physics lead to different predictions for low scale SUSY. The inverse is also true, if low scale supersymmetry is discovered, and we know that is broken at some higher scale, then it may be possible to learn about this additional hidden sector. In the introduction to this thesis it was emphasized that one shortcoming of the Standard Model is that it may not lead smoothly into its successor, but if supersymmetry is realized, we not only have the next step, but also we may have some insight into physics beyond the MSSM as well. This is usually celebrated in the literature in how supersymmetry can be embedded in string theories, but it is not emphasized enough that this possible connection to higher scale physics is far more generic and not just limited to string theories.

With supersymmetry and dark matter briefly reviewed, the last ingredient needed for axino phenomenology is a basic understanding of axions, which comes in the next chapter.

\chapter{Axions}
\label{ch:axions}

\section{Motivations and Models for Axions}
\label{sec:axmotive}
 Axions as a dark matter candidate are appealing because their original motivation has nothing to do with dark matter. The
Standard Model (SM) QCD Lagrangian allows for CP violation via the
term
\begin{equation}
{\cal L}_{\theta}=\bar{\theta}\frac{\alpha_s}{8\pi}F_{b}^{\mu\nu}\tilde{F}_{b\mu\nu} \;,
\end{equation}
where $F_{b\mu\nu}$ denotes the gluon field strength tensor,
$\tilde{F}_{b\mu\nu}$ its dual, and $b$ is a color index. The gluon field strength tensor is defined as

\begin{equation}
\label{eq:gluonfield}
{F_{\mu\nu}^{a}=\partial_{\mu}A_{\nu}^{a}-\partial_{\nu}A_{\mu}^{a}\mp g_{s}f^{abc}A_{\mu}^{b}A_{\nu}^{c}},
\end{equation}
where $A$ is the gluon field, $ g_{s}$ is the strong coupling constant, and $f^{abc}$ is the structure constant for SU(3). The
parameter $\theta$ is constrained to be $<10^{-9}$ from measurements
of the neutron EDM \cite{Peccei:2006as}. The existence of such a
small parameter is known as the strong CP problem. As shown by Peccei
and Quinn, ${\cal L}_\theta$ can be made to vanish naturally by
introducing a pseudoscalar field, the axion field ($a$), and
requiring the SM QCD Lagrangian to be invariant under a global
$U(1)_{PQ}$ symmetry, which is spontaneously broken
\cite{Peccei:1979}:
\begin{equation}
\label{eq:pqaxion}
{\cal L}_{QCD}={\cal L}_{QCD,SM}+{\cal L}_\theta-\frac{1}{2}\partial_{\mu}a \partial^{\mu}a+{\cal L}_{\mathrm{int}}[\partial^{\mu}a/f_{a};\Psi]+\xi\frac{a}{f_{a}}\frac{\alpha_s}{8\pi}F_{b}^{\mu\nu}\tilde{F}_{b\mu\nu}
\end{equation}
The term
$\xi\frac{a}{f_{a}}\frac{\alpha_s}{8\pi}F_{b}^{\mu\nu}\tilde{F}_{b\mu\nu}$
acts as the axion potential. The axion field obeys a shift symmetry,
$a_{\mathrm{phys}}=a-<a>$, with the potential being minimized by
$<a>=-\frac{f_{a}}{\xi}\bar{\theta}$, so that ${\cal L}_{\theta}$ is
canceled when ${\cal L}_{QCD}$ is expressed in terms of
$a_{\mathrm{phys}}$. The strength of the axion's interactions is suppressed by
the Peccei-Quinn scale $f_{a}$. When $f_{a}$ is sufficiently large the
axion becomes a viable dark matter candidate. As mentioned in chapter 2, the axions role as dark matter candidate is very different from stable WIMPs.  For the axion to be a dark matter candidate the scale $f_{a}$ must be small enough that its average lifetime is longer than the age of the universe, but it does not strictly have to be stable.
	
The mass of the axion is determined by QCD instanton effects, and is given by 

\begin{equation}
\label{eq:axmass}
m_{a}=\frac{\sqrt{z}}{1+z}\frac{f_{\pi}m_{\pi}}{f_{a}}
\end{equation}
where $f_{\pi}$ and $m_{\pi}$ are the pion decay constant and mass
respectively and $z$ is the ratio of light quark masses $\frac{m_{u}}{m_{d}}$. Since $f_{a}$ is the only free parameter here, one can determine the upper bound on axion mass for stable dark matter to be ~24 eV (\cite{Baer:2014eja}). This may make it seem as though theories of axions are one parameter models, where determining the Peccei-Quinn scale will give all the information needed for phenomenology, but this is not so. In addition to the coupling to gluons in the axion potential there also model dependent couplings which appear in the explict form of ${\cal L}_{int}$.

The main source of variability in axion
models is in the choice of what other fields have a charge under the
$U(1)_{PQ}$ symmetry. The properties of these other particles with the
Peccei-Quinn (PQ) charge determine the model dependent factors in the couplings. The
two models most widely studied for axion dark matter are the
Dine-Fischler-Srednicki-Zhitnitsky (DFSZ) axion
\cite{Dine:1981rt,Zhitnitsky:1980tq} and the
Kim-Shifman-Vainshtein-Zhakharov (KSVZ) axion
\cite{Kim:1979if,Shifman:1979if}. In the DFSZ model, standard model fermions and an additional Higgs
boson carry the new charge, while in the KSVZ model there are one or
more new heavy quarks that carry the charge. Other generic QCD axions
are possible that have both new Higgs bosons and new quarks with PQ
charge. 

All of the interactions of the axion, including the coupling to gluons come about from exchanging these PQ charged particles, so the couplings can be in theory be very different between models, though all interactions are suppressed by  $f_{a}$ . The usual rule of thumb (which will be dropped in the next chapter) is to assume the model dependent factors are of order unity. Under the assumption of this rule, these axion models do simplify to one parameter theories and the only model space to explore is over different ranges of  $f_{a}$ .

\section{Searches for Axions}
\label{sec:axsearch}

 A summary of all the current constraints on
$f_{a}$ is shown in Fig.~\ref{fig:axconstraints}~\cite{Agashe:2014kda}. As can be seen in the figure, a very wide range of  $f_{a}$ has already been explored, though there are gaps.  While an important exception is discussed in the next chapter, the usual allowed range for  $f_{a}$ is considered to be $10^{9}\mbox{ GeV}<f_{a}<10^{14}$~GeV, resulting in a very light and extremely weakly coupled axion. The earliest searches for axions in the MeV range were only motivated by the solution to the strong CP problem, as such axions would have decayed away too quickly to dark matter candidates.  These were quickly ruled out in collider experiments, such as beam dump experiments such as those in \cite{Riordan:1987aw}.  Beyond these early collider experiments, the majority of axion searches have primarily probed the axion coupling to photons.  At the other extreme of  very weakly coupled axions,  the Axion Dark Matter eXperiment (ADMX)\cite{Rybka:2010zza} look for axion photon conversion in a cavity of very high strength magnetic field (8 Tesla). This type of search clearly does not fit with any of the main categories of methods described in chapter 2.  The intermediate range limits on  $f_{a}$  all come from cosmology and astrophysics, but again, most of these look for effects that depend on the coupling to photons. The slow decay of axions to two photons should give predictable monochromatic lines at energies characterized by the axion's mass, this is the exclusion range listed as ``telescope". The presence of extra light degrees of freedom could allow for stellar objects  to cool more rapidly. These gives the exclusion ranges from the burst duration part of SN1987A limits and the white dwarf limits in particular look at the coupling to of axions to electrons. The globular cluster limits also come from a possible cooling effect as well. Enhanced cooling would mean that the length of different stages of a the stars lifetime would be effected differently. Stars burning helium will have their lifetimes affected far more significantly than those still burning primarily hydrogen, and the difference in relative lifetimes means that the relative populations would be re-weighted in the presence of axions \cite{Ayala:2014pea}. The ``Hot-DM" bound from figure ~\ref{fig:axconstraints} comes from the consideration that there cannot be too much of a hot dark matter species in the early universe as dicussed in chapter 2 on dark matter. This bound shows that axions heavier than 0.02 eV would constitute too much of a hot relic. Naively, the hot dark matter bound is a lower bound on mass when considering other dark matter candidates such as WIMPs; without stronger interactions to slow it down, dark matter requires gravity to rob it of kinematic energy (cool it) and sufficiently light particles will not cool enough. This is not the case for axions because of the relationship between coupling and mass. For larger axion masses the coupling becomes larger (the suppression scale shrinks). On the surface this is still counter-intuitive, a larger coupling, a stronger interaction, should allow the axion dark matter to cool more efficiently through interactions, but this not the correct way to interpret this limit. Even below the hot dark matter bound in the figure the axions are too light to be thermally produced as cold dark matter, but the catch is that at those coupling strengths they are not thermally produced in a large enough quantity to constitute a hot relic anyway, and so axion dark matter requires non-thermal production mechanism with some tiny population of axions left over. When the axions interactions become too strong this hot population is sizable enough that it complicates the cosmology.

\begin{figure}
\centerline{\includegraphics[width=4.5in]{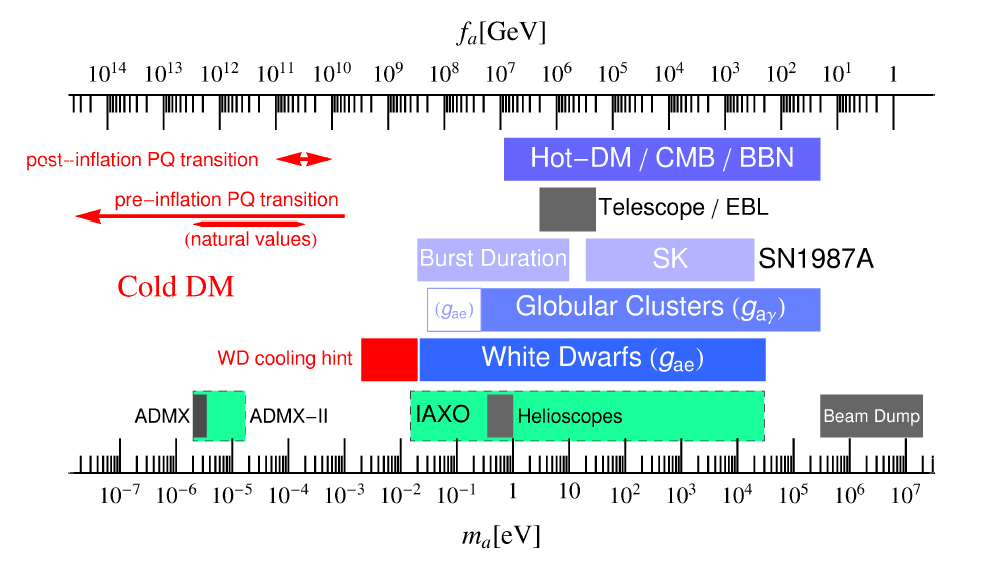}}
\caption{Exclusion ranges for the Peccei-Quinn scale $f_{a}$ from
  various constraints as described in~\protect\cite{Agashe:2014kda}. \label{fig:axconstraints}}
\end{figure}
\section{Cosmology of Axions}
\label{sec:axsearch}

The most common way to generate a cold population of axions in the literature is by what is called the misalignment mechanism. In the early universe, when the temperature of the primordial plasma was above QCD scale, the then massless axion field takes random values around the minimum of its potential. When the universe cools below the QCD scale the potential becomes ``tilted" from QCD instanton effects, causing the axion field to now oscillate around the minimum. As the axion field relaxes into its minium the potential energy produces the relic abundance of cold dark matter axions as a bosonic condensate. The amount of dark matter produced this way depends on both the size of  $f_{a}$ and the intial value of the field when oscillations begin. There are several possible complications to this production mechanism which depend on more details of the cosmology \cite{Baer:2010kw}.  One obvious consequence of this form of production is that the cold dark matter forms a Bose-Einstein condensate which can lead to non-trivial (and observable) structures in dark matter halos \cite{Banik:2015sma}. Another complication from this form of axion production is that it can lead to topological defects in the form of cosmic strings and domain walls \cite{Sikivie:1999sy}. These are expected to arise because the intial value of the axion field will not be the same everywhere in the universe. In KSVZ axion models these domain walls are unstable and should harmlessly decay into a smaller populations of cold axions. This problem is complicated further in DFSZ models where there are multiple degenerate vacua for the axion potential, and the domain walls are expected to be stable and particular implementations of this cosmology will need to find a way around this domain wall problem.

\section{SUSY with Axions}
\label{sec:axsusy}

The role of the axion in a SUSY model can be more complicated than
simply complementing the neutralino relic abundance. Embedded in a
SUSY model, the axion is a member of a supermultiplet, joined by a
neutral, $R$ parity odd, Majorana fermion, the axino, and the
$R$ parity even scalar saxion \cite{Kim:1983ia}.  The neutralino, axion
and axino are all valid dark matter candidates; which particle (or set
of particles) actually plays the role of dark matter in a given model
depends on their mass hierarchy and the cosmology, so there are a
variety of possible scenarios. These scenarios have been studied
extensively
in~\cite{Baer:2009vr,Baer:2011hx,Bae:2013hma,Baer:2011uz,Baer:2010wm},
where it has been pointed out that these are attractive possibilities,
accommodating dark matter, solving the strong CP problem, and
sometimes with other benefits such as Yukawa unification
\cite{Baer:2012cf} or being embedded in a full GUT theory
\cite{Baer:2012jp}. These scenarios, that add the axion supermultiplet
to the MSSM, often have an easier time accommodating naturalness in
SUSY, if for nothing else then because an extended model has more
parameters and is more flexible. This lower tuning is usually achieved
by allowing the lightest ordinary SUSY particle (LOSP) to be the
Higgsino, so that the SUSY Higgs mass parameter, $\mu$, can be small
\cite{Baer:2013wza,Baer:2012cf}. For DFSZ axion models there is an even more obvious way to alleviate tunings issues in SUSY by invoking the Kim-Nilles mechanism ~\cite{Kim:1983dt}. As a pseudo scalar goldstone boson the axion already resembles an additional Higgs, and in the DFSZ model it couples to the Higgs directly. Through this coupling and by extension its supersymmetric version the scales $f_{a}$ and $\mu$ can be related by $\mu\sim\frac{f_{a}^{2}}{M_{p}}$. For the range of $f_{a}$ usually favored by the constraints discussed above, this results in a  $\mu$ scale around the weak scale. The precise value of $f_{a}$ will determine the degree to which  the  $\mu$ problem is alleviated, but as already stated the degree of tuning which is ``bad" in a model can be a subjective matter.

\chapter{Axinos at the Large Hadron Collider}
\label{ch:background}

\section{Difficulties and Existing Searches}
\label{sec:diffsearch}

In the previous chapters the theories of axions and supersymmetry were separately motivated. Both can alleviate fine tuning in the Standard Model in a natural way and both can introduce dark matter candidates. Together in one theory, supersymmetry and axions can have many prospective benefits which have been explored in the literature. The possible tuning of the dark sector of SUSY can be relaxed with the addition of the axion and the axino. The way the dark matter solution is realized in these combined models can be very different than either case alone, as both the LSP and the axion can provide some of the relic abundance. The LSPs abundance will also be modified by the extended SUSY spectra of the combined model. An unfortunate drawback to combining these ideas is that it adds very little to the phenomenology. Some SUSY spectra may be preferred in scenarios that additionally have axions and axinos, but there is no reason these spectra necesarily require that particular extension to the dark sector. Separately detecting the axion in future experiments may help to resolve the situation, but the work here is primarily concerned with the possibility of models with more direct evidence where the axino itself has consequences for collider phenomenology. There already is some work in this field for particular models, and it will be reviewed briefly in this chapter, but the focus of this thesis is to explore a class of models with an axino signature that has been overlooked up to this point.
The usual wisdom that prevents axinos from being considered for
collider phenomenology is their extremely weak coupling. This
extremely weak coupling, though often considered a defining
characteristic of axions, is not actually a necessity, and there are
viable models where the coupling suppression is not as severe. All the
couplings of the axion are suppressed by the scale $f_{a}$, which in
theory can take any value. Axions as dark matter candidates are
usually only considered with
$10^{9}\mbox{ GeV}<f_{a}<10^{14}$~GeV~\cite{Kim:2008hd}, with string theory
axions typically towards the higher end of this range
\cite{Svrcek:2006yi}.  For an axion to solve the strong CP problem,
the important ingredient is the axion potential of
Eq.~\ref{eq:pqaxion}, which comes from the coupling between an axion
and gluons, so that $f_{a}$ is a free model parameter, only determined
by experimental and observational constraints.  First proposals in
\cite{Weinberg:1977ma,Wilczek:1977pj} of a particle axion associated
with the Peccei-Quinn mechanism, posited that the scale $f_{a}$ was
connected to the weak scale, but such a value was quickly ruled out
\cite{Fairfield:1988du}. In
practice the axion-gluon coupling is difficult to measure, so most
constraints come from various measurements of the couplings in ${\cal
  L}_{int}$ above, which does depend on $f_{a}$, but also contains
model dependent factors.

	The model-dependent nature of these limits can be exploited to
conceive models appropriate for collider phenomenology, but first it
is worth considering why the typical limits are so restrictive. If the
usual lower bound of $f_{a}>10^{9}$~GeV is taken at face value then
collider studies for axions (or equally suppressed axinos) are very
limited.  Direct production rates of axions/axinos are simply too
small with such a suppressed coupling, but an incredibly weak coupling
may still be probed in certain cases, usually by taking advantage of
$R$ parity.  If an extremely weakly coupled particle (such as an axino
or gravitino) is the lightest particle with an odd $R$ parity then their
appearance at the end of a SUSY decay chain is inevitable, regardless
of the coupling size. One way to take advantage of this has already
been mentioned: if the NLSP is charged then its delayed decay to the
suppressed LSP will leave a track. These are referred to as heavy
stable charged particle (HSCP) searches when the charged NLSP has a
long enough lifetime.  While a charged LOSP cannot be the LSP for
cosmological reasons, there is no such constraint when there is a
lighter SUSY particle to decay to. This type of search can be used for
both gravitinos and axinos as LSPs. In gauge mediated SUSY breaking
(GMSB) models, there is a charged NLSP for a wide range of parameter
space \cite{Giudice:1998bp}, but for other SUSY breaking mechanisms
this may not be the case. Furthermore, because HSCP searches are so
effective, many models with a charged NLSP and a very weakly coupled
LSP are already constrained. Even with neutral NLSPs, $R$ parity can
still be exploited in much the same way to look for extremely weakly
coupled particles, but with a less spectacular signal. $R$ parity still
requires the LSP to be at the end of any SUSY decay chain, and while
there is no longer a charged track, the visible decay products in the
last leg will be displaced. Even though $R$ parity forces a branching
fraction of one for the LSP, the very weak coupling of the LSP still
has an effect: in determining the width of the NLSP which translates
to how displaced the last leg of the SUSY cascade will be. If the
suppression factor is great enough, then the displaced decay occurs
completely outside the detector, and so the scenario is
indistinguishable from one without the extra particle in the final leg
at all. The solution proposed in this case is to add additional detector material outside of the rest to catch these extra long lived NLSPs. 

	If these types of searches are taken seriously for gravitinos
(and they are \cite{Chatrchyan:2012ir,Aad:2014gfa}), one would hope
that this could be exploited for axinos, but there are a couple of
technical differences that make this difficult. Naively one would
assume these searches are even harder for gravitinos since their
interactions are suppressed by the Planck scale, which dwarfs even the
higher values of $f_{a}$ that are considered in the literature.  While
the gravitino's interactions are suppressed by the Planck scale, they
are also inversely proportional to the gravitino mass and so the
suppression is not as great as one may naively expect
\cite{Ambrosanio:1996jn}. The partial width of a neutralino decaying to a gravitino and a photon, for example, is given by \cite{Ambrosanio:1996jn}
\begin{equation}
{\Gamma_{\tilde{\chi}_{1}^{0}\rightarrow\gamma\tilde{G}}=\frac{\kappa}{48\pi}\frac{m_{\tilde{\chi}_{1}^{0}}^{5}}{M_{P}^{2}m_{\tilde{G}}^{2}}} \;,
\end{equation}
where $\kappa$ contains the relevant mixing information for the neutralino, and the masses in the denominator are the Plank mass and the gravitino mass. In effect, the coupling strength can be
tuned to any value provided there is freedom in choice of the
gravitino mass (which can take a wide range of values depending on the
model). For extremely small gravitino masses, searches for displaced
or even prompt decays of NLSPs become possible. In displaced decays to
axinos, the effective coupling is relatively insensitive to the axino
mass and only depends strongly on $f_{a}$ and any axino with
$f_{a}>10^{9}$~GeV is expected to be completely invisible at
colliders, as any decaying NLSP will always leave the detector
\cite{covi:11:moriondEW}. Very recently an exception to this common
wisdom was explored in \cite{Barenboim:2014kka}, where the authors
showed that a Higgsino NLSP can decay to an axino LSP inside the
detector easily even with $f_{a}>10^{9}$~GeV, when there is a direct
coupling between Higgsinos and axinos (and an appropriate mass
spectrum) as in the case of DFSZ axions.

For DFSZ axions, searches for displaced decays of a neutral NLSP are possible because
of the coupling between axinos and Higgs bosons/Higgsinos, but for the
other main class of axion model, the KSVZ axion, it seems that 
collider studies are only possible if the constraints on $f_{a}$ can 
be relaxed. Finding a scenario with a lower $f_{a}$ does not in effect fix the problem with these existing scenarios, but rather it is simply identifying another equally likely part of parameter space that just has the advantage of being testable. One assumption implicit in the constraints summarized
by figure~\ref{fig:axconstraints} is that the mass of the axion is determined by the suppression
scale, given by equation~\ref{eq:axmass}.
Changing the relation between mass and coupling for the axion does
indeed open a new window of lower coupling, but with an explicit mass
term, or even a radiatively induced mass, so that the shift symmetry of the
axion field is spoiled, and the solution to the strong CP problem
is lost. Axions such as this are usually set aside in the broader
category of ALPs, and their collider phenomenology has recently be
proposed in \cite{Mimasu:2014nea}. Aside from losing the solution
to the strong CP problem, these more strongly coupled ALPs are not
guaranteed to be dark matter candidates, as they are no longer lighter
than the SM particles they couple to, and they do not have to be protected
from decays by any symmetry. 

\section{Evading Axion Constraints}
\label{sec:constraints}

Retaining the mass relation, and along with it the solution to the
strong CP problem and the possibility of dark matter, makes it more
difficult to avoid the constraints on $f_{a}$. Limit plots, such as
figure~\ref{fig:axconstraints} where all the constraints are given in terms of
$f_{a}$, often do not easily reveal the underlying model-dependent
assumptions in extracting these constraints.  After closer inspection,
it turns out that in the KSVZ model there is the intriguing
possibility of evading most of the constraints in a way not possible
in the DFSZ model. In the KSVZ model the axion coupling to leptons is
vanishing at tree level, and the effective coupling to leptons at one
loop has been shown to be non-constraining
\cite{Srednicki:1985xd,Kaplan:1985dv}. As a result, a whole category
of constraints, the limits from white dwarf cooling, are irrelevant
for the KSVZ axion. In the DFSZ model, the coupling to leptons is
always non-vanishing. Aside from the white dwarf limits, the coupling
most often tested is the axion coupling to photons, given by
\begin{equation}
{\cal L}_{a\gamma\gamma}=\frac{\alpha}{4\pi}K_{a\gamma\gamma}\frac{a_{\mathrm{phys}}}{f_{a}}F^{\mu\nu}\tilde{F}_{\mu\nu} \; .
\end{equation}
where $F^{\mu\nu}$ is the photon field strength tensor and 
$K_{a\gamma\gamma}$ parameterizes the model-dependent axion-photon coupling. This too
can be made to vanish in the KSVZ model (but not the DFSZ). In KSVZ
the coupling to photons is determined by 
\begin{equation}\label{eq:kgg}
K_{a\gamma\gamma}=K_{a\gamma\gamma}^{0}-\frac{2(4+z)}{3(1+z)} \; ,
\end{equation}
and the difference can be thought of as a balancing between short and
long-range effects. The short-range effect $K_{a\gamma\gamma}^{0}$
comes from the chiral anomaly depending on the electromagnetic
charge(s) of the new heavy quark(s). The second term in
Eq.~\ref{eq:kgg} comes from the axion's mixing with light mesons and
depends on the value of $z$, the ratio of light quark masses, which comes with some uncertainty. For an
appropriate choice of charge(s) for the new quark(s) these two terms
can cancel and $K_{a\gamma\gamma}$ can be made to vanish
\cite{Kaplan:1985dv,Srednicki:1985xd}. It should be noted that to
avoid the existing constraints $K_{a\gamma\gamma}$ does not have to
vanish exactly, but only be so small that it is consistent with the
limits on the photon coupling.

For a KSVZ axion with no photon or lepton coupling, nearly all constraints
are irrelevant. The only constraints in figure~\ref{fig:axconstraints} that are truly inescapable
are those coming from supernova 1987a \cite{Turner:1989vc,Raffelt:1990yz}.
Supernova bounds are useful for testing virtually any new light species,
as all couplings and decays are relevant in events that energetic.
In practice, the new light species do not even have to be detected
from these events, but their properties can be constrained from the
burst duration of the supernova, and the number of particles detected
from the light species of the Standard Model (neutrinos). In the case
of QCD axions this is especially interesting because it directly tests
the otherwise elusive gluon-axion coupling, the only coupling necessary
to solve the strong CP problem, and the only coupling free from model
dependent factors. As seen in figure~\ref{fig:axconstraints}, there are actually two separate
regions of bounds from SN1987a, corresponding to two regimes of coupling
strength. The upper bound range comes from when the axion is so weakly
coupled that it is free streaming after it is produced in the supernova;
had axions like this been present in SN1987a they would shorten the
bust duration as they would assist the neutrinos in carrying away
energy. The lower bound range is from axions that still have interactions
with nuclear material on there way out of the supernova; which would
have affected the number of neutrinos measured on Earth. While supernovae
can be powerful tools in constraining new light species, they must
be close enough to Earth that there is an adequate flux of neutrinos.
To this date SN1987a has yielded the most detected neutrinos of any
supernova observed. Larger, more energetic supernova have been observed,
but either at a greater distance, or before humans had appropriate
instrumentation. As a result of the scarcity of proximal events, the
supernova bounds on axions come from a mere 24 neutrinos \cite{Vissani:2010zi}. While even
this small number of detected neutrinos allows constraints to be made,
it does leave a window between the two regimes of free streaming and
interacting axions in supernovae. This window allows a range of lower
suppression scale $3\times10^{5} {\rm GeV}<f_{a}<3\times10^{6} {\rm GeV}$, known
as the hadronic axion window, which has been examined in the literature
in the past, particularly in the context of axions as hot dark matter
and its cosmological implications~\cite{Moroi:1998qs,Chang:1993gm}. This coupling range still has axions weakly coupled enough that they have lifetimes comparable to the age of the universe and so they are still dark matter candidates.
Though hot dark matter is now greatly disfavored, axions in this window,
while still relatively suppressed, should have coupling strengths
such that their partners, the axinos, can be studied at colliders,
via the general strategy for gravitinos and displaced decays described
above.

\section{Costs of a Scenario with a Smaller Peccei-Quinn Scale}
\label{sec:costs}

Assuming a KSVZ axion with suppression scale $f_{a}$ in the hadronic
axion window does make collider studies a possibility, as will be
discussed in the next chapter, but there are some costs and considerations
for being set in this scenario that should be addressed. Choosing
a KSVZ axion forfeits the use of the Kim-Nilles mechanism to resolve
the little hierarchy problem in SUSY (this explicitly requires new
Higgs bosons with PQ charge as in the DFSZ model). The mechanism would not be attractive at this scale for  $f_{a}$ in any case because the scale of $\mu$ in the Kim-Nilles mechanism is set to be $\mu\sim\frac{f_{a}^{2}}{M_{p}}$. While this puts $\mu$ near the weak scale for the usual extremely weakly coupled axions, the smaller  $f_{a}$ in the hadronic axion window will make the Kim-Nilles mechanism less impressive. There exist other possiblites to alleviate the $\mu$ problem though, most of which require full knowledge of the mechanism of SUSY breaking, so this study will take the route of agnosticism. A possible trade-off to be considered here is that while Kim-Nilles is forfeited by being forced into KSVZ models, this also means that possible domain wall issues, as mentioned in chapter 4, are now avoided.
	
	Another issue is that having the photon coupling vanish in this scenario requires
the cancellation of two terms, so it is easy to expect that this introduces
a new source of tuning, whereas a large motivation of this scenario
was to avoid tuning in both the electroweak and QCD sectors. A simple
way to avoid (but not resolve) this issue is to say the full UV theory
must be known before tuning can be quantified accurately. Two specific
examples of natural models are given in \cite{Chang:1993gm} showing
that these can be easy to build.  Taking a cue from that paper
it should be noted that the amount of tuning here is highly model
dependent, and this level of model dependence does not effect any
of the collider phenomenology that follows, so there is no reason
not to be agnostic.

	 Perhaps more glaring than the possibility of tuning
is that in the literature on the hadronic axion window, the dark matter
in this scenario is always hot, which is now in severe conflict with
simulations of structure formation and CMB measurements. This corresponds
to the ``Hot DM'' bound shown in figure~\ref{fig:axconstraints}, which is independent
of the photon or lepton couplings and clearly overlaps with the hadronic
axion window. This hot dark matter bound is why the hadronic axion
window is usually not considered in the literature anymore, but there
are reasons why this should not be overly concerning. In a SUSY scenario
such as this, the axions will not be the only contribution to the
dark matter abundance, with $R$ parity conservation they are joined
by the LSP axino. Regardless of whatever other components there are
to the dark matter, the hot DM bound states that the model simply
predicts too much of a relativistic species. Although the hot dark
matter bound is model independent with regards to the axions themselves,
it is model dependent as far as the cosmology is concerned. A dangerous hot
thermal relic in one cosmology can be made safe in another
cosmology with a lower reheat temperature. Standard Model neutrinos
were shown to be a viable warm dark matter candidates with this method in
\cite{Giudice:2000dp}, as mentioned in chapter 2, and the same principle has been applied to
axions to alleviate constraints \cite{Grin:2007yg}.  By lowering the reheat temperature sufficiently thermal relics freeze out while the universe is still undergoing inflation, and the relic abundance will be diluted.

	Lowering the reheat temperature does not come without its own consequences.
A low reheat temperature makes leptogenesis in SUSY much more difficult
\cite{Fukugita:1986hr} and losing this attractive possibility can
be considered another ``cost'' of this scenario. While Leptogenesis
is a very popular way to generate a baryon asymmetry in SUSY models,
it is not the only way. Affleck-Dine (AD) Baryogenesis \cite{Affleck:1984fy}
is an attractive alternative to thermal Leptogenesis, that is well
suited for a scenario with axinos in the hadronic axion window for
several reasons. AD Baryogenesis is compatible with low reheat cosmologies
and can generate the baryon asymmetry via RPV couplings \cite{Higaki:2014eda}.
While the hadronic axion window requires a low reheat temperature,
as in the AD scenario, the RPV couplings of AD in turn require that
the SUSY model have some kind of supplemental dark matter since neutralinos
are expected to decay, and so axions/axinos are a good fit. There
is even further reason to find these two scenarios to be an interesting
match. The AD scenario with RPV requires the size of the RPV couplings
to be within a certain range. As will be shown in chapter~\ref{ch:results}, the scale
of RPV in AD, and the size of $f_{a}$ in the hadronic axion window
will make it so it is not unreasonable to expect the NLSP to have
a comparable branching fractions between RPV decays and decays with
axinos. This means that AD Baryogenesis is not only a consistent choice
for axinos in the hadronic axion window, but it may also be a testable
choice, with both an axino signal and an RPV signal observable at
the LHC.

	 One may worry that with RPV the axino LSP is no longer a viable dark matter candidate. The decays of axinos via RPV is explored thoroughly in \cite{Poletanovic:2010yba} for DFSZ models. From the results for DFSZ axinos the expected decay rates for KSVZ axinos can be extrapolated. There are in principle three ways the axino may decay by RPV: by coupling to a Higgs/Higgsino, by mixing with a neutrino or neutralino, and by explicit dependence on both $f_{a}$ and the  RPV coupling. As emphasized earlier, the KSVZ axion/axino lacks the Higgs/Higgsino couplings of the DFSZ at tree level, so these decays will be supressed. By the same reasoning the second class of decays will be supressed as axino/neutralino/neutrino mixing only comes about in the presence of a Higgs/Higgsino coupling with RPV. The last class of decays, that explicitly depend on both couplings are possible, but these should be non-threatening to the cosmology as RPV couplings are constrained to be small, so its product with 1/$f_{a}$  should be adequately suppressive for a range of axino masses.

With these assumptions laid out, the scenario to be studied at the LHC
should be clear: KSVZ axinos with a neutral NLSP, with only a QCD
coupling, and the suppression scale, $f_{a}$, to be considered lying
in the range given by the hadronic axion window:
\[3\times10^{5} \, {\rm GeV}<f_{a}<3\times10^{6} \, {\rm GeV} \]
This scenario is motivated by being perhaps
the only KSVZ model that is testable at a collider without a charged
NLSP. It also may be one of the only ways to probe $f_{a}$ in the hadronic axion window, short of waiting for another supernova close enough to the earth. This scenario also has the possibility of having low tuning
and an interesting cosmology with its own testable consequences. The next chapter explores what the most promising signals for this scenario are at the LHC.
\chapter{Signal and Benchmark}
\label{ch:signal}

\section{Signals for Axino Searches}
\label{sec:signals}

In studying neutral NLSPs decaying to axinos there are two
possibilities for the NLSP: neutralinos and sneutrinos. While
sneutrino LOSPs are constrained by direct detection in models without
extra dark matter candidates \cite{Falk:1994es}, this is no longer the
case when adding a lighter weakly interacting particle. For the sake
of simplicity this study focuses on scenarios with a neutralino
NLSP. Unlike the assumptions made in the previous chapter, which were
necessary for a collider study, this choice is made for convenience. A
sneutrino NLSP is perfectly valid, and is studied in the case of
gravitino LSPs \cite{Ellis:2008as}. This scenario could potentially  requires a large
number of possible search channels to be considered. In contrast, the
restriction to neutralino NLSPs and only the QCD coupling being
allowed in the hadronic axion window, makes for a very predictive
scenario. While the detailed study with event simulation is performed for neutralino NLSPs only, there are some comments at the end of this chapter concerning the possibility of sneutrino decays to axinos.

The supersymmetric version of the axion-gluon coupling is the axino-gluino-gluon coupling~\cite{Baer:2011hx},
\begin{equation}
\label{eq:agg}
{\cal L}_{\tilde{a}\tilde{g}g}=i\frac{\alpha_{s}}{16\pi f_{a}}\bar{\tilde{a}}\gamma_{5}[\gamma^{\mu},\gamma^{\nu}]\tilde{g}_{b}F_{b\mu\nu}
\end{equation}
and is the only coupling available to produce axinos in this
scenario. Here,  $\tilde a$ and $\tilde g_b$ denote the axino and gluino
field respectively.   In the literature this is often referred to as the ``tree level" coupling for the axino, but it is in fact the result of the effective field theory with the new heavy KSVZ quarks (denoted by Q) and squarks integrated out. The loop diagrams that lead to this vertex in the effective field theory are given in figure 6.1. This ${\cal L}_{\tilde{a}\tilde{g}g}$  vertex and the resulting diagrams using this will still be referred to as  ``tree level"  for this work in agreement with the literature to avoid confusion.   Even within the window of lower $f_{a}$, the
suppression is still too great to expect production of axinos at the
LHC unless they follow the NLSP in a decay chain so that there are no
other less suppressed options for decay. Once a neutralino is produced
there is only one dominant topology for its decay to an axino at tree
level (figure.~\ref{fig:ax3jet}), via an off shell squark and an off shell
gluino, resulting in missing transverse energy (MET) and three
displaced jets, (plus whatever SM particles were produced in
association with the neutralino). This topology allows decays to heavy
quarks, but the decay width should be relatively small compared to
that of decays to light quarks, provided the neutralino is not too
massive. At tree level this is the only topology that leads to four
decay products and there are no topologies with a smaller
multiplicity.  Any other decay path from neutralino to axino involves
more final state particles and possibly more massive off shell
sparticles in the decay chain, and so such processes are even more greatly
suppressed to the point where it is negligible compared to the three
displaced jets and MET channel. It is very important however to
consider one loop effects here. The vertex correction to squarks
decaying to axinos, shown in figure~\ref{fig:squarkvert}, provides an
effective squark-quark-axino coupling, ${\cal L}_{\tilde{a}q\tilde{q}}$ ,  that can provide the dominant
decay channel for neutralinos for large swathes of SUSY parameter
space. Note that in the UV complete theory with the KSVZ heavy quarks and squarks included that the ${\cal L}_{\tilde{a}q\tilde{q}}$ vertex is actually a two loop effect, but again, in agreement with convention this will be denoted as the one loop effect of the effective theory with the ${\cal L}_{\tilde{a}\tilde{g}g}$  vertex. This effective coupling was first explored in
\cite{Covi:2002vw}, and with the heavy SUSY states integrated out, this
interaction takes the form
\begin{equation}
\label{eq:aqq}
{\cal L}_{\tilde{a}q\tilde{q}}=-g_{eff} \tilde{q}_{j}^{L/R}\bar{q}_{j}P_{R/L}\tilde{a} \; ,
\end{equation}
where $m_{\tilde g}$ is the gluino mass, $\bar{q}_{j}$ and
$\tilde{q}_{j}^{L/R}$ is the quark and (left or right-handed) squark
field respectively, and $P_{R/L}=(1\pm \gamma_5)/2$, and
the effective coupling given by
\begin{equation} \label{eq:geff}
g_{eff} \simeq \frac{\alpha_{s}^{2}}{\sqrt{2}\pi^{2}}\frac{m_{\tilde{g}}}{f_{a}}\log(\frac{f_{a}}{m_{\tilde{g}}})\; .
\end{equation}
With this effective coupling considered, there is the possibility of neutralino
decay to an axino and two jets (figure~\ref{fig:ax2jet}), and this is
the decay channel we focus on here. With a smaller final state multiplicity the ${\cal L}_{\tilde{a}q\tilde{q}}$  vertex can actually dominat despite the loop supression. The relative strength of ${\cal
  L}_{\tilde{a}\tilde{g}g}$ and ${\cal L}_{\tilde{a}q\tilde{q}}$ was
explored in \cite{Covi:2002vw} with regards to squark decays, where it
was shown that the ${\cal L}_{\tilde{a}q\tilde{q}}$ decay dominates
unless $m_{\tilde{q}} \gg m_{\tilde{g}}$. This also holds true here,
where neutralino decays are mediated by an off-shell squark. In practice, for neutralino decays,  ${\cal L}_{\tilde{a}\tilde{g}g}$ is only dominant once the squark mass is several times the gluino mass. The decay width for $\tilde \chi_j^{0} \to q \bar q \tilde a$ is discussed in more detail in chapter ~\ref{ch:width}.

\begin{figure}
\centerline{\includegraphics[scale=0.3]{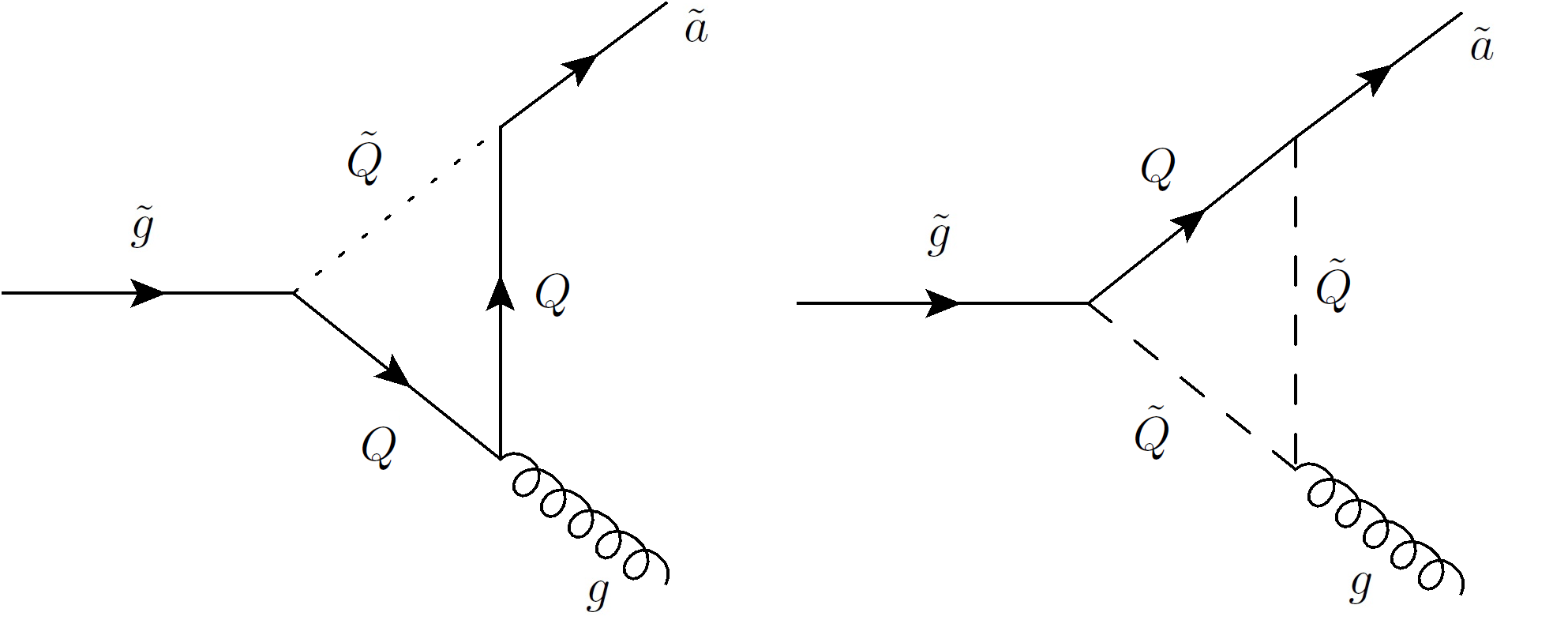}} \caption{The vertex corrections that lead to the effective gluino-gluon-axino interaction, ${\cal L}_{\tilde{a}\tilde{g}g}$ of Eq.~\ref{eq:agg}.\label{fig:gluvert}}
\end{figure}

\begin{figure}
\centerline{\includegraphics[scale=0.3]{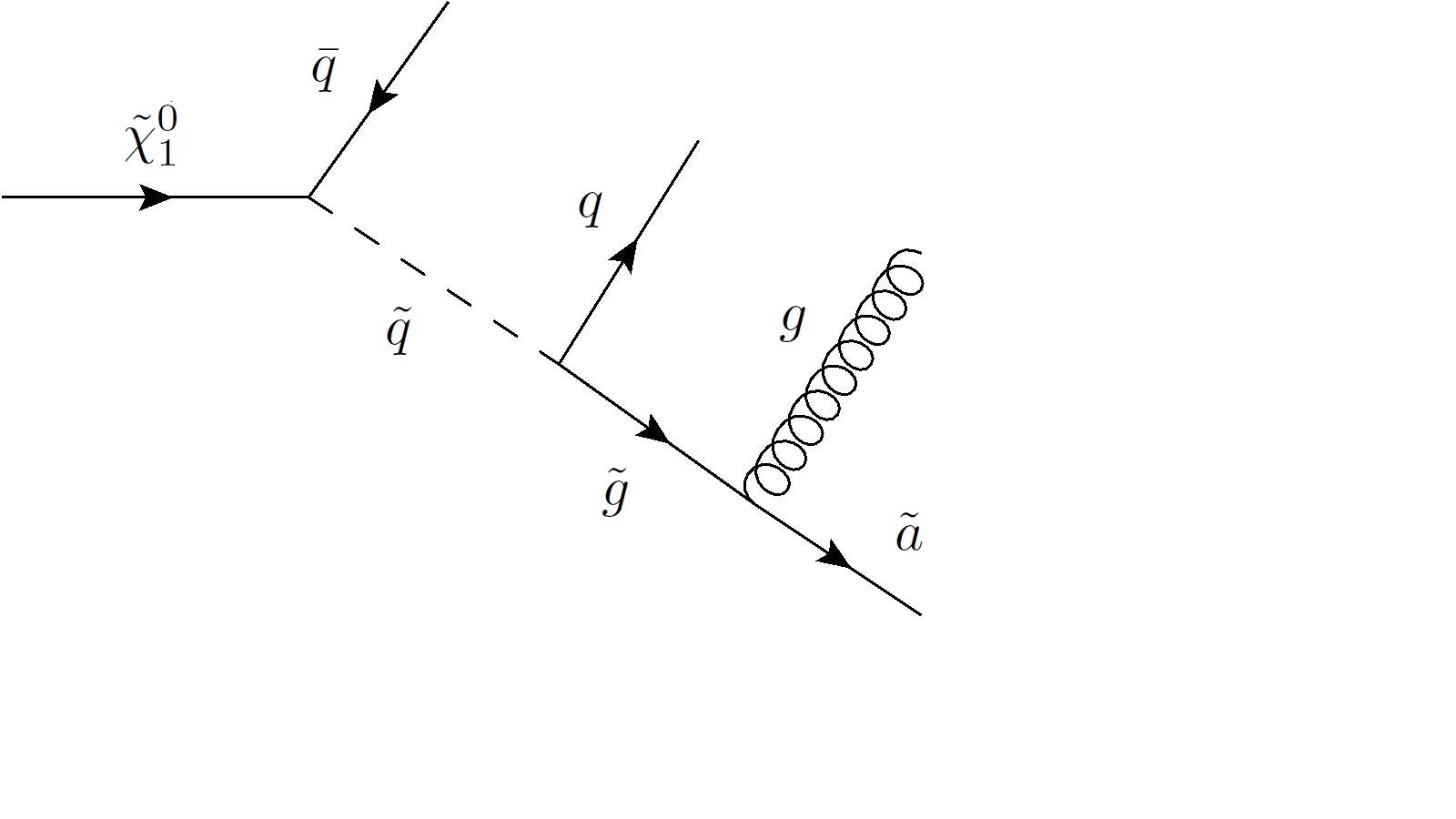}} \caption{Neutralino decay to three jets and an axino via ${\cal L}_{\tilde{a}\tilde{g}g}$ of equation ~\ref{eq:agg}.\label{fig:ax3jet}}
\end{figure}

\begin{figure}
\centerline{\includegraphics[scale=0.3]{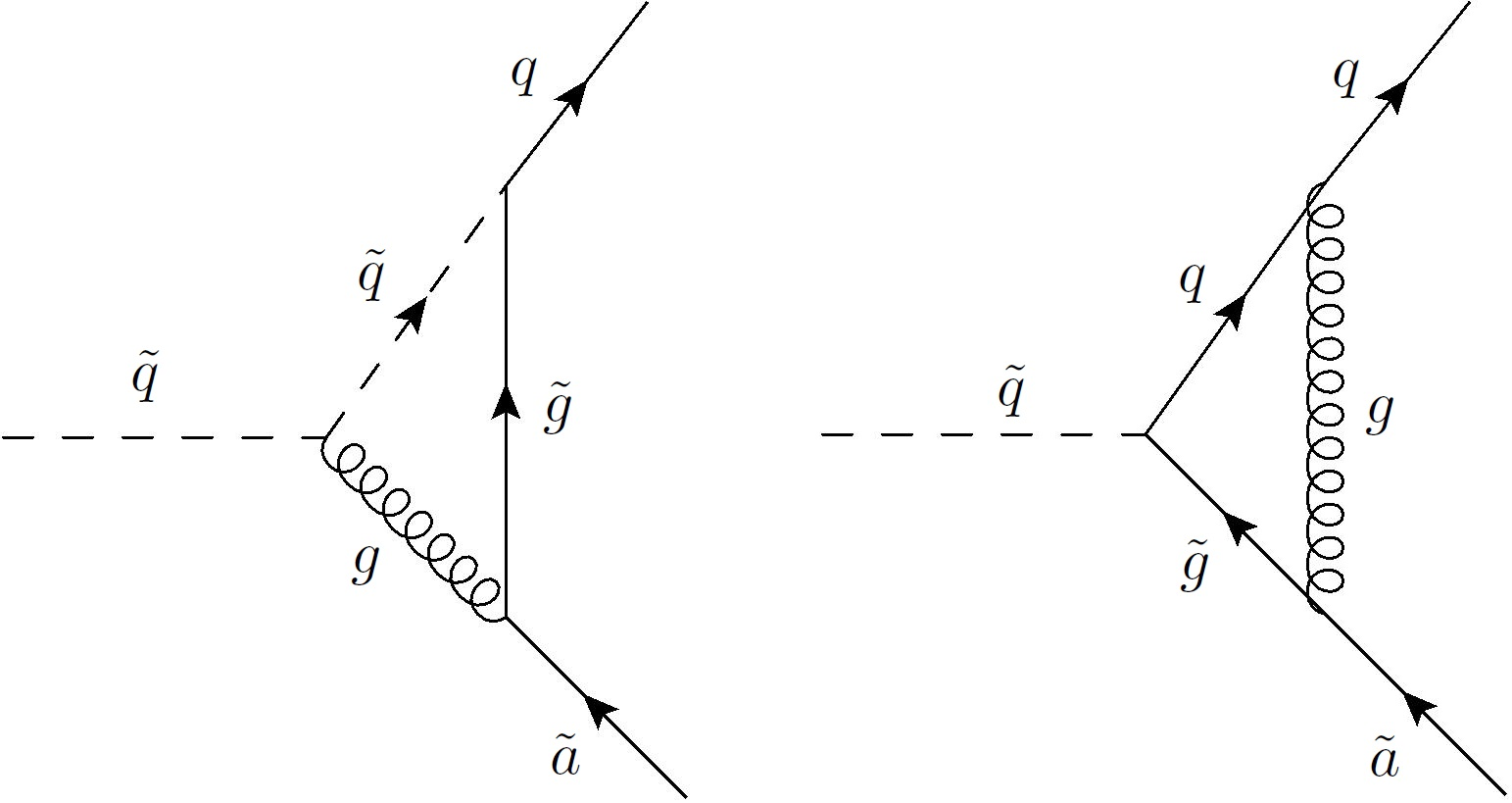}}
\caption{The vertex corrections that lead to the effective squark-quark-axino
interaction, ${\cal L}_{\tilde{a}q\tilde{q}}$ of Eq.~\ref{eq:aqq}.\label{fig:squarkvert}}
\end{figure}

\begin{figure}
\centerline{\includegraphics[scale=0.3]{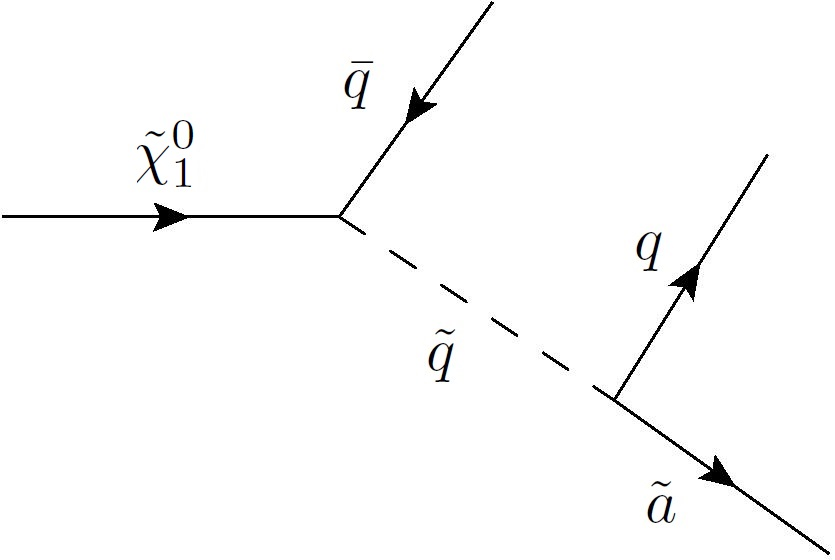}} \caption{Neutralino decay to two jets and an axino via ${\cal L}_{\tilde{a}q\tilde{q}}.$ of Eq.~\ref{eq:aqq}. \label{fig:ax2jet}}
\end{figure}

For two produced neutralinos decaying to axinos, the signal is always
multi-jets and MET, and depending on the SUSY spectra the dominant decay
will either be two or three jets per decay leg (before showering/clustering).
Multi-jets and MET is by far the most commonly studied signal for prospective
new physics at the LHC, but if the jets are displaced enough, then
the signal for KSVZ axinos with a neutralino NLSP can become rather
unique. If the jets are displaced enough when neutralinos decay then
the SM background will become negligible, and the only
competing or alternative source for such a signal would come from
other new physics. Two such alternative sources are displaced decays
of neutralinos to gravitinos and displaced decays of neutralinos to
neutrinos via RPV. These two alternative sources may not only arise
in alternative models, but could all exist consistently in one model,
i.e, a model with axinos, light gravitinos and RPV couplings all at
once is allowed. There would have to be a coincidence of scales for
there to be a sizable branching fraction for the neutralino to each
of these, instead of one mode dominating. Distinguishing between these
sources of highly displaced jets is left to the next chapter, but
for now it should be noted that these types of searches have already
been considered in the literature for gravitinos~\cite{Meade:2010ji} and RPV~\cite{Graham:2012th}, and these
studies can be used as a guide for what can be done with axinos. Besides
removing the SM background, highly displaced jets also allow the SUSY
production channel and the signal to be discussed independently. Regardless of how neutralinos
are produced, either in a simple two to two process, or at the ends
of various long cascade chains, the six jets signal is relatively
unchanged, so long as the displaced jets are what is triggered on.
This means that optimistically, the rate of the displaced jets signal
can be taken as the inclusive SUSY production rate for a given benchmark.
There are however, a few ways the production mechanism will affect
the signal, even for highly displaced jets. Exclusive neutralino pair
production will produce the most highly boosted jets, with longer
and longer decay chains reducing the amount of boost, though this
is likely a small effect due to convolution with parton distribution functions (PDFs). In addition to this distribution of boosts,
the rest of the SM particles produced in decay chains must be considered
when determining the MET of the whole event, and the MET resolution
may vary between decay chains. Another effect to consider is that the
triggers for highly displaced objects usually have isolation requirements,
so that the production channel for neutralinos must not produce calorimeter
activity in a region that points to the displaced decay.

While there are these advantages to considering highly displaced jets,
the drawback is that jet measurements may be difficult in the outer
parts of the detector. The degree to which detailed reconstruction
of jets in the outer detector is possible is beyond the scope of this
work, but at least it should be noted that in similar searches, such
as displaced decays to gravitinos \cite{Meade:2010ji}, the strategy
is to make use of triggers developed for hidden valley searches at
ATLAS \cite{Aad:2013txa}. The hope is that these same triggers could
be used for displaced decays to axinos. ATLAS has an advantage here
simply because of the detector geometry: a larger detector has a chance
to detect particles with longer decay lengths. A hidden valley can
produce displaced jets very similar to gravitinos or axinos, but the
hidden valley is not a particular model, or even frame work of models,
but rather a feature that can arise in various settings, so no attempt
is made in this work to make a direct comparison between a hidden
valley signal and other neutralino decays. A hidden valley is basically a dark sector, a collection of fields that may interact with each other with some strength, but whose interactions through the Standard Model are suppressed, sometimes only accessible through one particular mediator \cite{Strassler:2006qa}. The axino and the other neutralino decay products are not strictly speaking part of a hidden valley, but there phenomenology is very similiar so it seems reasonable to think that triggers developed for the hidden valley would be applicable in this scenario.

\section{Choice of Benchmark Spectra}
\label{sec:benchark}

With the expected signal identified as displaced jets and MET, the
SUSY spectra must be specified to obtain more quantitative results.
An appropriate benchmark SUSY model should meet a few criteria. Two
such benchmark models are chosen here, so that in
chapter~\ref{ch:results} the effect of varying kinematics on the
distributions can be explored. Model 188924 and 2178683, both proposed
as PMSSM benchmarks for Snowmass 2013 \cite{Cahill-Rowley:2013gca} are
appropriate and appealing for several reasons. The spectra of these
models are given in figures~\ref{fig:lightbm} and~\ref{fig:heavybm}. The
important difference between these two models is that model 188924 has a
lighter LOSP neutralino, a bino near 200 GeV, and here it will be
referred to as the ``lighter'' benchmark, while model 2178683 also has an LOSP
bino, but a bit heavier, closer to 500 GeV in mass and will be
referred to as the ``heavier'' benchmark.  Both models have colored
sparticle masses all between 1~TeV and 4~TeV.  These masses are the
relevant model parameters to the topology in figure~\ref{fig:ax2jet},
along with the neutralino mixing and the PQ scale, $f_{a}$.

 As described in chapter 3, the PMSSM is agnostic to high
scale physics and in addition this, this particular set of benchmarks was chosen by the authors as
being testable at a 14 TeV LHC and has been checked thoroughly so that
they evade the gauntlet of existing searches up to this point, including much of the 8 TeV LHC analysis. Many
SUSY models can evade existing constraints, but these model do not
implement any special considerations to do so, they are simply the result
of a scan of the large PMSSM parameter space, and so can be thought of
as  ``generic'' SUSY models that may be realized in the next run of
the LHC. All the models in this collection are stated to have possible
dark matter candidates, in that they do not over saturate the relic
abundance, but this point is irrelevant in this context since the neutralino
LOSPs will all decay to axinos in the scenario here. In addition to
these features which are common to all of the PMSSM benchmarks
described in \cite{Cahill-Rowley:2013gca}, the benchmarks for this
specific scenario of neutralino decays to axinos requires a few more
features. The total neutralino width should be in a range such that
the decay is clearly displaced from the primary vertex, but still
within the ATLAS detector. The range considered appropriate for this
is between 0.1~m and 10~m. The analysis is also easier if only one of
the two possible decays (two jet or three jet per leg) is clearly
dominant, so that there is no issue of double counting and matching
with the number of jets. For both the lighter and heavier benchmarks
chosen here with a gluino heavier than most of the squarks, the 3 jet (${\cal L}_{\tilde{a}\tilde{g}g}$)
channel is suppressed by several orders of magnitude compared to the
2-jet channel (${\cal L}_{\tilde{a}q\tilde{q}}$), so that the only coupling that needs to be considered
is ${\cal L}_{\tilde{a}q\tilde{q}}$ of Eq.~\ref{eq:aqq} and the only
relevant topology is that shown in figure~\ref{fig:ax2jet}. Also, with
neutralinos in this mass range, the branching fraction to heavy quarks
is greatly suppressed so that the neutralino branching fraction to two
light jets and an axino is very nearly one. There should also be an
adequate rate for a signal, which optimistically can be taken as the
inclusive SUSY rate. At the 14 TeV LHC the total inclusive SUSY cross
section for the lighter model benchmark is $\sigma_{\mathrm{SUSY}}=5.4$~fb, and for the
heavier one is $\sigma_{\mathrm{SUSY}}=23$~fb, as obtained at leading-order
with MadGraph/MadEvent \cite{Alwall:2014hca}. Several of the
benchmarks in this collection actually satisfied all of these
criteria, and the lighter benchmark model 188924 was only chosen because it
has the added appeal of relatively light sparticles, especially with
relatively light Higgsinos near 270 GeV, indicating that this
benchmark may have lower fine tuning. The heavier benchmark was simply
chosen because a heavier neutralino will have an impact on the
kinematic distributions used to distinguish between different
neutralino decays (axino/gravitino RPV) as will be shown in
chapter~\ref{ch:results}. A discussion of the phenomenology's sensitivity to the choice of benchmark is explored in the next chapter.

\begin{figure}
\centerline{\includegraphics[scale=0.275]{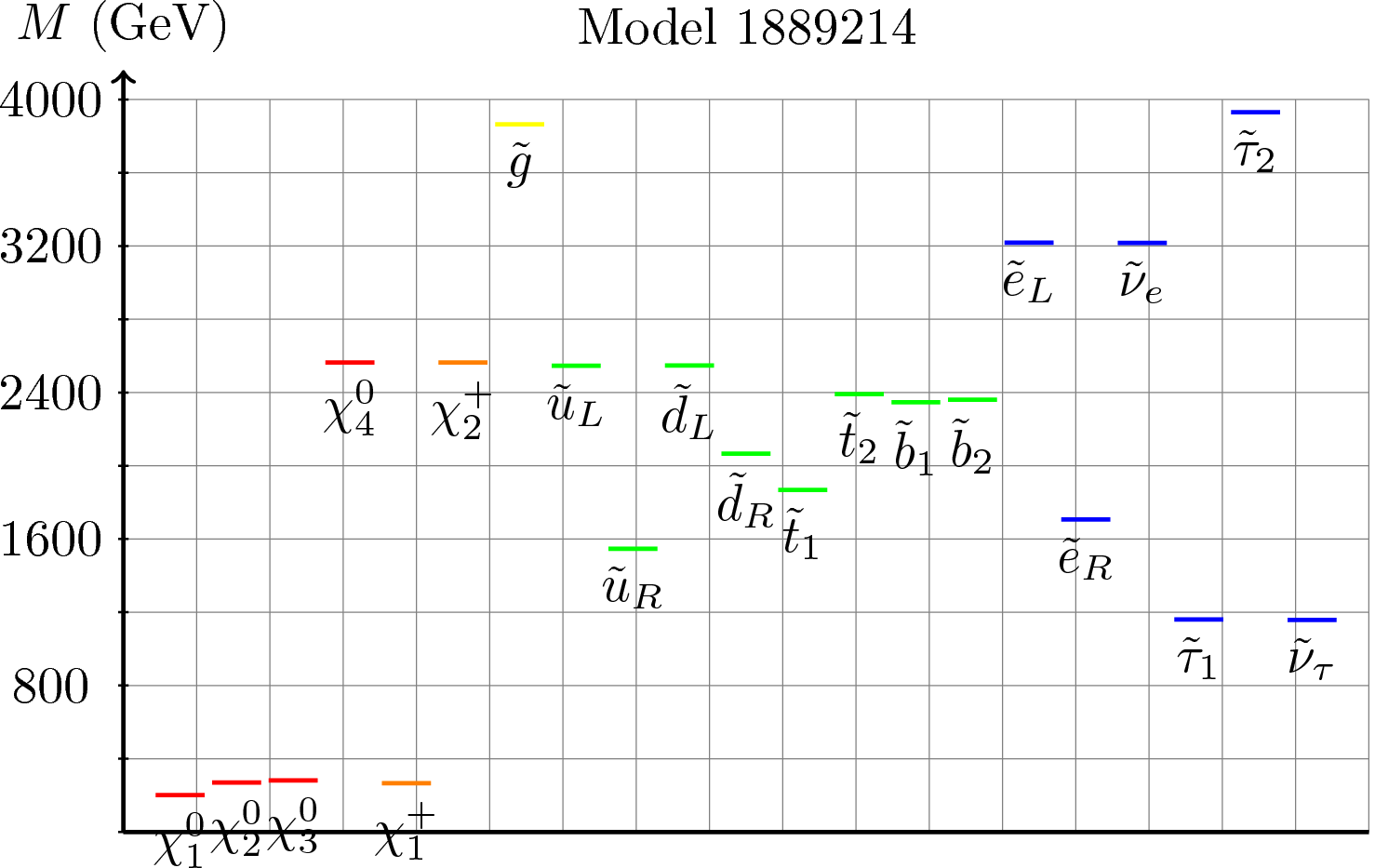}}
\caption{The ``lighter'' benchmark with an LOSP bino at $\approx 200$~GeV, taken from \protect\cite{Cahill-Rowley:2013gca}. \label{fig:lightbm}}
\end{figure}

\begin{figure}
\centerline{\includegraphics[scale=0.275]{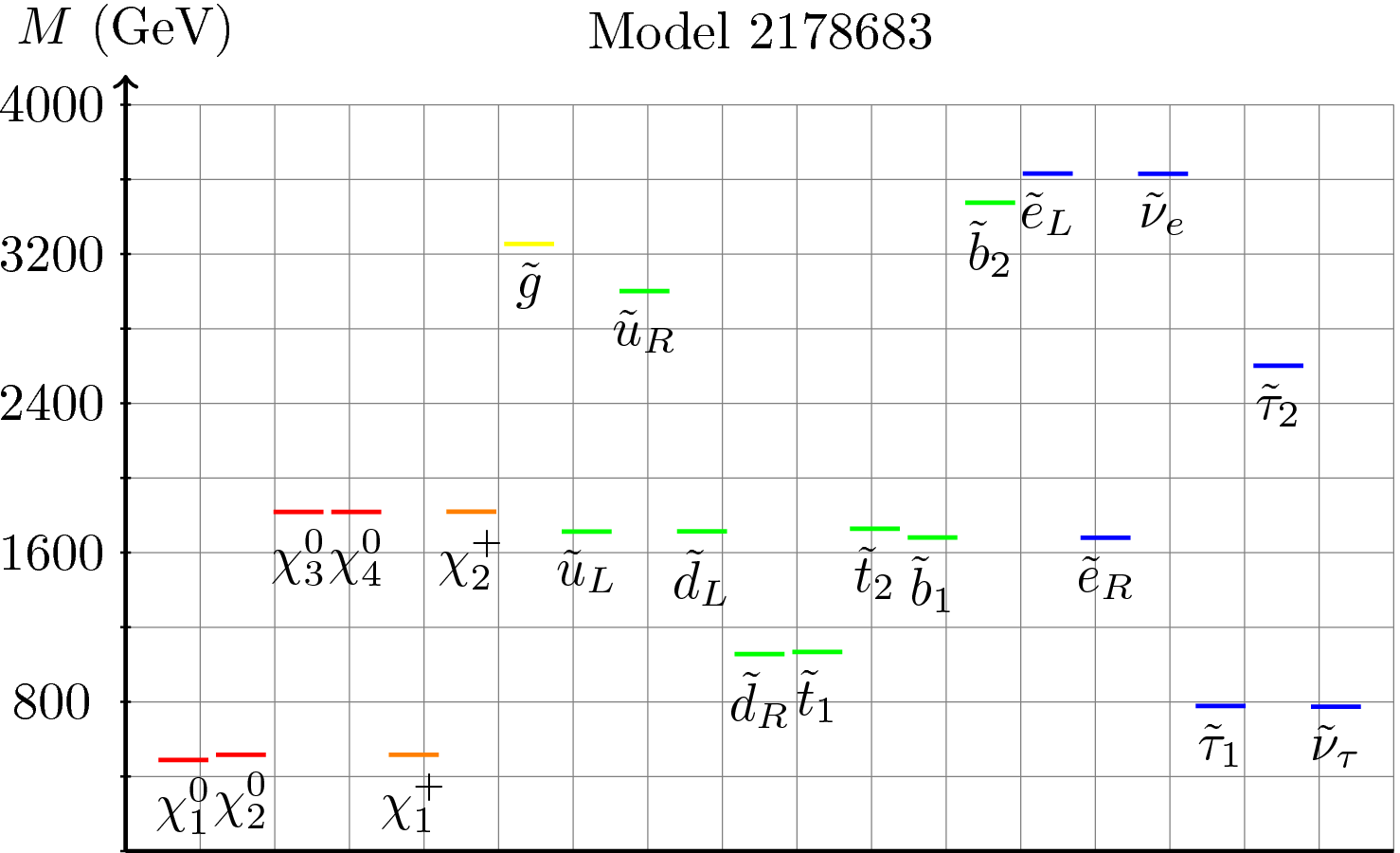}}
\caption{The ``heavier'' benchmark with an LOSP bino at $\approx 500$~GeV, taken from \protect\cite{Cahill-Rowley:2013gca}.\label{fig:heavybm}}
\end{figure}

\section{Other Parameters}
\label{sec:parameters}
	Several SUSY masses and parameters are not specified in these benchmarks.
The gravitino, for the purposes of this study, will be assumed to
be heavy enough that it does not effect the collider phenomenology.
Note that this does not have to be very heavy, as only very light
gravitinos are expected to be relevant to collider phenomenology. Gravitinos heavy
enough to decouple from collider physics can still have a large effect
on the cosmology. The only remaining masses are those from the axion
supermultiplet. The KSVZ axions mass is directly determined by the
scale $f_{a}$ so in the hadronic axion window these axions are still
very light, with a mass of approximately 10 eV. The axion is still
too weakly coupled to have an effect on collider studies, and since
it is $R$ parity even, there are no tricks to apply as in the case of
the axino. The scalar saxion's mass is model dependent, but it is
not expected to effect collider phenomenology, because like the axion,
it has even $R$ parity. Like the gravitino, the saxion can still greatly
affect the cosmology without changing collider studies. Finally, the
object of interest, the axino does not yet have a specified mass.
Theoretically the axino mass is highly model dependent and a large
range of values are explored in the literature, so it can be taken
as a free parameter here. As an LSP, lighter axinos are preferred
so they are not over-produced in the early universe \cite{Baer:2010gr},
but this will be somewhat alleviated by assuming a low reheat cosmology
as mentioned in the previous chapter. The signal of displaced jets
and MET is expected to be insensitive to the axino mass for relatively light
axinos. As the axino becomes heavy enough the width of the NLSP
will be affected, which will be explored in more detail in chapter ~\ref{ch:width}. 

\section{Sneutrino NLSPs}
\label{sec:sneutrino}

	Before moving on to a more detailed study of the phenomenology with a neutralino NLSP, this is an appropriate place to discuss what may happen with a sneutrino NLSP instead. Figure \ref{fig:sneunlsp} shows the two sneutrino decay processes with the smallest final state multiplicity when the ${\cal L}_{\tilde{a}q\tilde{q}}$ coupling is dominant.  As with neutralino NLSPs the limited options in couplings means the decay is restricted to only this topology, with the only variance coming from which neutralinos/charginos/squarks are contributing.  There are two basic cases here, the sneutrino decays to a neutralino and neutrino, and then the decay proceeds the same as the topology in figure~\ref{fig:ax2jet}, or the sneutrino decays to chargino and a charged lepton, with the chargino decay still following a similiar topology as the neutralino decay. Which of the two decay paths is dominant will depend on the SUSY mass spectra, in particular, which is lighter, the charginos or neutralinos, and it is not unreasonable in this case to expect spectra where both channels contribute significantly. Contribution from both channels will make the phenomenology more difficult for a number of reasons. NLSP decays with gravitinos or RPV both have some branching fraction to all jets and MET, but also some branching fraction to other final states with leptons and photons. In the case where the NLSP is the neutralino, those alternative decay channels with leptons and photons can be used to discriminate between scenarios as the axino case should never have displaced leptons or photons.  With a sneutrino NLSP this discriminatory power is diminished as the axino case now also has multiple decay paths. This does not mean the problem is insurmountable, but simulated events in this work are restricted to only the neutralino NLSP case, so this is left for future investigation. Even in the case where the lepton plus jets channel is suppressed (for charginos much heavier then neutralinos) the all jets plus MET channel is still more difficult for sneutrinos as now there are multiple sources of real MET in each decay leg, which makes kinematic analysis more difficult. Besides the two decays shown in figure  ~\ref{fig:sneunlsp} there are other possible decays with more final state particles. The remaining ways for the sneutrino cascade decay to start are via a slepton and a guage boson or a slepton and a Higgs. Both of these involve longer decay chains with more final state particles, so they should not have a significant branching fraction unless there is an unusual SUSY spectra, such as a spectra with a sneutrino NLSP and a slepton NNLSP with the rest of the sparticles very heavy. 

\begin{figure}
\centerline{\includegraphics[scale=0.3]{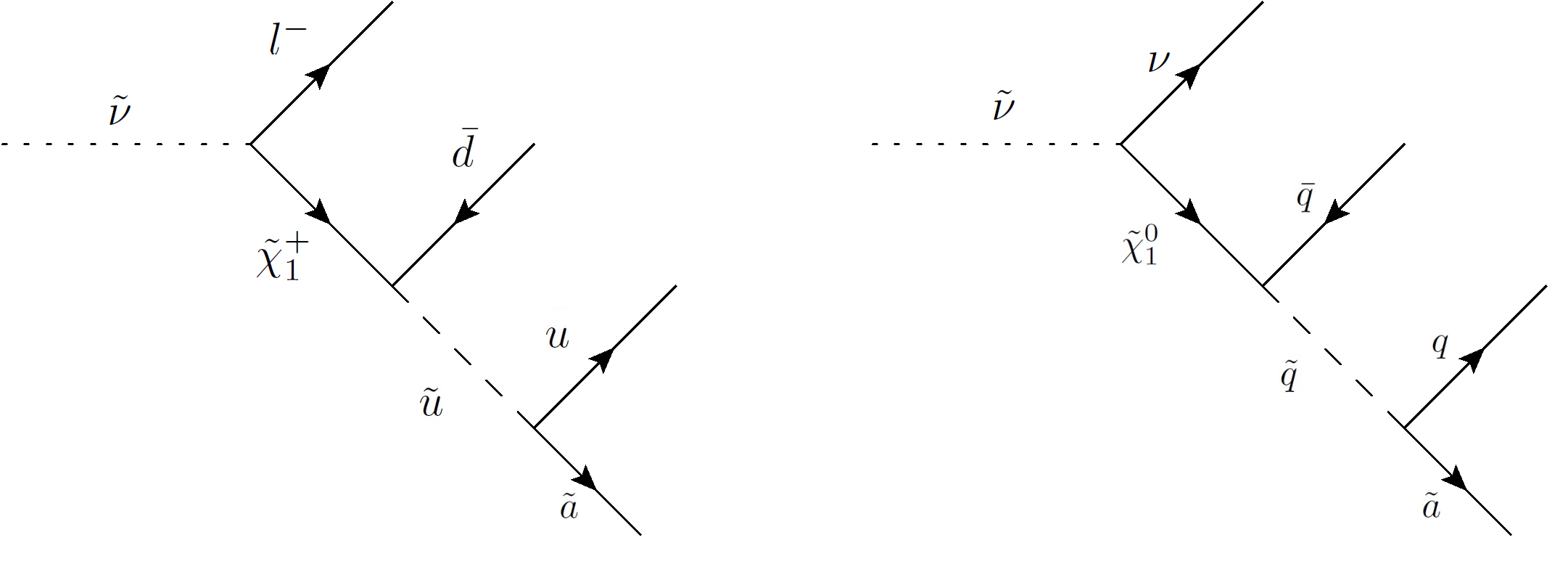}} \caption{Examples of decay topologies with a sneutrino NLSP to final states with additional neutrinos or charged leptons. \label{fig:sneunlsp}}
\end{figure}

	The topologies, and the qualitative conclusions drawn in this section are not modified greatly if the ${\cal L}_{\tilde{a}\tilde{g}g}$ coupling is dominant instead of ${\cal L}_{\tilde{a}q\tilde{q}}$. The nature of the axino couplings means any cascade, for either NLSP will end with a squark decay to jets and axino, and which coupling is dominant will just add or remove an additional jet at the end ot the decay.The signal for KSVZ axinos with a neutralino NLSP at the LHC is very predictive in that there are only two couplings to consider, each
providing one dominant topology. For this first study, the benchmarks
chosen have spectra that seems appropriate in that they are relatively
generic (the result of scans and not specifically ``engineered'')
and they are expected to give decay lengths in an appropriate range.
Beyond these particular benchmarks, the mixing of the NLSP neutralino,
the amount of compression of the spectra, and the mass of the axino
may affect the signal to some degree and it is worth testing. Though
there are no SM backgrounds to compete with very displaced
jets, there are other possible decays for the neutralino, including
decays to gravitinos and decays via RPV. Now that these qualitative
aspects of the signal have been summarized, in the next section results
are presented for simulated events, for decays to axinos and the alternatives.
The impact of the possible effects described above are explored and
the degree to which these neutralino decays can be distinguished is
tested. 

\chapter{Distinquishing Neutralino Decays}
\label{ch:results}

\section{Tools for Simulation and Analysis}
\label{sec:tools}

To investigate how predictive the axino multi-jet and MET signal is,
we simulated events for the LHC at 14 TeV. The primary tool
used to generate Monte Carlo events was MadGraph/MadEvent \cite{Alwall:2014hca}. The PDF set used with MadEvent was CTEQ6L1. The renormalization and factorization scales were allowed to run and were determined by MadEvents default settings with the scale for decay events set to the mass of the decaying parent particle. We added the axino field and its couplings to the MSSM using
FeynRules \cite{Alloul:2013bka,Degrande:2011ua,Duhr:2011se}. FeynRules takes a human readable Lagrangian as input and outputs model files in a
format useable by MadGraph. The
FeynRules implementation of the axino was validated for ${\cal
  L}_{\tilde{a}\tilde{g}g}$ of Eq.~\ref{eq:agg} by comparing the tree
level decay of a heavy axino to the analytical result, and for ${\cal
  L}_{\tilde{a}q\tilde{q}}$ of Eq.~\ref{eq:aqq} by comparing the squark
decay width to the results in \cite{Covi:2002vw}. We also calculated the
neutralino decay width for $\tilde \chi^0_j\to q_i \bar q_i \tilde a$ analytically and
confirmed the result of \cite{Barnett:1987kn} with the appropriate adjustments (see Eq.~\ref{eq:width}), 
and used this analytic form of the decay width to
verify the results obtained with MadGraph/MadEvent.

Existing model files in the FeynRules data base were used when
generating comparison events for the cases with gravitinos
\cite{Christensen:2013aua} and RPV couplings \cite{Fuks:2012im}. Mass
spectra were generated using SoftSusy \cite{Allanach:2001kg} and
checked with SuSpect \cite{Djouadi:2002ze}. These spectrum generators take as input the high
scale values for model paramters and then run the renormalization group equations to output all the
model parameters at the weak scale. Jet clustering was done
with FastJet \cite{Cacciari:2011ma} using $kT$ jets with $D = 0.4$ as the distance measure between jets \cite{Cacciari:2005hq},
and parton showers were generated by PYTHIA \cite{Sjostrand:2006za}.
The analysis is done in Mathematica with the Chameleon package
\cite{Thaler:2006:Online} as a base, but with plenty of modifications
and extensions. The Chameleon package can read LHE files (the standard output for MadGraph) and defines basic functions for kinematic variables. Modifications include the ability to calculate various kinematic functions from the base functions provided, implementing cuts and reading weighted events correctly. Examples of Mathematica notebooks for event analysis with the Chameleon package including these modifications can be found at \cite{Redino:2015:Online}. Events are generated at tree level, but the vertex
correction of figure \ref{fig:squarkvert} is captured in the effective
coupling in ${\cal L}_{\tilde{a}q\tilde{q}}$ of Eq.~\ref{eq:aqq}.  Other loop effects may be important for production of sparticles, depending on the channel, but as a first look at this type of collider search for KSVZ axinos, we restrict ourselves to tree level.

The only tool required for this study which is less common was evchain
\cite{Kim:2013ivd} which acts as a ``MadGraph manager'' to combine
separate subprocess runs, and is especially useful for decay chains
which are difficult for MadGraph to manage alone. In this scenario
with axinos, MadGraph has difficulty because of the extremely narrow
decay width of the neutralino. As described in the previous chapter,
there is effectively only one topology by which the neutralino can
decay, i.e. to two jets and the axino when ${\cal L}_{\tilde{a}q\tilde{q}}$ is the only dominant axino interaction. While the branching fraction to
two jets and the axino is very close to one, the width is still extremely
small because of both the suppression from the presence of $f_{a}$
in the denominator and because of the heavy off-shell squark required
for the decay. MadGraph can generate the decays of the neutralino
just fine, but to include these decays in a larger event is problematic.

\section{Neutralino Width and Signal Cross-section of Benchmarks}
\label{sec:widthandcross}

Looking at just the neutralino decay alone, the decay width can be calculated and the expected decay length $c\tau$ in the detector can be determined, so the assumption
that there are plenty of highly displaced jets can be tested. For
the lighter benchmark of figure~\ref{fig:lightbm} with a light (taken to be massless) axino, the width
of the lightest neutralino varies between $\Gamma_{\tilde \chi^{0}_1}=7.3\times10^{-16}$~GeV
and $\Gamma_{\tilde \chi^{0}_1}=1.7\times10^{-17}$~GeV over the window $3\times10^{5}{\rm \mbox{ GeV}}<f_{a}<3\times10^{6}{\rm \mbox{ GeV}}$.
This corresponds to a mean decay length range between roughly $c\tau=0.26$~m
and $c\tau=11.6$~m. For the heavier model (figure~\ref{fig:heavybm}) over the same range in $f_{a}$
the neutralino width spans the range $\Gamma_{\tilde \chi^{0}_1}=1.7\times10^{-13}$~ GeV and $\Gamma_{\tilde \chi^{0}_1}=1.2\times10^{-15}$~ GeV
or a length range of $c\tau=0.0012$~m and $c\tau=0.16$~m. This range is a very
appropriate size for the ATLAS detector, allowing for a sizable number
of events that are displaced enough to realize the advantages described
in the previous chapter: negligible SM backgrounds, and
the ability to separate particles from production and particles from
neutralino decay. This width is also insensitive, i.e. within statistical errors, to the
axino mass in the range $0\le m_{\tilde a} < 10$ GeV. The hope is
that the axino signal would be trigger-able at this depth in the ATLAS
detector using the hidden valley triggers discussed in \cite{Aad:2013txa}.
No serious attempt is made here at determining the efficiency of such
triggers for this model, as adjusting for instance, detector simulation tools  for displaced jets is non-trivial
work, and not readily available in off-the-shelf tools, but Meade et al do make an estimate of the efficiency of some of these triggers for highly displaced jets in \cite{Meade:2010ji}.

For the remainder of events analyzed the axino is assumed to be very light (effectively massless)  and $f_{a}=10^{6}$~GeV,
corresponding to a neutralino decay width of
$\Gamma_{\tilde\chi^{0}_1}=1.1\times10^{-16}$~GeV for the lighter benchmark and
$\Gamma_{\tilde \chi^{0}_1}=7.5\times10^{-15}$~GeV for the heavier benchmark.

If the
trigger can actually be agnostic to the production mechanism, then all
SUSY channels can contribute to the signal cross section, and for the
benchmark this gives an inclusive SUSY rate of $\sigma_{SUSY} \sim
5$~fb for the lighter benchmark and $\sigma_{SUSY}\sim 23$~fb for the
heavier one. In the much more pessimistic case where one only attempts
to look for events with neutralino decays only, i.e. neutralino pair
production, then the rate is only $\sigma_{\tilde \chi^{0}_1 \tilde \chi^{0}_1}\sim 30$~ab for
the light benchmark and$\sigma_{\tilde \chi^{0}_1 \tilde \chi^{0}_1}\sim 14$~ab for the heavy
one, possibly providing just a few events with ${\cal L}=300 \, {\rm
  fb}^{-1}$, if they survive the efficiency of the triggers (note that
the HL-LHC is designed to reach ${\cal L}=3 \, {\rm ab}^{-1}$).

\section{More Event Simulation Details}
\label{sec:details}

Looking at simulated events for the decay alone is still useful for
studying the shapes of the kinematic distributions. Even though the
neutralino decays actually will occur after some production process
with Lorentz boosted momentum and convolution with parton distribution
functions (PDFs), the distributions from decay-only events are still
physical in that they show the relevant observables in the neutralino
rest frame. These rest-frame distributions can provide interesting
hints as to how the lab-frame distributions may be distinguished
between different neutralino decays (axino/gravitino/RPV). More
optimistically, these rest-frame distributions may be directly
accessible, if the neutralino momentum in the lab frame can be
reconstructed then the appropriate Lorentz boost on the lab-frame observables
can be made. Such a boosting is not a simple task for partially
invisible decays, and no explicit algorthim is provided here, but
similiar reconstruction for partially invisible decays has been done for instance
in the context of top decays \cite{Guillian:1999jh}. Cleaner distributions (without PDF convolution) would also be accessible at a lepton collider, since this signal channel is indifferent to the SUSY production mechanism.

When looking at the full event, with SUSY production and the full
decay, with such a small decay width, MadEvent fails to sample an appropriate
phase space and the results of the Monte Carlo integration are unreliable. This can
be illustrated as follows: in the narrow-width approximation  
\begin{equation}\label{eq:nw}
d\sigma_{tot}=d\sigma_{prod}\frac{\Gamma_{decay1}}{\Gamma_{total}}\frac{\Gamma_{decay2}}{\Gamma_{total}}
\end{equation}
where here BR$=\Gamma_{decay1/decay2}/\Gamma_{total}\sim 1$ for both
decays and a very narrow neutralino decay width
$\Gamma_{decay1/decay2} \ll 1$, the cross section of neutralino
pairs should be the same, regardless of whether or not their decays
are included, that is $d\sigma_{tot}=d\sigma_{prod}$, which is not
found with MadEvent when including the neutralino decays in this
model. The way evchain circumvents this limitation is in a way by
implementing the narrow-width approximation ``by hand''. The
production process for neutralinos is done in one run of MadGraph
(either by direct pair production or via any SUSY cascade) and the
decay of the neutralinos is done in another, separate run. The
resulting LHE event files from these two separate runs are combined by
evchain (with the appropriate Lorentz boosts being made), and the cross
section calculated from production events are scaled by the branching
fraction to the decay events, as per the narrow width approximation of
Eq.~\ref{eq:nw}.  In the case of the axino LSP, no scaling is
necessary since the branching is effectively one, but when similar
events with gravitinos and RPV decays are generated for comparison,
the appropriate scaling has to be applied.

A minimal set of loose default generation cuts are implemented in MadEvent (with
any other cuts done during the analysis part). This set of cuts is the same for
all three models (axino/gravitino/RPV). It should be noted that these
cuts themselves, are ``boosted'' by evchain as well, for example
a small minimum $p_T$ requirement on a jet will actually cut events at
a higher $p_T$ after evchain boosts the events. This effect should be
negligible however, as in all the events generated the $p_T$ of jets
is rather large (also good news for triggering) because of the mass
of the neutralino NLSP.

The comparison models are intended to be as similar as possible to
the benchmark cases for axinos so that distinguishing between events
here can be thought of as a ``worst case'' scenario, where distinguishing
models is the most difficult. This also means that the axino is taken
to be very light, like the gravitinos or neutrinos (from RPV) that
appear in the alternative neutralino decays. A heavier axino will
provide more handles for distinguishing neutralino decays via the
kinematic distributions. Of course the ability to distinguish between
distributions is dependent on the ability of the detector to measure
such features when the jets are highly displaced, and again, no detailed
detector simulation is attempted in this work. Aside from the axino
mass and the gravitino mass, the rest of the SUSY spectra is identical
between the models, so the production rates above are the same for
all three alternatives (axino/gravitino/RPV). In addition to the similar
spectra, the width of the lightest neutralino to 2 jets and MET is
made to be very close to the axino benchmarks so that the similar
signal would appear in the same region of the detector (though not
necessarily at the same rate, since the total width does not have
to be the same as the axino benchmark). It is reasonable to assume
that other comparison models could produce better ``impostor'' signals,
by having a higher rate (from a different SUSY spectra) and a different
decay length, but with a comparable number of events in the same part
of the detector. Comparison of the axino benchmark signal to these
two comparison models follows, with gravitinos first, and then with the RPV signal.

\section{Axinos Vs. Gravitinos}
\label{sec:gravitinos}

In the case of gravitinos, to have a neutralino decay length for the
2-jet and MET signal similar to the axino model, the gravitino mass
is taken to be $500$~eV for the lighter benchmark and $750$~eV for
the heavier one . Unlike the axino, the gravitino has numerous couplings,
and is not restricted to the 2-jet+MET channel. One obvious consequence
of this is that gravitinos can be distinguished from axinos simply
by looking for these other decay channels, e.g. anything with leptons
or photons that is highly displaced. There is plenty of literature
describing how to search for gravitinos with leptons or photons, (see, e.~g., ~\cite{Aad:2014gfa}), but
this is not enough to resolve the issue definitively. While it would
require a coincidence of parameters, neutralino decays to gravitinos
and axinos could co-exist in a model, so for the sake of being thorough,
the 2-jet+MET signals can be compared in hopes of distinguishing
them based on the shapes of kinematic distributions alone. By the
same reasoning, the presence of alternative decay channels for the
gravitino means that the branching fraction will be less than one,
and so for identical SUSY spectra the gravitino model will have the
multi-jet+MET signal occurring at a smaller rate.  Unlike the axino
case, there are several topologies to produce two jets and a gravitino
(see figure~\ref{fig:grv2jet}). While the same topology exists as in the axino case (figure~\ref{fig:grv2jet} left), it is not dominate here, as there are also topologies without
off-shell colored sparticles that contribute with greater strength
(figure~\ref{fig:grv2jet} right). Diagrams like the right one in Figure~\ref{fig:grv2jet}  give two hints as to how the
scenarios can be distinguished. Since the dominant topologies have the
gravitino radiated at the beginning of the decay chain, rather than
the end, one should expect the MET to recoil differently between the
axino and gravitino models, with the gravitino MET being harder. Conversely,
the visible jets should be harder for the axino case, and softer for
the decays to gravitinos. These kinematic differences are subtle when
looking at neutralino decays for this particular benchmark, and when
these decays are simulated for a SUSY cascade the effect is washed
out by the Lorentz boosts and smearing by the PDFs. For a model with a much
heavier neutralino this difference is more noticeable. 
Results comparing
the MET for axino events versus gravitino events are shown in figure~\ref{fig:grvmet}
and results comparing the total jet $p_T$ (the $H_T$) are shown in figure~\ref{fig:grvht}.
Jet $p_T$ and MET being relatively larger or smaller between decays
with axinos or gravitinos is not a very useful handle by itself. If
only one type of decay is actually measured, it begs the questions:
more or less $p_T$ compared to what? The difference in weighting between
visible and invisible $p_T$ can be seen by plotting $H_T$ against the MET,
as shown in Figure~\ref{fig:grvhtmet}. In all of these kinematic plots the effect
is more noticeable for larger neutralino masses, but it is still smeared
away in the full event, i.~e. when including both production and neutralino decays. This unfortunate smearing effect may
be circumvented if the neutralino momentum can be reconstructed from decay products, in which
case the appropriate Lorentz boost can be made and the unsmeared distributions should be measurable.

\begin{figure}
\centerline{\includegraphics[scale=0.3]{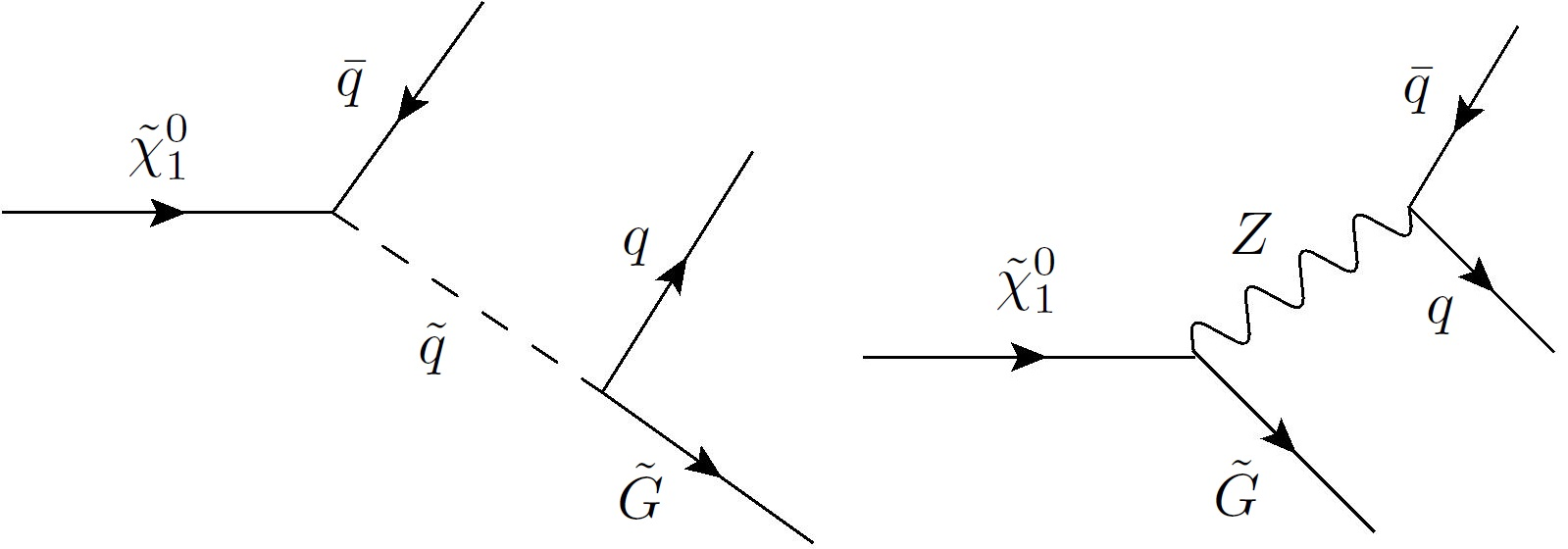}} \caption{Two example
  diagrams that contribute to neutralinos decaying to two jets and a
  gravitino. \label{fig:grv2jet}}
\end{figure}

\begin{figure}
\centerline{\includegraphics[width=5.5in]{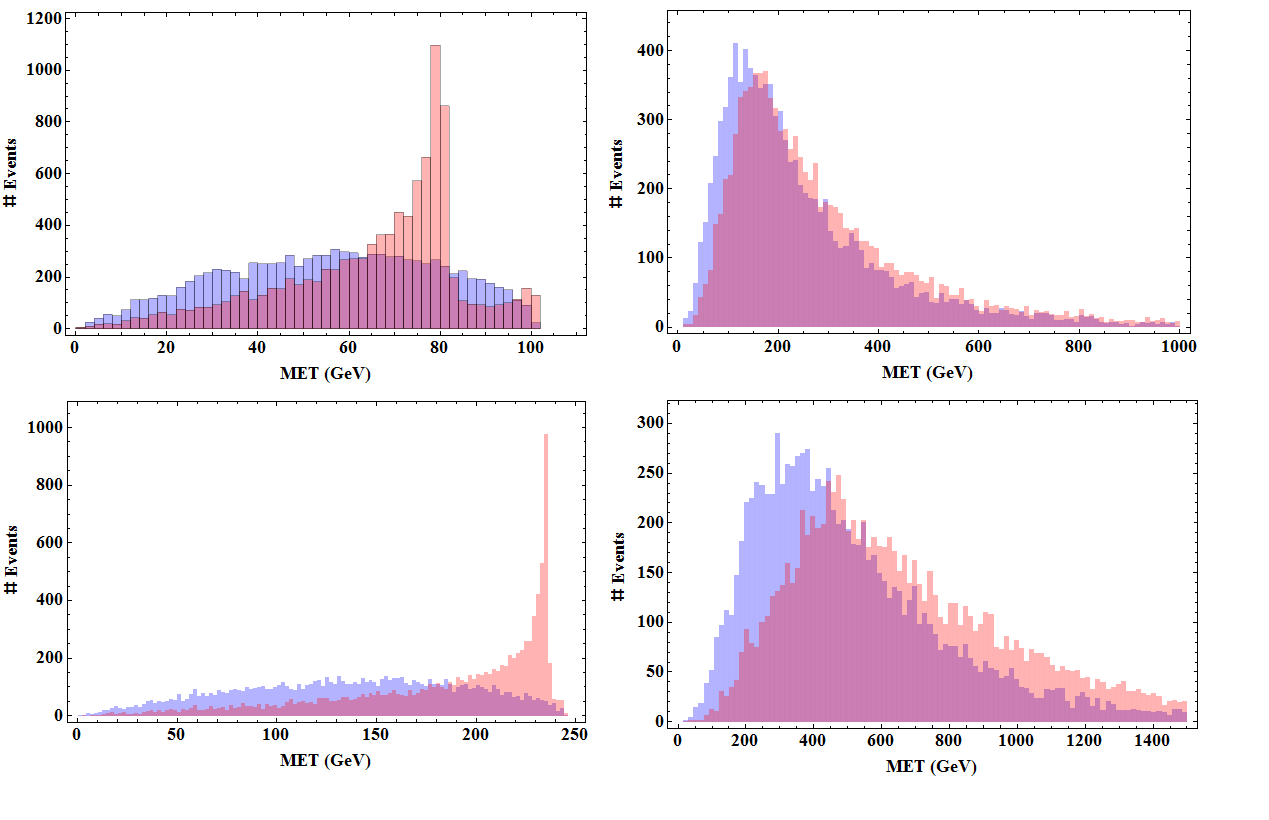}}
\caption{Distributions of the MET from neutralino decays to axinos (blue) and
gravitinos (red).  Events are simulated with minimal generation cuts only, and at parton level (no showering/clustering). The left plots  consider the neutralino decay alone, while the right plots are in the lab frame
of the whole event at 14 TeV at the LHC, i.e. when including both production and decay via evchain. The upper
  plots are for the lighter benchmark, and the lower plots for the
  heavier benchmark. \label{fig:grvmet}}
\end{figure}

\begin{figure}
\centerline{\includegraphics[width=5.5in]{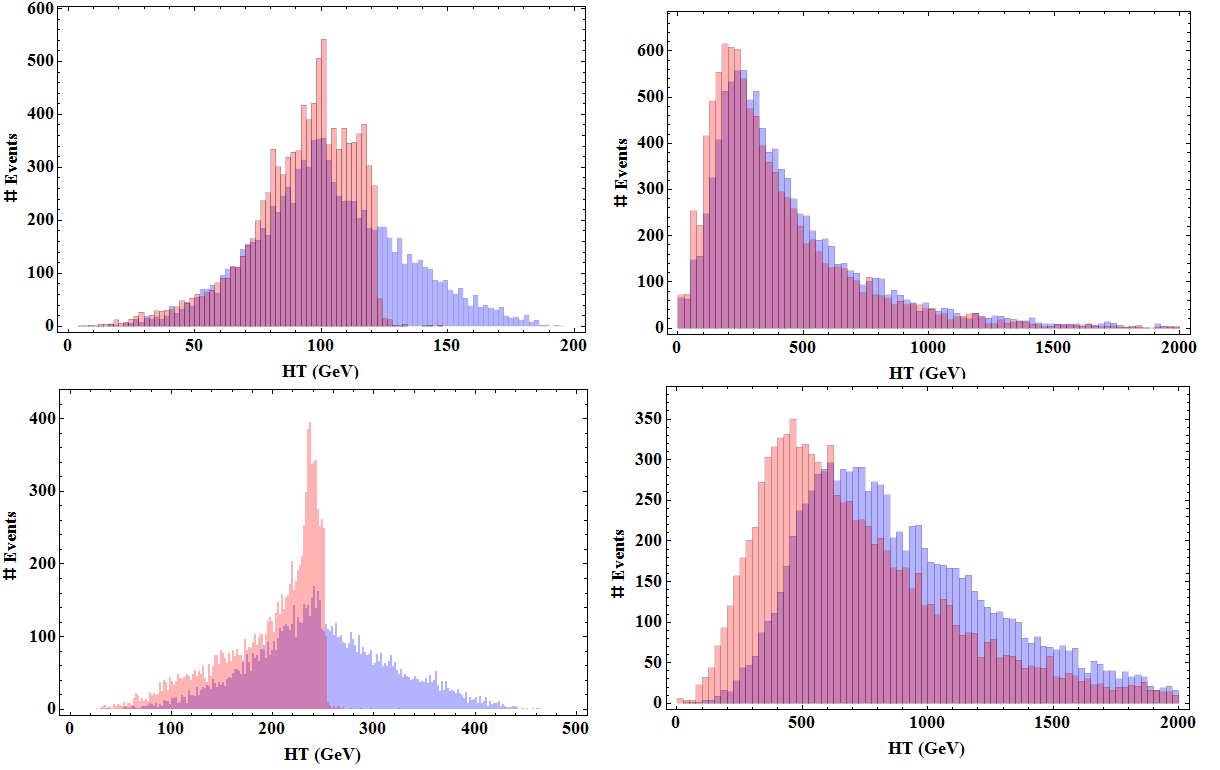}}
\caption{Distributions of the scalar sum of jet transverse momenta (the $H_T$) from
  neutralino decays to axinos (blue) and gravitinos (red). Events are simulated with minimal generation cuts only, and at parton level (no showering/clustering). The left plots  consider the neutralino decay alone, while the right plots are in the lab frame
of the whole event at 14 TeV at the LHC, i.e. when including both production and decay via evchain. The upper
  plots are for the lighter benchmark, and the lower plots for the
  heavier benchmark. \label{fig:grvht}}
\end{figure}

\begin{figure}
\centerline{\includegraphics[width=5.5in]{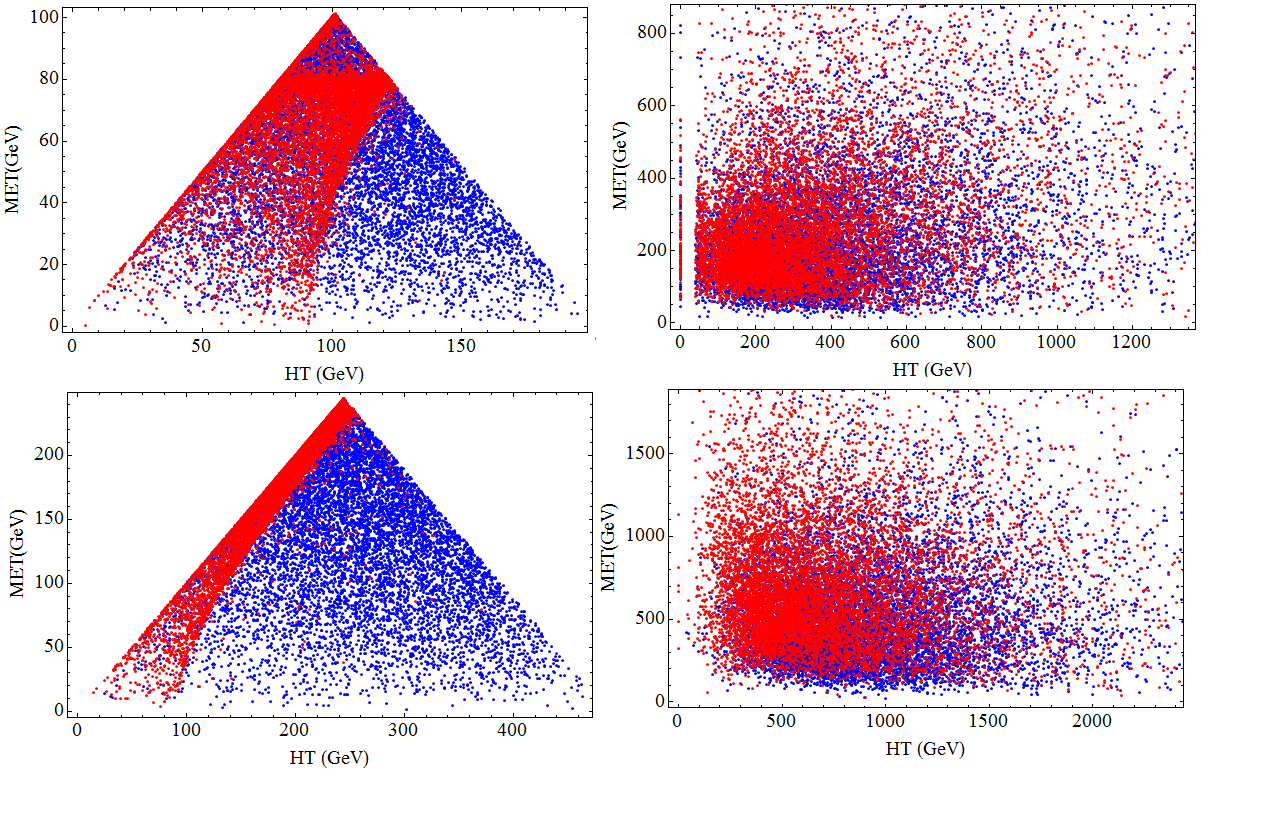}}
\caption{$H_T$ versus MET from neutralino decays to axinos (blue) and
  gravitinos (red).  Events are simulated with minimal generation cuts only, and at parton level (no showering/clustering). The left plots  consider the neutralino decay alone, while the right plots are in the lab frame
of the whole event at 14 TeV at the LHC, i.e. when including both production and decay via evchain. The upper
  plots are for the lighter benchmark, and the lower plots for the
  heavier benchmark.  \label{fig:grvhtmet}}
\end{figure}

Perhaps a simpler way to distinguish these two models is to just count
the displaced jets. It is reasonable to think that after parton showering with PYTHIA and clustering one could expect more jets from the axino case as these
jets are expected to be harder (based on the kinematic distributions
above) and are more likely to radiate, and this is reflected in the
generated events as shown in figure~\ref{fig:grvjets}. Applying stronger jet $p_T$ cuts
to satisfy triggers~\cite{Aad:2013txa} ($p_T>40 {\rm GeV})$ will remove the softer jets from both
samples, in addition to a pseudo-rapidity cut ($|\eta|<2.5)$. This does not
just reduce the total number of jets after showering, but it also
shifts events between different bins for numbers of jets. 

\begin{figure}
\centerline{\includegraphics[width=5.5in]{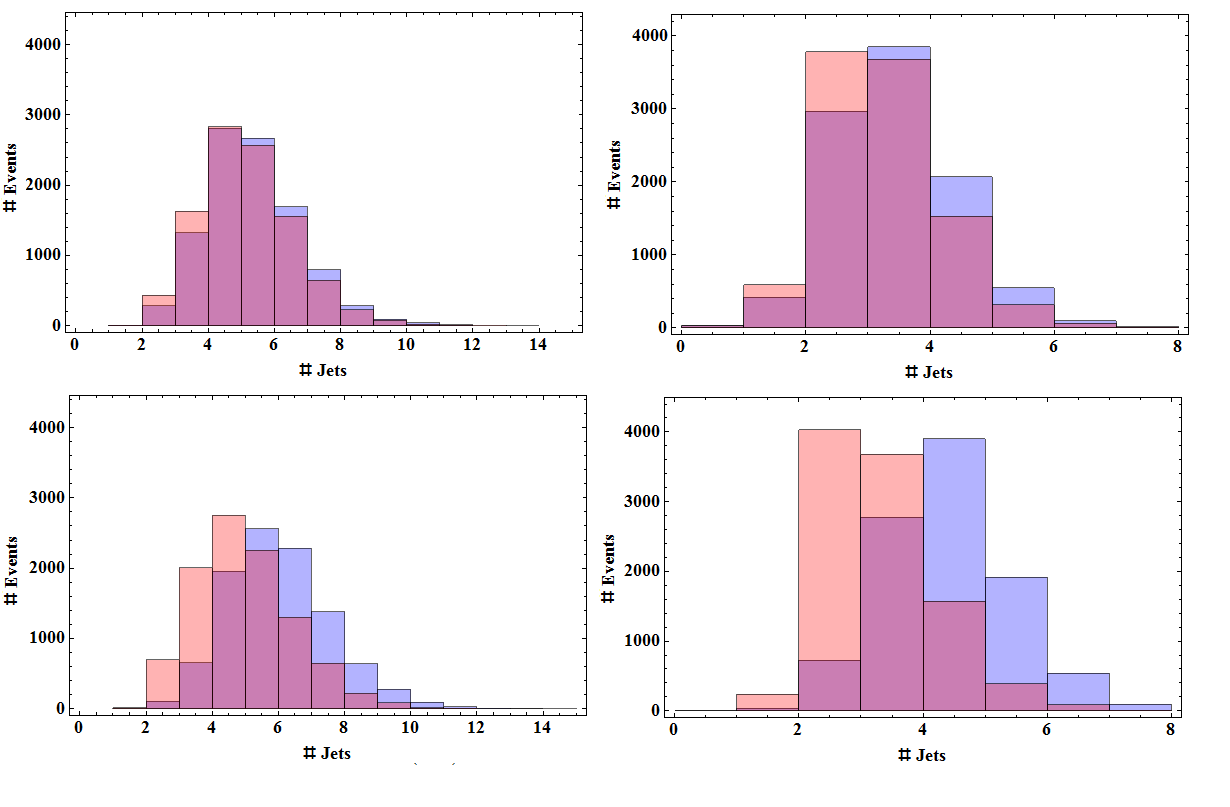}}
\caption{The number of jets from neutralino decays to axinos (blue) and gravitinos
(red). All plots are for the full event (production and decay via evchain) at 14 TeV with showering done by PYTHIA and jet clustering from FastJet using kT jets with D=0.4. The left plots are from events generated with loose generation cuts and the right plots
are obtained after applying more restrictive cuts, $p_T>40$~GeV and $|/eta|<2.5$. The
upper plots are for the lighter benchmark, and the lower plots for
the heavier benchmark. \label{fig:grvjets}}
\end{figure}

Even though a photon or a Higgs boson could take its place in diagrams similar
to the right diagram in figure~\ref{fig:grv2jet}, the presence of the $s$-channel Z has a significant effect
on the distributions and the $Z$ resonance can be reconstructed. For
event samples from just the decays, the $Z$ resonance simply comes from
the invariant mass of both jets (Figure~\ref{fig:grvjim}). When looking at the full
event with showering there is the question of which jet combination
to take. Including the full combinatoric background (all combinations
of jets), the $Z$ resonance is buried, but since the particular resonance
is known in this case, it is easy enough to just take those combinations
of jets which are closest to the known $Z$ mass. This method has the
drawback that it can create an artificial bump in the jet invariant
mass distribution. This artificial bump is much more pronounced when
the parent neutralino is lighter, so again, like the other methods
of discrimination, it is more difficult for lighter neutralinos. Reconstructing
the $Z$ resonance is a much more powerful way to distinguish the gravitino
and axino cases, as unlike the kinematic distributions discussed earlier it is
invariant under Lorentz boosts and somewhat insensitive to the sparticle masses.
A veto on the invariant mass of jet pairs in separate halves of the
detector (pairs from separate parent neutralinos) can cut the majority
of gravitino events and of the methods described here such a veto
is considered the best way to determine if neutralino decays contain
events with axinos, gravitinos, or both.  There are some subtleties here, but they can be addressed easily enough.  If there are only a small number events available then there might not be an adequate  sample size with exactly four jets, and other combinations should be considered (invariant mass of triplets, quadruplets), but again looking for these sets of jets in opposite halves of the detector so as to separate them by their decay parents. Even in the case where there is a sufficient sample size for events with exactly four jets, reconstruction the Z by jet pairs does not work on an event by event basis, as there is some probability that all of the decay products from one neutralino cluster to a single jet, and the other parent clusters to three jets.  This type of issue can be resolved either by more careful selection criteria or by considering singlet and triplet jet invariant mass combinations in addition to the pairs.

\begin{figure}
\centerline{\includegraphics[width=5.5in]{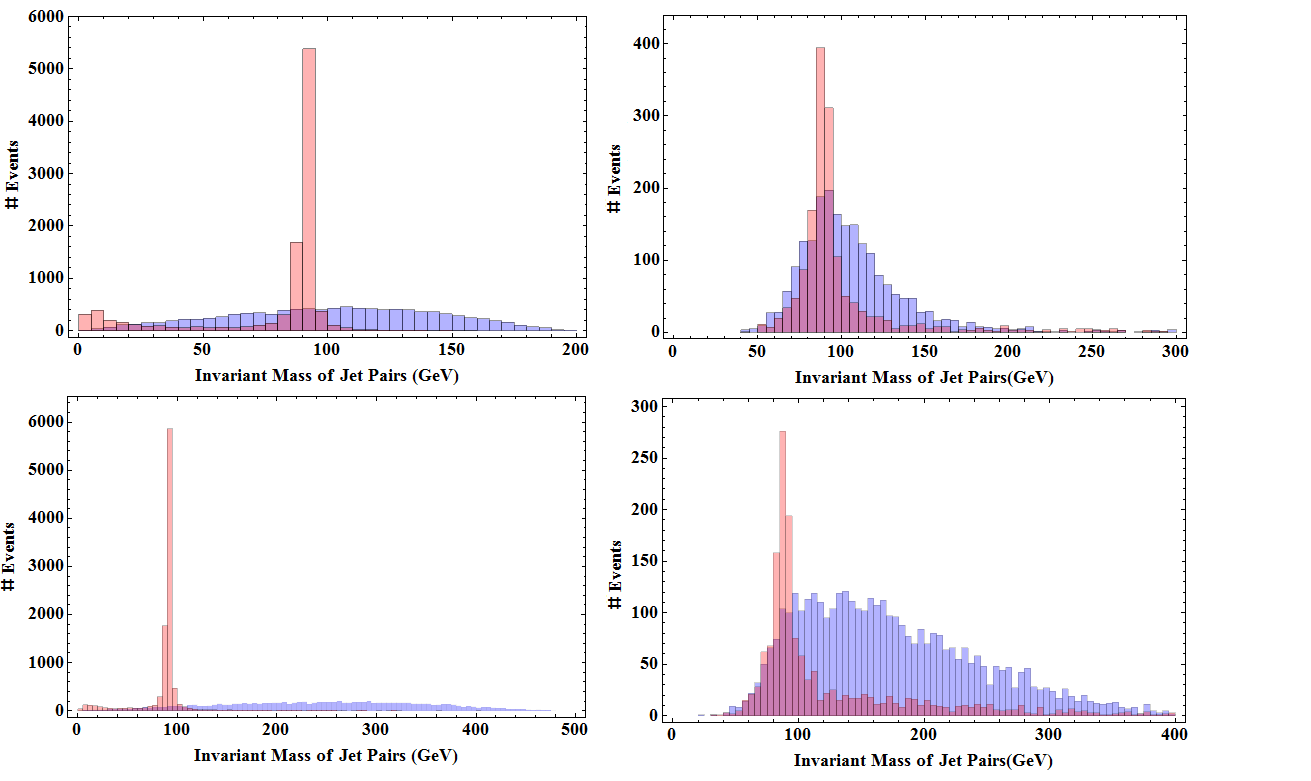}}
\caption{The invariant mass of jet pairs for neutralino decays to axinos (blue)
and gravitinos (red). The left plots are for a single decaying neutralino in
its rest frame, while the right is in the lab frame for the whole
event (in a sample selected to have exactly four jets) at 14 TeV with showering done by PYTHIA and jet clustering from FastJet using kT jets with D=0.4. The upper plots
are for the lighter benchmark, and the lower plots for the heavier
benchmark. \label{fig:grvjim}}
\end{figure}

\section{Axinos Vs. RPV}
\label{sec:RPV}

Like gravitinos, RPV scenarios can also produce a signal of displaced
jets and missing energy, and RPV decays can co-exist in a model with
decays to axinos, so a comparison of these similar signals is warranted.
There are many possible signatures of RPV as there are several possible
couplings, coming from both the super potential and also from soft
SUSY breaking terms. The couplings from the $R$ parity violating super
potential are given by~\cite{Barbier:2004ez} 
\[W_{RPV}=\mu_{i}H_{u}L_{i}+\frac{1}{2}\lambda_{ijk}L_{L}^{i}\centerdot L_{L}^{j}E_{R}^{k}+\lambda'_{ijk}L_{L}^{i}\centerdot Q_{L}^{jl}D_{Rl}^{k}+\frac{1}{2}\lambda''_{ijk}\epsilon^{lmn}U_{Rl}^{i}D_{Rm}^{j}D_{Rn}^{k} \; ,\]
where $i,j,k$ are flavor indices and $l,m,n$ are color indices.
The first three terms all violate lepton number, while the last term violates
baryon number. While all these couplings are possible, they are constrained
by the non-observation of certain processes. To avoid running into
bounds from unobserved processes, such as proton decay, the constraint
is not on the size of these couplings directly, but rather of their
products (for example proton decay requires B and L to be violated).
Because of this it is not unreasonable to assume that there could
be just one dominate RPV coupling, that is itself relatively small.
The UDD coupling can produce three displaced jets from neutralino decay
\ref{fig:rpvudd}, which may look like the axino signal after showering/clustering
but any missing energy would have to come from detector/trigger inefficiencies,
or jet mis-measurement. It is difficult to estimate how much ``fake
MET'' there could be in such events without performing a detailed
detector simulation, but it is expected that such events would be
distinguishable from axino events for any model similar to the benchmarks,
because the NLSP neutralinos are massive enough and the axino will
carry away a large portion of this energy, as shown in the MET plots
in figure \ref{fig:grvmet}. Also, due to the absence of true MET the jets themselves
will be harder than in the axino case.

\begin{figure}\label{fig:rpvudd}
\centerline{\includegraphics[scale=0.3]{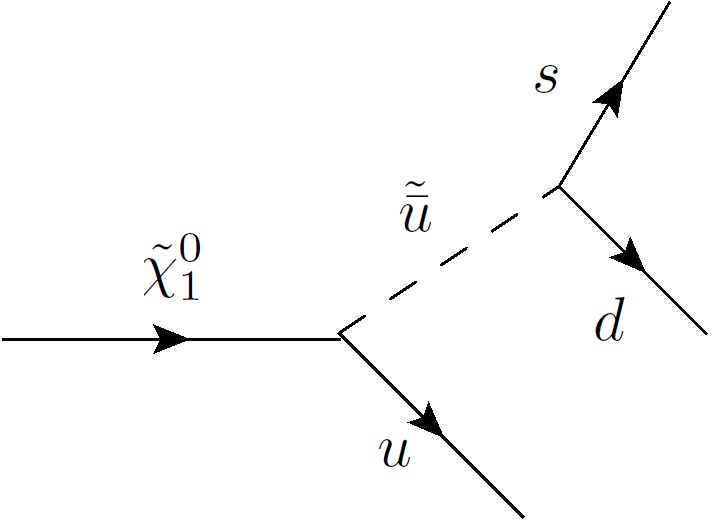}} \caption{RPV topology for UDD coupling.}
\end{figure}

In the case where the dominant RPV coupling is of the LQD type (with
$\lambda'\approx7.5\times10^{-6}$ for the lighter benchmark and $\lambda'\approx2.5\times10^{-6}$
for the heavier benchmark), the two jet signal can look very much
like the axino case when the neutralino decays to two jets and a neutrino.
As in the gravitino case, this RPV coupling also allows for channels
with photons and charged leptons in the final state. Again, the presence
of other channels will mean the rate of the 2 jet signal is less than
in the axino case, but a difference of rates is not helpful without
a priori knowledge of the SUSY spectra to calculate these rates. Discovery
of displaced photon or charged channels would imply there are neutralino
decays not involving the axino, but again, as was mentioned with the
gravitino case, this does not exclude the possibility that both decays
exist in the same model. While it was stated that it would require
a coincidence of parameters to have competitive neutralino decays
to gravitinos and axinos in the same model, it would be less surprising
in the case of RPV with trilinear couplings, such as the UDD or LQD
ones explored here. As mentioned in chapter~\ref{ch:background}, Affleck-Dine baryogenesis
with RPV couplings is a scenario that is attractively compatible with
a cosmology with LSP axinos in the hadronic axion window. This is
because with such a low Peccei-Quinn scale, $f_{a},$ the scenario
likely requires a very low reheat temperature, (which ADB can accomodate,
unlike thermal leptogenesis). Also, when ADB involves RPV couplings,
there must be another source of dark matter instead of the lightest
neutralino, which axions/axinos can accommodate. The coincidence of
scales required to make both RPV and axino decays competitive comes
from two independent sources. For the axinos, the window of lower
$f_{a}$ is set by the constraints mentioned in chapter~\ref{ch:background}, and for
ADB with RPV to be successful it requires a trilinear RPV coupling with
$\lambda\approx10^{-7}$\cite{Higaki:2014eda}, coincidently in the
same range to give similar width as the axino decays. Though it is
a distinct possibility that these channels co-exist in the same model,
the ``coincidence'' should not be overstated, as depending on the
value of $f_{a}$ the width to axinos actual varies over a couple
orders of magnitude, and with RPV, ADB can be accommodated
with $10^{-9}<\lambda<10^{-6}$, so while these correspond to the
same range of widths for decays, it is also possible that one process
dominates and the other will have a negligible rate.

Distinguishing LQD RPV from axino signals by the jet distributions
alone is much more difficult than the case with gravitinos as the
topologies contributing to the signal are now identical (figure~\ref{fig:rpvlqd}).
There is no massive resonance to distinguish the models as in the
case of gravitinos, and the MET and various jet variables are also
very similar between this case and the axino case. Some of the distributions
explored earlier are shown again in figure~\ref{fig:rpvdist} and figure~\ref{fig:rpvjets}, but now with LQD RPV
distributions as well. The only one that could potentially be used
as a tool to distinguish axino and RPV signals is the $H_T$, which shows
a peak at half the parent neutralino's mass for the axino case, but
not for the RPV case. This is only useful if the neutralino mass is
known or at least constrained (perhaps from analysis of the prompt
event seperately) and the distribution is only useful in the neutralino
rest frame, so its momentum must also be determined. If sufficiently
long lived RPV decays are discovered at the LHC, it
may be very difficult to rule out the possibility of a light axino
contributing to some of that signal.

\begin{figure}
\centerline{\includegraphics[scale=0.3]{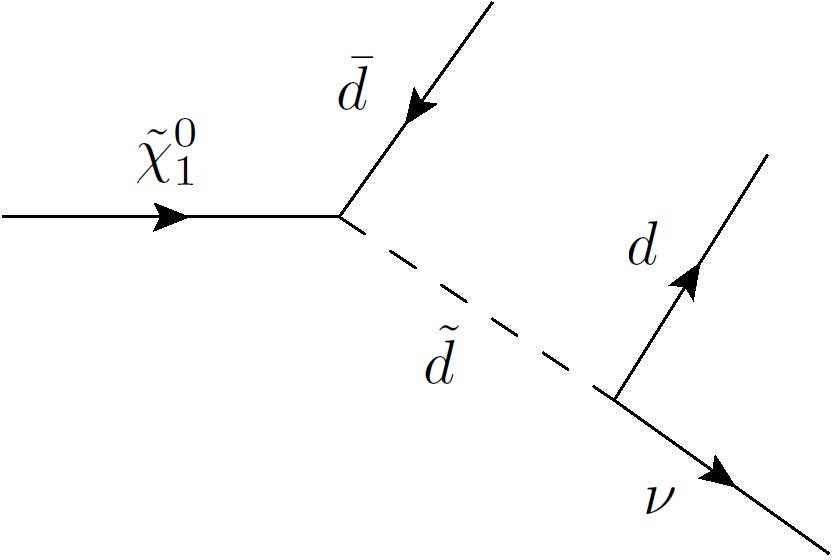}}
\caption{RPV topologies with LQD coupling. \label{fig:rpvlqd}}
\end{figure}

\begin{figure}
\centerline{\includegraphics[width=5.5in]{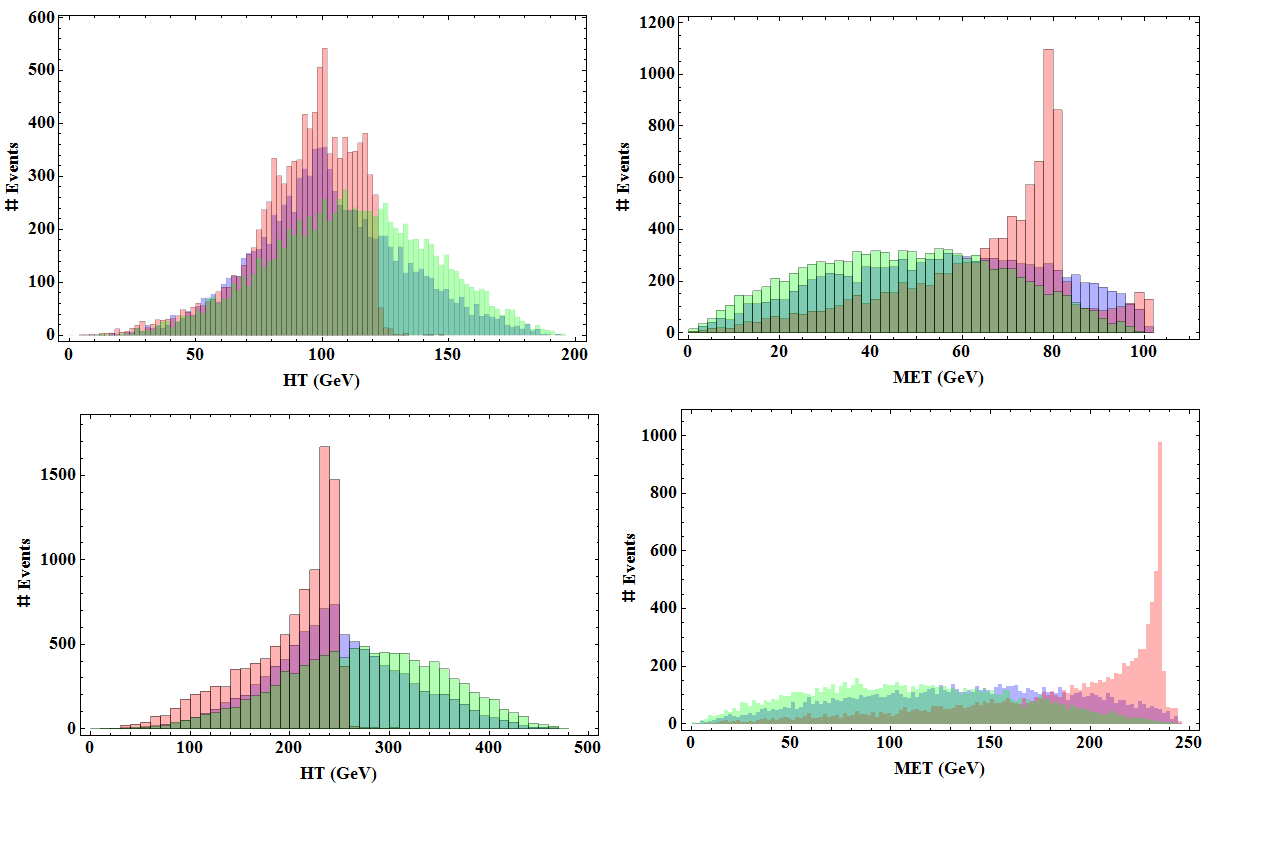}}
\caption{Summary of kinematic distributions for all three scenarios, axino
(blue), gravitino (red) and RPV with an LQD type coupling (green).  Events are simulated for the decay only with minimal generation cuts only, and at parton level (no showering/clustering). 
The top row is the lighter benchmark and the bottom row is the heavier
benchmark. \label{fig:rpvdist}}
\end{figure}

\begin{figure}
\centerline{\includegraphics[width=5.5in]{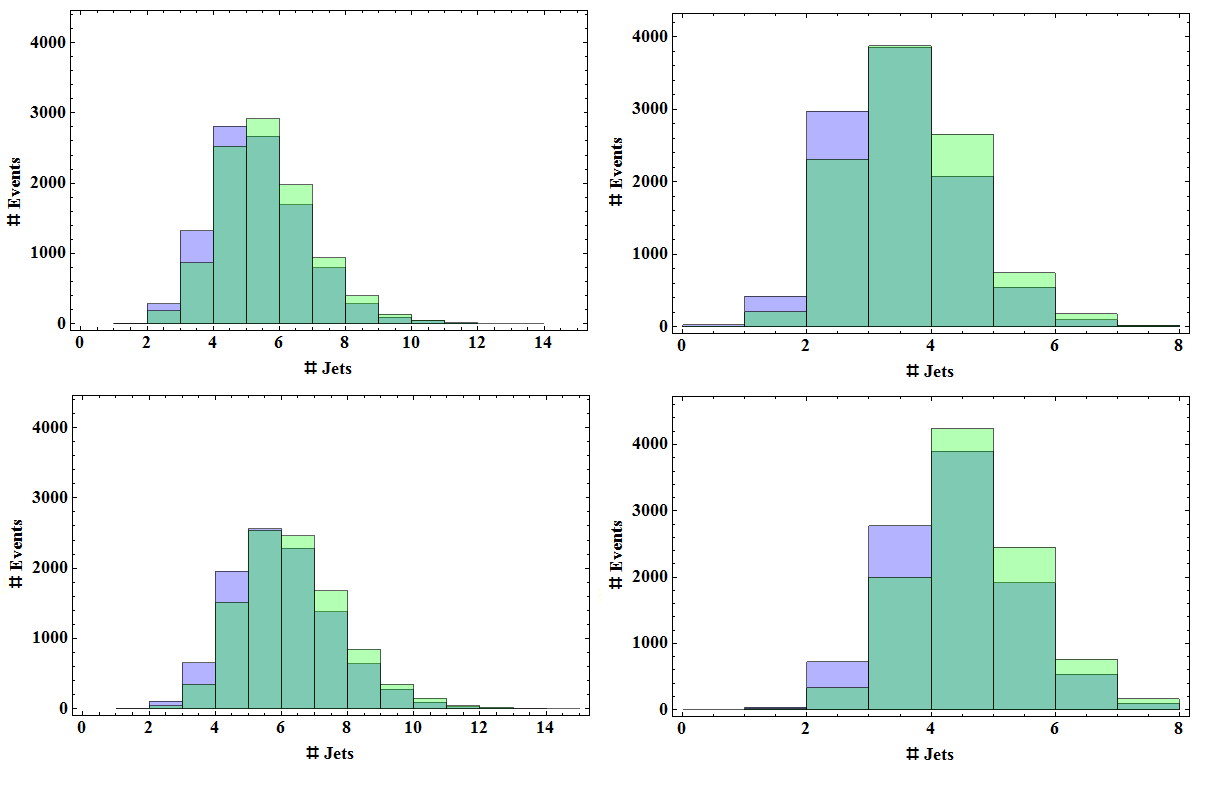}}
\caption{The number of jets from neutralino decays to axinos (blue) and neutralino decays via RPV with an LQD type coupling
(green). All plots are for the full event (production and decay via evchain) at 14 TeV with showering done by PYTHIA and jet clustering from FastJet using kT jets with D=0.4. The left plots are from events generated with loose generation cuts and the right plots
are obtained after applying more restrictive cuts, $p_T>40$~GeV and $|\eta|<2.5$. The
upper plots are for the lighter benchmark, and the lower plots for
the heavier benchmark. \label{fig:rpvjets}}
\end{figure}

When considered together, there are a few handles to distinguish between
gravitino and axino scenarios, and with varying success, some RPV
scenarios, but it should also be asked how strongly any of these results
depend on the choice of benchmark. Most of the distinguishing features
above come about simply as a consequence of the topology, and
should not be very sensitive to many aspects of the benchmark. The
parameter that is expected to have the greatest effect on the shapes
of these distributions is the mass of the lightest neutralino, which
has been demonstrated in our results for the kinematic distributions. The choice of neutralino
mixing, in this case a mostly Bino NLSP, does not have a large effect
on any of the axino distributions shown here, and the largest effect
this has on the gravitino signal is to change the branching ratio
to jets, but the rates are already very different from the axino case. 

\section{Decays via $L_{\tilde{a}\tilde{g}g}$}
\label{sec:3jetcase}
If the benchmark is substantially different, in particular if the squarks are much heavier than the
gluino then the $L_{\tilde{a}\tilde{g}g}$ coupling can become dominant and the main topology contributing to the decay will
be that of figure 6.2, and one can ask how much this will change the phenomenology. It is difficult
to say is this type of spectra is more or less likely to be seen at the LHC, as the parameter space
of SUSY is vast. To study the phenomenology when $L_{\tilde{a}\tilde{g}g}$ dominates, the $L_{\tilde{a}\tilde{q}q}$ coupling was simply
turned off in the model file rather then trying to find a separate benchmark for which  $L_{\tilde{a}\tilde{g}g}$ dominates.
Specifically the benchmark used is from the same set as the other two and also has a bino NLSP. The bino in this model is a bit heavier (850GeV) to make the kinematic differences between distributions more noticeable, but it is still light enough that the branching to heavy quarks is suppressed. This
methodology should not be overly concerning since it has been shown in the last sections how the
general distribution shape does not change much, with the general feature that heavier NLSP give
more distinct kinematic differences between models. 
	
	As in the study of $L_{\tilde{a}\tilde{q}q}$ , ``impostor" events are
generated for the alternative scenarios with gravitinos and RPV. In this case the competing events
are generated with the requirement that there are three jets per decay that pass generator level
cuts. This is equivalent to requiring that at least one extra jet from QCD radiation is sufficiently
hard. Examples of these topologies with extra radiation are shown in figures \ref{fig:grv3jet} and \ref{fig:rpv3jet}. The
possible issue of jet matching arises when generating additional radiation with MadGraph before
sending events to the parton shower generator. The issue is avoided here however, since there are
no other processes in either of the models that can generate three jets and MET. What might be
more worrisome is the issue of how the generator cuts behave in conjunction with evchain when
requiring additional radiation at the MadGraph level. Since the MadGraph cuts are Lorentz boosted by
evchain, and the amount of boosting changes event by event, in effect the requirement of a third
hard jet is inconsistent between events.  More sophisticated jet cuts may be able
to resolve this, or perhaps a future version of evchain will automate this complication. 

	Looking at the new diagrams which require three jets, the topology is largely unchanged, so one might expect
many of the strategies from the previous sections still hold now. In the $L_{\tilde{a}\tilde{g}g}$  case the gravitino is still emitted in the beginning of its decay chain, and the axino at the end of its own, so the generic kinematic features seen in figures \ref{fig:grvmet}, \ref{fig:grvht} and \ref{fig:grvhtmet} should still hold and this is reflected in \ref{fig:3jethtmet}. The RPV case  is still very difficult to distinguish from the axino case using these simple variables as can be seen in \ref{fig:3jethtmet}. The additional radiation in the gravitino case does nothing to effect the contribution of the Z resonance, so reconstructing the resonance is still a valid strategy, as shown in figure  \ref{fig:grvjim2}.

	The one distribution expected to change by switching to the  $L_{\tilde{a}\tilde{g}g}$ coupling is the number of jets. Since their are more jets at tree level for the axino case, there are expected to be more jets after showering/clustering as well. The additional jets required jets from radiation in the background scenario here must be sufficiently hard to pass generator cuts but they may still be soft enough, or close enough to the other jets that they are simply reabsorbed by the other jets during the jet clustering algorithm. Comparing figure \ref{fig:numjets} to figures \ref{fig:grvjets} and \ref{fig:rpvjets} the number of jets is still a good (if not better) handle for distinguishing axinos from gravitinos, but again RPV with an LQD type coupling remains difficult. 

	In requiring additional radiation in the RPV events there arises additional possibility for distinguishing these decays from similar decays with an axino. In the case with two jets per neutralino, figures \ref{fig:ax2jet} and ~\ref{fig:rpvlqd} the topologies were identical, but the additional radiation in the RPV events (figure \ref{fig:rpv3jet} can appear in multiple in multiple places, differentiating it from the sole topology of figure ~\ref{fig:ax3jet}. In attempt to see an effect from this, one can look for invariant mass combinations of specific sets of jets, as shown in figure ~\ref{fig:rpvjim}. The left two plots of figure \ref{fig:rpvjim} take invariant mass combinations of jets based on where they originate in the Feynman diagrams. This sort of jet ordering is known to MadGraph, but it is not something that can be physically detected, so instead, in the right plot the jets are ordered by their transverse momentum and shows the invariant mass combination of the hardest and softest (of a sample with exactly 3 three jets). That the right plot looks like a combination of the left two is not coincidence, the spiky left tail of the left most plot means those particular jets are likely to be the softest in the event, and the resulting distributions show a distinct shape difference between RPV and axino cases, the RPV being bimodal. This type of analysis was only done for neutralino decays only and not for the full event with showering and clustering. The good news is that such a strategy will be immune to Lorentz boosts just as the Z reconstruction strategy for distinguishing gravitinos. The bad news is that it in going to the full event with shower and clustering the jet selection criteria become more complicated, but this still seems to be a promising possibility.

The most important effect from varying the SUSY mass spectra and
mixings is in how it affects the width. The determination of the effect of the neutralino
mixing on the width is straight forward when looking at the Feynman
rules for diagrams like those in figure~\ref{fig:ax3jet} and
\ref{fig:ax2jet}.  In both cases, the neutralino decay chain begins with
an off-shell squark, but only the wino and bino components of the
neutralino will couple to the squark. The smaller these components,
the smaller the total width will become.  The next chapter will study in more detail 
the decay width for ${\tilde \chi^{0}_1} \to q \bar q \tilde a$.

\begin{figure}
\centerline{\includegraphics[scale=0.25]{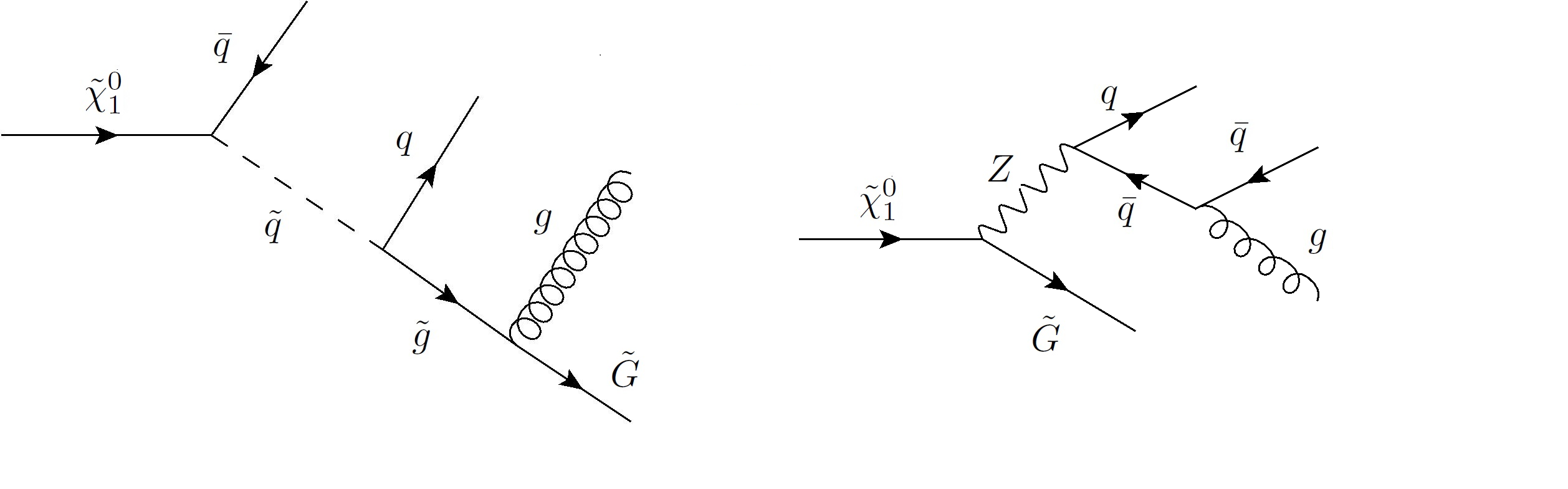}}
\caption{Examples of gravitino topologies for neutralino decay when three jets are required. \label{fig:grv3jet}}
\end{figure}

\begin{figure}
\centerline{\includegraphics[scale=0.3]{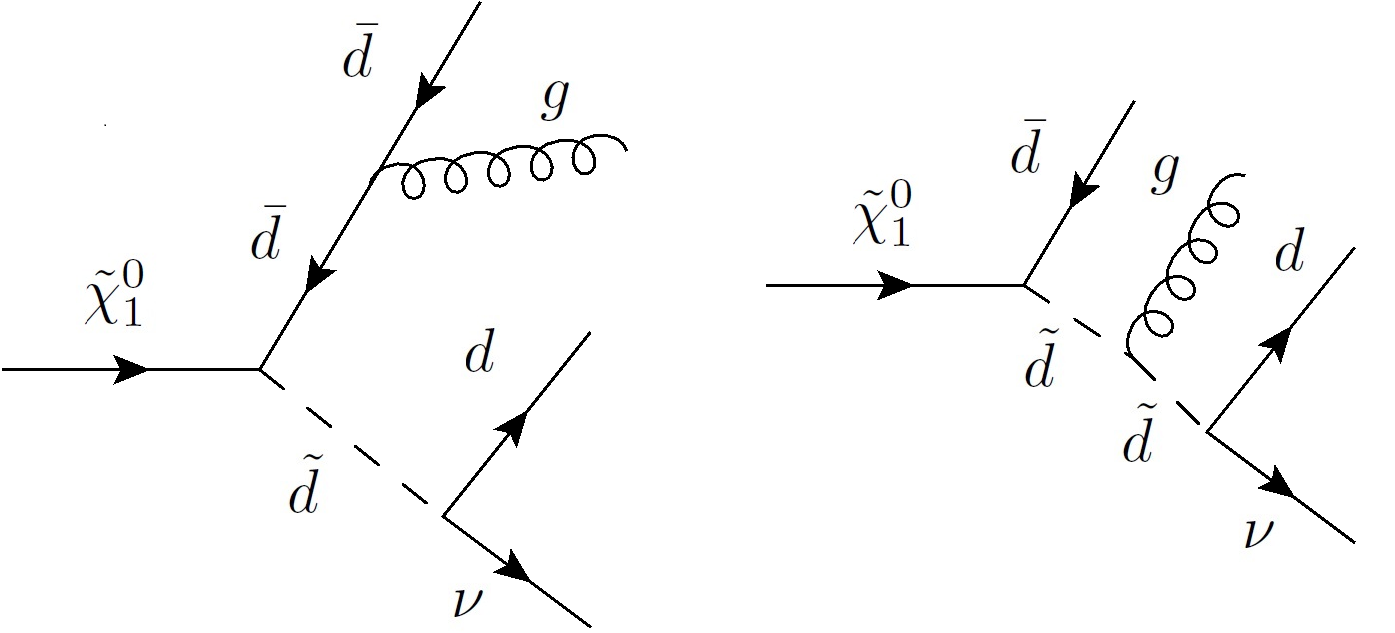}}
\caption{Examples of RPV topologies for neutralino decay with an LQD type coupling when three jets are required. \label{fig:rpv3jet}}
\end{figure}

\begin{figure}
\centerline{\includegraphics[width=5.5in]{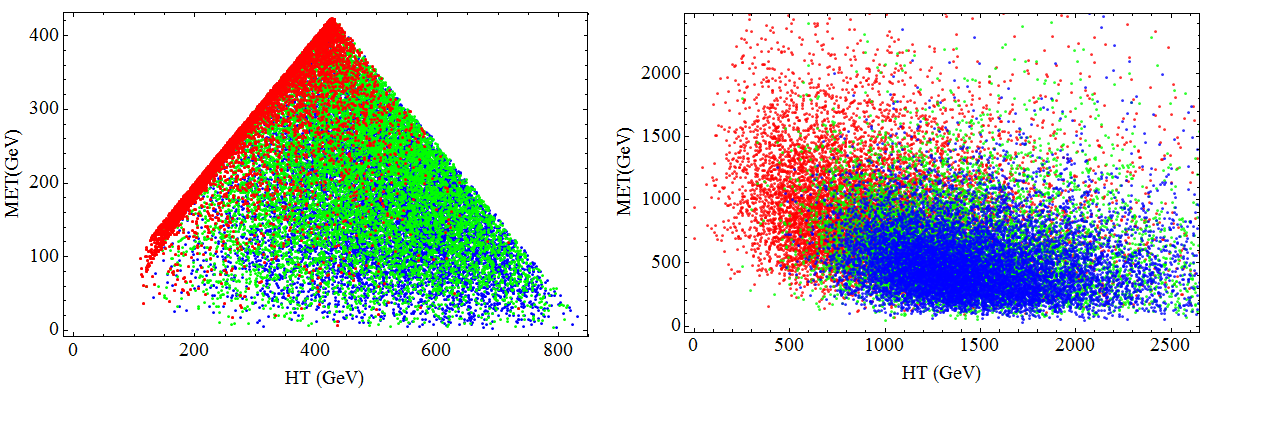}}
\caption{$H_T$ versus MET from neutralino decays to axinos (blue),
  gravitinos (red), and RPV (green).  Events are simulated with minimal generation cuts only, and at parton level (no showering/clustering). The left plots  consider the neutralino decay alone, while the right plots are in the lab frame
of the whole event at 14 TeV at the LHC, i.e. when including both production and decay via evchain.  Decays to axinos are produced only with the gluon/gluino coupling contributing. The gravitino and RPV comparison events are produced requiring exactly 3 hard jets per neutralino.  \label{fig:3jethtmet}}
\end{figure}

\begin{figure}
\centerline{\includegraphics[width=5.5in]{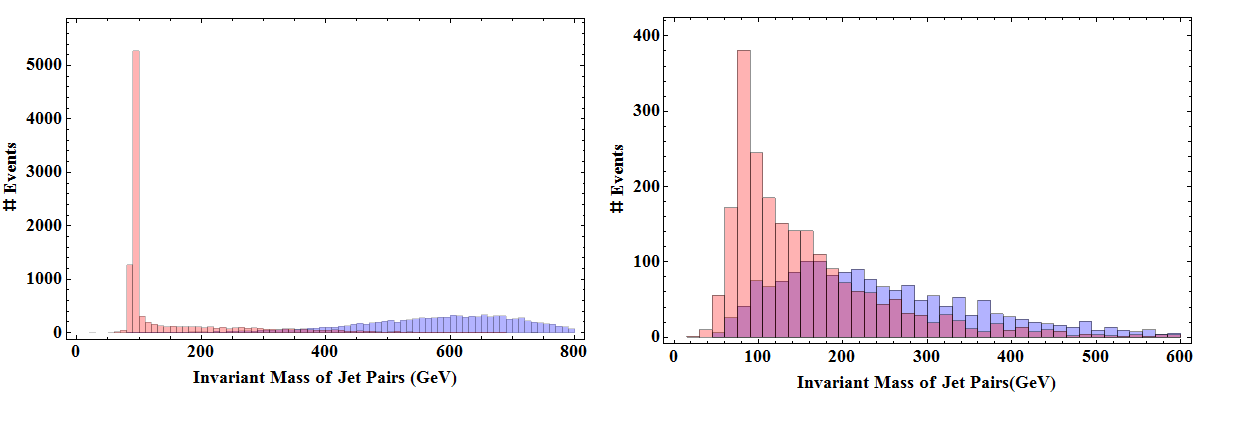}}
\caption{The invariant mass of jet pairs for neutralino decays to axinos (blue)
and gravitinos (red). The left plot is for a single decaying neutralino in
its rest frame, while the right is in the lab frame for the whole
event (in a sample selected to have exactly four jets) at 14 TeV with showering done by PYTHIA and jet clustering from FastJet using kT jets with D=0.4.  Decays to axinos are produced only with the gluon/gluino coupling contributing. The gravitino comparison events are produced requiring exactly 3 hard jets per neutralino (that pass generation cuts) at the MadGraph level before any clustering. \label{fig:grvjim2}}
\end{figure}

\begin{figure}
\centerline{\includegraphics[width=5.5in]{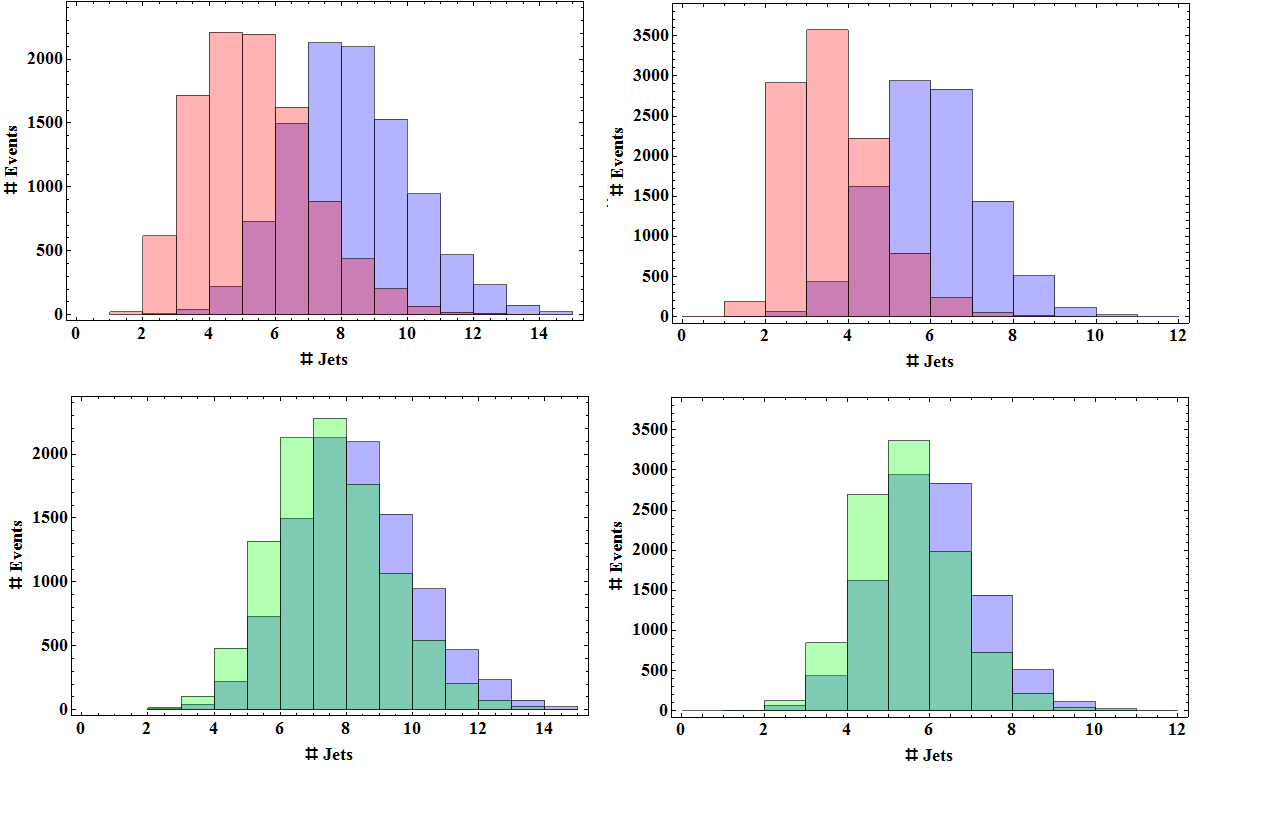}}
\caption{The number of jets from neutralino decays to axinos (blue), gravitinos
(red), and to neutrinos via RPV (green). Decays to axinos are produced only with the gluon/gluino coupling contributing. The RPV and gravitino comparison events are produced requiring exactly 3 hard jets  per neutralino (that pass generation cuts) at the MadGraph level before any clustering. All plots are for the full event (production and decay via evchain) at 14 TeV with showering done by PYTHIA and jet clustering from FastJet using kT jets with D=0.4. The left plots are from events generated with loose generation cuts and the right plots
are obtained after applying more restrictive cuts, $p_T>40$~GeV and $|\eta|<2.5$. \label{fig:numjets}}
\end{figure}

Overall,  we found that the KSVZ axino in the window of smaller $f_{a}$ has a rather
predictive signal. The multiple displaced jets and missing energy
signature is not unique, but can at least in principal be distinguished
from the more well studied alternatives for neutralino decays and
the signal is not particularly sensitive to the choice of the PMSSM benchmark
model. The same general strategies for distinquishing signals seem to be applicable regardless of whether the $L_{\tilde{a}\tilde{g}g}$ or  the $L_{\tilde{a}\tilde{q}q}$ coupling dominates, with the exception of one additional strategy for RPV in the case of $L_{\tilde{a}\tilde{g}g}$. Since the strategies are generally the same between the two regimes of couplings the same strategies may also be valid when both couplings contribute to neutralino decays with a significant branching fraction. The method of reconstructing the Z resonance will certaintly work in this case, but the other kinematic distributions would have to be looked at more carefully with a jet matching scheme to avoid double counting between MadGraph and PYTHIA.

\begin{figure}
\centerline{\includegraphics[width=6in]{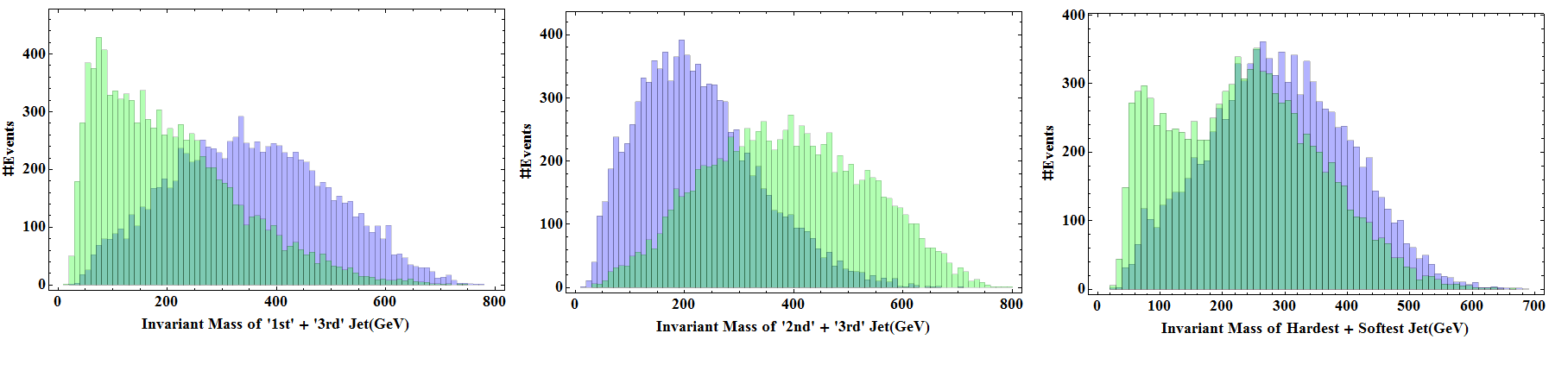}}
\caption{Invariant mass combinations that can be used to discriminate between neutralino decays to axinos (blue) and RPV with an LQD type coupling (green). Decays to axinos are produced only with the gluon/gluino coupling contributing. The left two plots use jet numbering  based on topology, which is not measurable by a detector. The right plot uses jets orderd by PT so a specific combinatoin of jet invariant mass in meaningful. Events are simulated for the decay only with minimal generation cuts only, and at parton level (no showering/clustering). The RPV comparison events are produced requiring exactly 3 hard jets per neutralino.
\label{fig:rpvjim}}
\end{figure}
\chapter{Decay Width of Neutralinos}
\label{ch:width}

\section{Analytic Expression of the Width}
\label{sec:analytic}

The neutralino decay width of the process in figure \ref{fig:ax2jet} 
for massless quarks and with universal squark masses $m_{\tilde q}=m_{\tilde q_L}=m_{\tilde q_R}$ is given by~\cite{Barnett:1987kn}
\begin{eqnarray} \label{eq:width}
\Gamma(\tilde\chi_j^{0} \to q \bar q \tilde a) &=& \frac{m_{\chi^{(0)}_j} \alpha g_{\mathrm{eff}}^2}{64 \pi^2 \sin^2\theta_w} 
\frac{3}{2} \sum_q 16 [(T_{3q} Z_{j2}+(Q_q-T_{3q}) Z_{j1} \tan\theta_w)^2 + Q_q^2 Z_{j1}^2 \tan^2\theta_w] 
\nonumber \\ 
& \times &  [g(m_{\tilde a}^2/m_{\tilde \chi^{(0)}_j}^2,m_{\tilde q}^2/m_{\tilde \chi^{(0)}_j}^2)+
h(m_{\tilde a}^2/m_{\tilde \chi^{(0)}_j}^2,m_{\tilde q}^2/m_{\tilde \chi^{(0)}_j}^2)] \; ,
\end{eqnarray}
where $Z_{ji}$ are the matrix elements of the matrix which
diagonalizes the neutralino mass matrix, $\theta_w$ is the weak
mixing angle, and $Q_q=(2/3,-1/3), T_{3q}=(1/2,-1/2)$ for (up,down)-type quarks. The effective coupling $g_{\mathrm{eff}}$ is given by Eq.~\ref{eq:geff} and the functions $g,h$ are provided
in~\cite{Barnett:1987kn}. The process studied in \cite{Barnett:1987kn} is actually the decay of gluinos, as in figure \ref{fig:gludecay} , but the topology and general structure of the couplings is the same and all that needs to be changed is the numerical coupling constants. As mentioned in chapter 7, this analytic expression for the width was used to validate the FeynRules implementation of the axinos interactions.

\begin{figure}
\centerline{\includegraphics[scale=0.3]{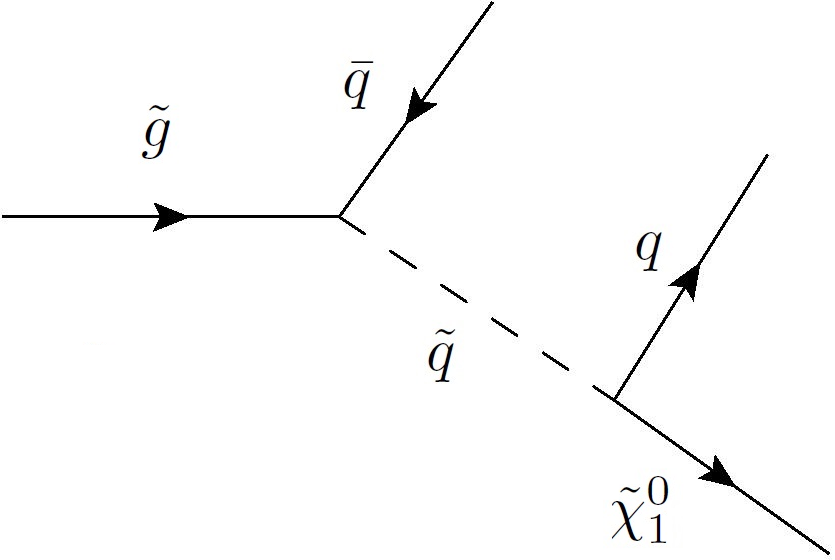}} \caption{Feynman diagram for gluino decay. This has the same topology and general form of couplings as the neutralino decay to two jets and an axino, aside from the constants.\label{fig:gludecay}}
\end{figure}

It was emphasized in chapter \ref{ch:signal} that the displaced multi-jet and MET signal
was the only one that needed be considered for decays to axinos, but
one can imagine that with the lightest neutralino as a sufficiently
pure Higgsino that diagrams like in figures~\ref{fig:ax2jet} and \ref{fig:ax3jet} would be suppressed
enough that another decay channel can dominate. While other axino
channels can have a larger partial width than the 2 and 3-jet channels
for a very pure higgsino, these channels are still very much suppressed
themselves, as they will contain additional final-state particles
or additional off-shell sparticles or both. An example of such a process is shown in figure \ref{fig:axlong}. This possibilty was explored
for the lighter benchmark only by varying the mixing parameters and
retaining the same mass spectrum. For this benchmark case the alternative
channels, containing additional gauge bosons or a Higgs boson, only began
to become competitive with the multijet channels once the decay length
was already several orders of magnitude larger than the detector (hundreds
of kilometers instead of meters). Even though this possibility was
only explored for a single benchmark, it seems unlikely that any choice
of spectra could reduce the decay length by enough that it would matter
to the phenomenology. The effect of varying the Peccei-Quinn scale $f_a$
over the allowed window is also relatively straightforward (see Eq.~\ref{eq:width}). How exactly
the neutralino width scales with $f_{a}$ will depend on which of
the two couplings is dominant, but in either case the width will vary
by about two orders of magnitude.

\begin{figure}
\centerline{\includegraphics[scale=0.3]{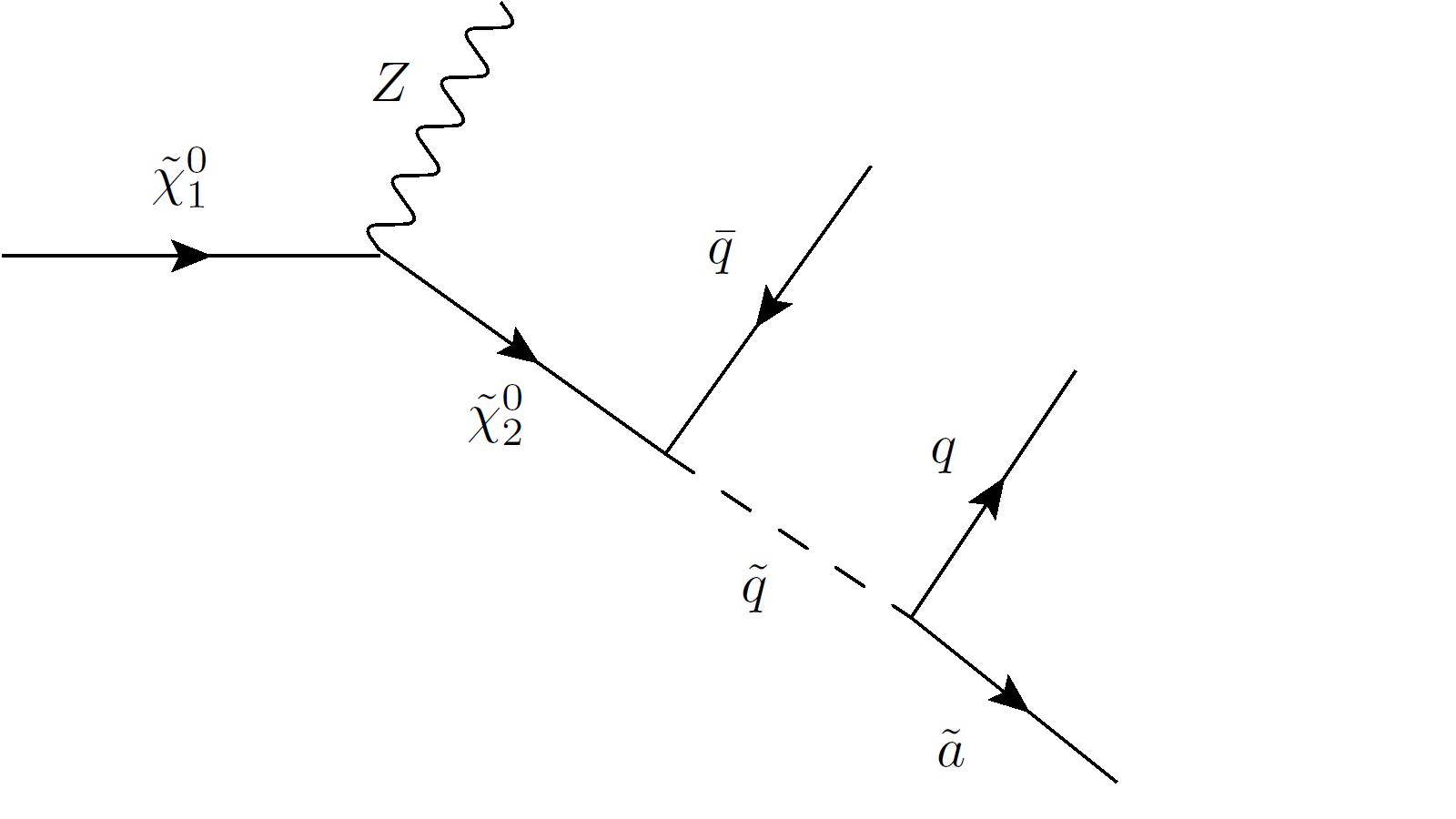}} \caption{An example of a longer decay path from neutralino to axino, which may become dominant for the correct neutralino mixings.\label{fig:axlong}}
\end{figure}

\section{Parameter Space for Collider Searches}
\label{sec:searchparam}

The other parameters affecting the width are all sparticle masses: The
axino mass, the neutralino mass, the squark mass and the gluino
mass. In chapter~\ref{ch:signal} it was stated that several of the
Snowmass PMSSM benchmarks from the collection in
\cite{Cahill-Rowley:2013gca} allowed for neutralino decays to axinos
with a decay length appropriate for searches at the LHC, even though
only two were chosen for simulation, and it seems as though this
scenario could be rather common for SUSY models with sparticle masses
in the range that is explorable in the near future. With the decay
width of neutralinos to axinos depending on so many variables it is
difficult to bound exactly what the model space is available to such
searches, but figures~\ref{fig:widthneutglu} through ~\ref{fig:widthglusquark}
make an attempt of demonstrating what range of SUSY parameters would
allow for this type of signal. Each is a plane in parameter space that
shows contours of equal neutralino decay length $c\tau$ for the 2-jet
plus axino signal only, as given in Eq.~\ref{eq:width}.  For very
heavy neutralinos the channel to heavy quarks opens up, and for
squarks much heavier than gluinos the 3-jet signal will start to
become competitive and eventually dominate and neither of these effects is reflected in these plots.  In each of these plots
the neutralino is taken to be a very pure bino and the Peccei-Quinn
scale $f_{a}$ takes its lowest value in the allowed window, so that
these planes of parameters space are already at there ``least
displaced'' for these parameters. These plots also show for what SUSY
masses prompt decays to axinos may be possible, though this signal
comes with its own set of challenges that are not discussed here.

\begin{figure}
\centerline{\includegraphics[width=4.5in]{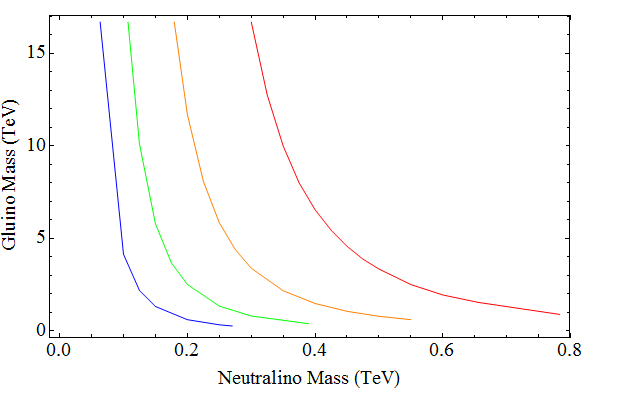}}
\caption{Contours of constant neutralino decay length $c\tau$ for decays to an axino and two
jets.  Only the coupling  ${\cal L}_{\tilde{a}q\tilde{q}}$ is considered here.  Red is 0.01m, yellow is 0.1m, green is 1m and blue is 10m. All
squarks are at a mass of 2 TeV and the axino is taken to be massless. \label{fig:widthneutglu}}
\end{figure}

\begin{figure}
\centerline{\includegraphics[width=4.5in]{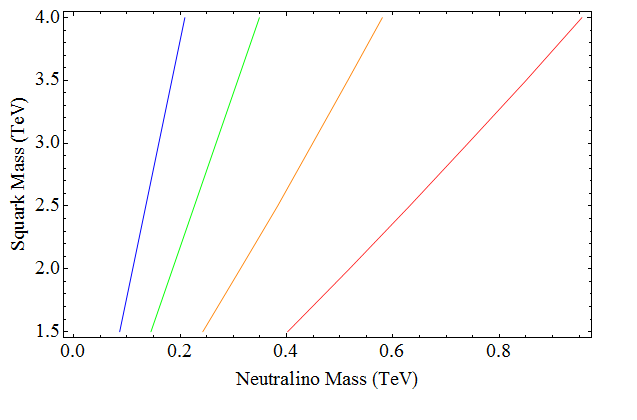}}
\caption{Contours of constant neutralino decay length $c\tau$ for decays to an axino and two
jets.  Only the coupling  ${\cal L}_{\tilde{a}q\tilde{q}}$ is considered here. Red is 0.01m, yellow is 0.1m, green is 1m and blue is 10m. The
gluino mass here is 3 TeV and the axino is taken to be massless. \label{fig:widthneutsquark}}
\end{figure}

\begin{figure}
\centerline{\includegraphics[width=4.5in]{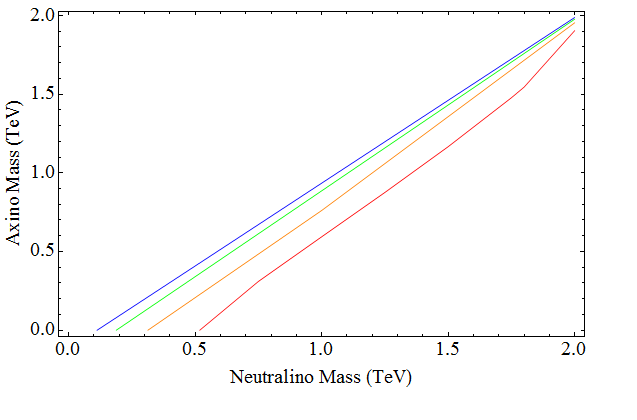}} \caption{Contours of constant neutralino decay length $c\tau$ for decays to an axino and two
jets. Only the coupling  ${\cal L}_{\tilde{a}q\tilde{q}}$ is considered here. Red is 0.01m, yellow is 0.1m, green is 1m and blue is 10m. The
gluino mass here is 3 TeV and all squark masses are at 2 TeV. \label{fig:widthneutax}}
\end{figure}

\begin{figure}
\centerline{\includegraphics[width=4.5in]{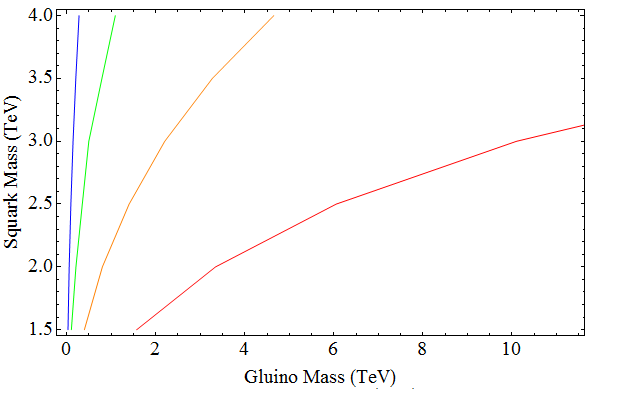}}
\caption{Contours of constant neutralino decay length $c\tau$ for decays to an axino and two
jets.  Only the coupling  ${\cal L}_{\tilde{a}q\tilde{q}}$ is considered here. Red is 0.01m, yellow is 0.1m, green is 1m and blue is 10m. The
neutralino mass here is 0.5TeV and the axino is taken to be massless. \label{fig:widthglusquark}}
\end{figure}

\section{Compressed Spectra}
\label{sec:compressed}

	A compressed spectrum (with a neutralino NLSP mass close to the gluino
mass) will make the off-shell decay to axinos easier, resulting in a
shorter mean decay length. The spectra can become very compressed and
the decay will still be displaced, (especially for larger values of
$f_{a}$), perhaps providing an easier discovery channel than otherwise
available for a compressed spectrum. For less compressed spectra, or
spectra with a lighter neutralino, the mean decay length increases.
When considering a larger parameter space of SUSY models (as the PMSSM
does), compressed spectra are not uncommon \cite{LeCompte:2011cn}.
Compressed spectra can evade many conventional SUSY searches because many of the kinematic variables that are normally triggered on will reduced greatly, such as the visible transverse momentum of jets or their scalar sum (the HT). While more compressed spectra can be very difficult to search for at
colliders \cite{Dreiner:2012gx}, in this case the primary effect on
the signal is on the total width of the NLSP, and typically more
compressed spectra will simply have the displaced jets closer to the
primary vertex (with some fraction still being more displaced). The width does change very quickly in this regime also, making it very sensitive to the gluino mass. This can be seen from the bottom of figure \ref{fig:widthneutglu}, where a line of compressed spectra with degenerate neutralino/gluino would be nearly flat at the scale of the plot sloping up and to the right.  In effect this search stategy for compressed spectra works because the axino is the true LSP and is not itself compressed with the rest of the spectra. It is not unreasonable to think the axino would be seperated from the rest of the spectra in this way since its mass is model dependent and because lighter axinos LSPs may be favored cosmologically. This
is an interesting scenario in of itself, implying that an otherwise
difficult to study spectrum at the LHC may have axino production as
its discovery channel.
\chapter{Possible Cosmologies}
\label{ch:cosmo}

\section{Challenges to a Working Cosmology}
\label{sec:Challenges}

There are also many unanswered questions concerning the cosmology of
such a model. In this work, it is only attempted to argue that there are
enough parameters that can be adjusted and that such a model has all
the right ingredients for a working cosmology, but this does not
guarantee such a cosmology exists. Besides simply getting the relic abundance of dark matter correct, a working cosmology should not spoil any of our other observations, such as measurements of the large scale structure requiring predominately cold dark matter, or measurements of the CMB that separately put constraints on the number of light relativistic species in the universe (often called the effective number of neutrinos). There is also the possibility that the decays of extra heavy states in the early universe can spoil big bang nucelosythesis (BBN). Spoiling BBN is a common issue with models that have an intermediate mass gravitino \cite{Asaka:2000zh}, either from their own late decays, or the late decays of neutralinos to a lighter gravitino. The problem is so common in models that it is simply referred to as the gravitino problem. Axinos can run into the same issue, where their presence allows late decays that spoil BBN. There are however also models that avoid a gravitino problem by introducing an axino \cite{Asaka:2000ew}. 
	It seems as though a low reheat temperature is required to make this scenario viable, otherwise there are too many hot axions thermally produced to be consistent with observations of large scale structure.  The process of inflation reduces the temperature in the early universe by many orders of magnitude and the process of reheating occurs through the weak interactions between the inflaton and the Standard Model particles. The reheat temperature is the temperature of the thermal plasma bath of the early universe, after the end of inflation and at the onset of the radiation dominated phase. In cosmologies with a lower reheat temperature, where radiation domination starts at a lower energies, thermal dark matter relics of any type will be reduced in abundance. A lower temperature will produce a smaller number density of thermal relics, but also these scenarios can often involve additional entropy generation prior to radiation domination, to further dilute the thermal relics number density. The recent BICEP
result~\cite{Ade:2014xna} may make such a cosmology more attractive (if at least
some of the signal is not dust ~\cite{Ade:2015tva}). If a larger tensor-to-scalar ratio
$r$ is found, then high $f_{a}$ axions become constrained unless there
is a low reheat temperature, as in this scenario\cite{Marsh:2014qoa}. One drawback of low reheat cosmology is that it makes some popular methods of baryogenesis very difficult, but there are alternatives, such as Affleck-Dine (AD) baryognesis. AD Baryogensis may be even more appealing in a scenario with collider detectable axinos for reasons discussed in earlier chapters. A class of AD baryogenesis schemes make use of RPV couplings, and the presence of RPV couplings means another dark matter candidate is required besides the neutralino. With a KSVZ axino in the hadronic axion window, axino LSPs will be stable even with the RPV couplings, and in addition to this a coincidence of scales make may RPV and axinos simultaneously observable in collider experiments.

\section{Flexiblity of Scenario}
\label{sec:flex}
Throughout this work, only those parameters necessary for the collider phenomenology were specified, and this leaves a number of parameters flexible which can greatly affect the cosmology.
Whether or not there are RPV couplings present can affect
both the axino and axion abundance. There are several options for what
types of RPV couplings there are (if any) and the size of each
coupling has a wide allowed range. The axion/axino cosmology is more
sensitive to the choice of RPV couplings in this case, because the
axion/axino couplings are restricted in the hadronic axion window, and
so there are not as many options for decay chains. If AD Baryogenesis
with RPV is implemented in the model, then this limits the
possibilities for the type and size of RPV couplings.
The size of axion and axino dark matter abundances depends on a number
of factors that have not been specified here, either because they do
not affect the collider signal, or the implications of their inclusion
require further study. The phenomenology here is relatively
insensitive to the axino mass, which will effect the size of its
abundance and how relativistic it is. Late decays of the saxion can
effect the size of both the axion and axino population, and can inject
extra entropy into the early universe to dilute these species, so the
role of the saxion is non trivial and requires further study in this
scenario.  The gravitino was assumed to be heavy enough not to affect
the collider phenomenology in this scenario, but it too could play a
more complicated role. A light enough gravitino can have a comparable
coupling with the LOSP as the axino (as shown in
chapter~\ref{ch:results}), but can also be coupled more strongly or
weekly depending on its mass. While only an LSP gravitino is likely to
impact the collider phenomenology discussed here, an intermediate mass
gravitino can still affect the cosmology with late decays to other
states.

\section{Towards a Realistic Model}
\label{sec:realistic}

 It is also possible that the gravitino is the LSP, but with a mass still too
large to affect the collider phenomenology, so that distributions of chapter 7 are still seen, but well
outside of the detector the axino eventually decays to a gravitino. Multiple low reheat scenarios with
an axino NLSP and a gravitino LSP close in mass have been proposd in \cite{Cheung:2011mg} and this may be the best starting
point for building realistic cosmologies that are consistent with the scenario proposed in this thesis.
The authors of \cite{Cheung:2011mg} also argue that the axino must be heavier then the gravitino unless there is fine tuning and if this argument is to be accepted then these cosmologies become much more attractive
if one wants supersymmetry with a Peccei-Quinn axion. The scenarios proposed in \cite{Cheung:2011mg} can provide
the correct amount of dark matter and evade cosmological bounds for gravitino and axino masses over a range of several orders of magnitude, including the axino mass range explored in this thesis.
Whether or not these types of scenarios are still viable for the much lower $f_{a}$ used here requires
further investigation.
 To properly take
into account all these considerations and assemble a working cosmology
from them is beyond the scope of this work.

\chapter{Conclusion}
\label{sec:conclusion}

Supersymmetric models with axions and axinos are very attractive
extensions of the SM since they can address issues of naturalness in
QCD and the electroweak sector, and with regard to dark matter.  The
one feature of these types of models that could be considered
disappointing is that the additional particles, the axion and the
axino, can be rather difficult to detect. The scenario proposed here
is a PMSSM model with a light LSP axino with only QCD couplings and a
neutralino NLSP. The signal studied here is the production of
neutralinos and their displaced decay to two jets and a KSVZ axino via
an effective squark-quark-axino coupling.  This scenario could be
detectable at the 14 TeV LHC provided the Peccei-Quinn scale can exist
in the smaller range $3\times10^{5}\, {\rm GeV}<f_{a}<3\times10^{6}\,
{\rm GeV}$ (hadronic axino window).  The possibility of  sneutrino
NLSPs still requires more study, in which case the topologies for NLSP decays becomes more
varied and can include photons and charged leptons, making the
phenomenology more complicated, especially in distinguishing from RPV
and gravitino scenarios.

The scenario of the hadronic axion window is not new, and its
cosmology has been discussed in the literature (see, e.g.,~\cite{Moroi:1998qs}), but the consequences of having this window in a SUSY model
have not been explored until recently, and there is still much
to learn.  This is not the only scenario that allows axinos to be
detectable at colliders, but to the author's best knowledge it is the
only way currently proposed to detect KSVZ axinos with a neutral
NLSP. This scenario gives a predictive collider signature due to its
limited couplings. It has been shown in this work  that such as model has the potential to be
distinguishable from similiar models with neutralino decay, and that
this signature is relatively insensitive to the choices of SUSY
parameters. While  there is potential for the LHC to be
sensitive to the scenario studied here, a detailed detector simulation
that implements for instance the triggers used in hidden valley
searches~\cite{Aad:2013txa} is needed to fully assess its observability.

It is interesting to probe the hadronic axion window via collider
searches for a variety of reasons. While it has been argued extensively
in the literature that there can be a variety of benefits to having
SUSY with axions, there are very few ways to test the axion
coupling $f_{a}$ independent of its photon or electron coupling, which
this scenario allows for. While there is still much to learn about
this scenario, there are tentative hints that it could have attractive
features beyond a detectable axino. It may also provide a discovery
channel for otherwise difficult to study compressed SUSY spectra, it may alleviate
some issues of fine tuning, and the cosmology it fits in may have other
interesting consequences, such as evading axion bounds for larger
values of $r$ or perhaps detectable RPV decays that are competitive
with decays to axinos. This scenario is still very new, both for collider
studies and for cosmology, and much more work is required to determine
its viability, detectability and consequences.
\end{ubmainmatter}

\begin{ubbackmatter}
\references{thesis}
\end{ubbackmatter}

\end{document}